\documentclass[]{aa}

\usepackage{natbib,color}
\usepackage[breaklinks=true]{hyperref}
\usepackage{times}
\usepackage{amsmath,amsfonts}
\usepackage{mathtools,url}
\usepackage{scalerel,stackengine}

\usepackage{graphicx}
\usepackage{txfonts}

\stackMath
\newcommand\reallywidehat[1]{
\savestack{\tmpbox}{\stretchto{
  \scaleto{
    \scalerel*[\widthof{\ensuremath{#1}}]{\kern-.6pt\bigwedge\kern-.6pt}
    {\rule[-\textheight/2]{1ex}{\textheight}}
  }{\textheight}
}{0.5ex}}
\stackon[1pt]{#1}{\tmpbox}
}
\def\Bonn{1}
\def\Leiden{2}
\def\UBC{3}
\def\UCL{4}
\def\Edinburgh{5}
\def\NRCC{6}
\def\AIM{7}
\def\CADC{8}
\def\Bochum{9}
\def\Princeton{10}
\def\SantaCruz{11}
\def\Rio{12}
\def\BuenosAires{13}
\def\LERMA{14}
\def\JPL{15}
\def\Oxford{16}
\def\EPFL{17}

\defcitealias{shh15}{S15}
\defcitealias{mhb06}{M06}

\begin{document}
\title {Tightening weak lensing constraints on the ellipticity of galaxy-scale dark matter haloes}
\titlerunning{Constraining galaxy halo ellipticity with weak lensing}
   \author{Tim Schrabback\inst{\Bonn,*}
     \and
     Henk Hoekstra\inst{\Leiden}
     \and
     Ludovic Van Waerbeke\inst{\UBC}
     \and
     Edo van Uitert\inst{\UCL}
     \and
     Christos Georgiou\inst{\Leiden}
     \and
     Marika Asgari\inst{\Edinburgh}
     \and
     Patrick C\^ot\'e\inst{\NRCC}
     \and
     Jean-Charles Cuillandre\inst{\AIM}
     \and
     Thomas Erben\inst{\Bonn}
     \and
     Laura Ferrarese\inst{\NRCC}
     \and
     Stephen D.J.~Gwyn\inst{\CADC}
     \and
     Catherine Heymans\inst{\Edinburgh,\Bochum}
     \and
     Hendrik Hildebrandt\inst{\Bochum}
     \and
     Arun Kannawadi\inst{\Leiden,\Princeton}
     \and
     Konrad Kuijken\inst{\Leiden}
     \and
     Alexie Leauthaud\inst{\SantaCruz}
     \and
     Martin Makler\inst{\Rio,\BuenosAires}
     \and
     Simona Mei\inst{\LERMA,\JPL}
     \and
     Lance Miller\inst{\Oxford}
     \and
     Anand Raichoor\inst{\EPFL}
     \and
     Peter Schneider\inst{\Bonn}
     \and
     Angus Wright\inst{\Bochum}
           }
   \institute{Argelander-Institut f\"{u}r Astronomie, Universit\"{a}t Bonn, Auf dem
     H\"{u}gel 71, 53121, Bonn, Germany\\
    $^*$:           \email{schrabba@astro.uni-bonn.de}
               \and
 Leiden Observatory, Leiden University, Niels Bohrweg 2, NL-2333 CA Leiden, The Netherlands
              \and
Department of Physics and Astronomy, University of British Columbia, 6224 Agricultural Rd, Vancouver BC, V6T 1Z1, Canada
\and
Department of Physics and Astronomy, University College London, Gower Street, London WC1E 6BT, UK
\and
Institute for Astronomy, University of Edinburgh, Royal Observatory, Blackford Hill, Edinburgh, EH9 3HJ, U.K.
\and
National Research Council of Canada,
Astronomy and Astrophysics Research Centre,
3071 West Saanich Road,
Victoria BC V9E2E7,
Canada
\and
AIM, CEA, CNRS, Universit\'e Paris-Saclay, Universit\'e Paris Diderot,
Sorbonne Paris Cit\'e, Observatoire de Paris, PSL University
F-91191 Gif-sur-Yvette Cedex, France
\and
Canadian Astronomy Data Centre, National Research Council, 5071 West Saanich Road, Victoria, BC, V9E 2E7, Canada
\and
Ruhr-University Bochum, Astronomical Institute, German Centre for Cosmological Lensing, Universit\"atsstr. 150, 44801 Bochum, Germany
\and
Department of Astrophysical Sciences, Princeton University, 4 Ivy Lane, Princeton, NJ 08544, USA
\and
Department of Astronomy and Astrophysics, University of California, Santa Cruz, 1156 High Street, Santa Cruz, CA 95064 USA
\and
Centro Brasileiro de Pesquisas F\'isicas, Rua Dr. Xavier Sigaud 150, Rio de Janeiro, RJ, 22290-180, Brazil
\and
International Center for Advanced Studies \& ICIFI (CONICET), ECyT-UNSAM,
Campus Miguelete, 25 de Mayo y Francia,
(1650) Buenos Aires, Argentina
 \and
Universit\'{e} de Paris, F-75013, Paris, France, LERMA, Observatoire de Paris, PSL Research University, CNRS, Sorbonne Universit\'e,  F-75014 Paris, France
\and
Jet Propulsion Laboratory and Cahill Center for Astronomy \& Astrophysics, California Institute of Technology, 4800 Oak Grove Drive, Pasadena, California 91011, USA
\and
Department of Physics, University of Oxford, Denys Wilkinson Building, Keble Road, Oxford, OX1 3RH, UK
\and
Institute of Physics, Laboratory of Astrophysics, Ecole Polytechnique F\'{e}d\'{e}rale de Lausanne (EPFL), Observatoire de Sauverny, 1290 Versoix, Switzerland
}
   \date{Received February 05, 2020; accepted October 16, 2020}
   \abstract{Cosmological simulations predict that galaxies are embedded into triaxial dark matter haloes,
  which appear approximately elliptical in projection.
     Weak gravitational lensing allows us to constrain these halo shapes and thereby test the nature of dark matter.
     Weak lensing has already provided robust detections of the signature of halo flattening at the mass scales of groups and clusters, whereas results for galaxies have been somewhat  inconclusive.
     Here we combine data from five weak lensing surveys (NGVSLenS,  KiDS/KV450, CFHTLenS, CS82, and RCSLenS, listed in order of most to least constraining) in order to tighten observational constraints on galaxy-scale halo ellipticity for photometrically selected lens samples.
     We constrain $f_\mathrm{h}$,
 the average ratio between the aligned component of the halo ellipticity  and the ellipticity of the light distribution,
 finding  \mbox{$f_\mathrm{h}=0.303^{+0.080}_{-0.079}$} for red lens galaxies and  \mbox{$f_\mathrm{h}=0.217^{+0.160}_{-0.159}$} for blue lens galaxies when assuming elliptical Navarro-Frenk-White
 density profiles and a linear scaling between halo ellipticity and galaxy ellipticity.
     Our constraints for red galaxies constitute the currently most significant ($3.8\sigma$) systematics-corrected detection  of the signature of halo flattening at the mass scale of galaxies.
Our results are in good agreement with expectations from
the Millennium Simulation that apply the same analysis scheme and incorporate models for  galaxy--halo misalignment.
Assuming
these misalignment models and the analysis assumptions stated above are correct, our measurements imply an average dark matter halo ellipticity for the studied red galaxy samples of
\mbox{$\langle|\epsilon_\mathrm{h}|\rangle=0.174\pm 0.046$}, where
\mbox{$ | \epsilon_\mathrm{h}|=(1-q)/(1+q)$}
relates to the ratio \mbox{$q=b/a$} of the minor and major axes of the projected mass distribution.
Similar measurements based on larger upcoming weak lensing data sets can help to calibrate models for intrinsic galaxy alignments, which constitute an important source of systematic uncertainty in cosmological weak lensing studies.
}


   \maketitle

   \section{Introduction}
  According to the current  cosmological model, galaxies, galaxy groups, and galaxy clusters are embedded in large haloes dominated by   invisible dark matter.
  Based on simulations of cosmological structure formation, we expect that the average density profiles of these haloes
   should closely follow the Navarro-Frenk-White (NFW) profile \citep[][]{nfw96,nfw97},
   and that their shapes are roughly triaxial \citep[e.g.][]{jis02,vyg17}, appearing  elliptical in projection.
   Several approaches have been used to test this prediction and constrain halo shapes as well as the relative alignment of galaxies with their haloes observationally; these approaches have included
   the use of baryonic tracers such as
   satellite galaxies \citep[e.g.][]{ojl09,agb10,hac14,scj18,gce19}
   and studies of stellar and gas kinematics in polar ring  \citep{kmk14}  or edge-on galaxies \citep{pka17}.
   Such measurements allow us to test predictions from hydrodynamical simulations, which aim to  model galaxy formation and evolution, including  the interplay, relative distribution, and alignment of the baryonic and dark matter components \citep[e.g.][]{tmm14,tmd16,lpc15,dbr15,vcs15,cjc18,cpv19,bcc20,bct20}.

   An alternative route to constrain halo ellipticity is provided by gravitational lensing, which is directly sensitive to the projected mass distribution \citep[e.g.][]{skw06}.
   Strong lensing images of highly distorted or multiply imaged background galaxies or quasars
   probe the inner  halo shapes of  galaxies \citep[e.g.][]{shk12,brc16},  clusters \citep[e.g.][]{lms13,cgr16,psc18,crg19},
   and cluster member galaxies \citep{dbb15,jhm18}.
    However, baryons have a significant impact on the mass distribution in the inner regions of galaxies and clusters.
   To study the outer, dark matter-dominated  halo shapes instead, the gravitational potential must  be probed at larger projected radii.
   This is possible with weak gravitational lensing  \citep[e.g.][]{bas01}.
   In this regime, typical distortions
   are small compared to the noise caused by the dispersion of intrinsic galaxy ellipticities.
   Therefore, the signal  can only be detected statistically by combining shape information from many background galaxies.
   For mass distributions that appear approximately elliptical in projection, the weak lensing tangential shear is stronger at a given radius along the direction of the
major axis of the  ellipse compared to the  minor axis \citep[e.g.][]{nar00,brw00}.
For massive clusters and  deep weak lensing data,
this effect can be used to constrain halo ellipticities for individual targets  \citep[e.g.][]{oto10,htj19}, yielding robust (\mbox{$\gtrsim 5\sigma$})
detections  of
halo ellipticity from joint analyses of larger samples \citep[e.g.][]{oguri12,ust18}.

 When less massive clusters or galaxies act as lenses, detections can no longer be obtained for individual targets, but one has to rely on stacking.
In order to constrain halo ellipticity via stacked weak lensing measurements,
 the shear fields first have to be aligned according to the orientations of the (mostly dark) matter haloes on the sky.
Unfortunately, these orientations are unknown, which is why one has to rely on proxies for
the halo orientation, such as the orientation of the major axis of the light distribution in case of galaxy-scale lenses \citep[e.g.][]{hyg04,phh07}. For group- and cluster-scale lenses the orientation of the brightest cluster (or group) galaxy (BCG) and the major axis of the distribution of satellite galaxies have been employed as proxies, yielding  \mbox{$\sim 3$--$5\sigma$} detections of the signature of halo ellipticity \citep[][]{evb09,clj16,vanuitert17}.
Here it is important to realise that
the measurements are only sensitive to the aligned component of the halo ellipticity,
while misalignment between the true halo orientation and the orientation proxy washes out the signal.

An important parameter for the analysis of galaxy-scale lenses
is given
by the average\footnote{In weak lensing estimates of \mbox{$f_\mathrm{h}$}, weights that depend on  $|\epsilon_\mathrm{g}|$ are typically applied in the averaging, see Sect.\thinspace\ref{se:constrain-ani}.}
aligned ratio
\begin{equation}
f_\mathrm{h}  = \langle
  \cos(2\Delta\phi_\mathrm{h,g})|\epsilon_\mathrm{h}|/|\epsilon_\mathrm{g}| \rangle
\end{equation}
of the ellipticities of
the projected halo mass distribution $\epsilon_\mathrm{h}$ and the projected lens light distribution $\epsilon_\mathrm{g}$,
with $\Delta\phi_\mathrm{h,g}$ indicating the misalignment angle between their major axes.
In the case of perfect alignment  (\mbox{$\Delta\phi_\mathrm{h,g}=0$})
\mbox{$f_\mathrm{h}$} would reduce to
the actual ellipticity ratio.
However, both numerical simulations  \citep[e.g.][]{tmm14} and studies that approximate halo shapes from the distribution of satellites
\citep[e.g.][]{ojl09} suggest that  misalignment should have a significant impact at the mass scale of galaxies.
This reduces \mbox{$f_\mathrm{h}$} and makes the detection of halo ellipticity with weak lensing challenging \citep{bet12}.

 Indeed, a robust weak lensing detection of the signature of halo ellipticity at the mass scale of galaxies is still lacking. For example,   \citet[][\citetalias{mhb06} henceforth]{mhb06} obtained \mbox{$f_\mathrm{h}=0.60 \pm 0.38$}
 for red lenses and  \mbox{$f_\mathrm{h}=-1.4^{+1.7}_{-2.0}$} for blue lenses using SDSS data and  assuming elliptical NFW density profiles.
 Similar tentative signals with significances at the \mbox{$\lesssim 2\sigma$} level were reported by
 \citet[][]{phh07,uhs12} and \citet[][\citetalias{shh15} henceforth]{shh15}.
 The formally most significant detection has so far been reported in early work by \citet{hyg04}, who found  \mbox{$f_\mathrm{h}=0.77^{+0.18}_{-0.21}$} using magnitude-selected lenses in  RCS data.
 They employed a maximum likelihood fit to the 2D shear field, which can extract more information \citep{dzh19}, but does not correct for
spurious signal that is introduced for the halo shape estimation if other effects cause an extra
alignment of lens and source galaxies.
 In particular,  \mbox{$f_\mathrm{h}$} can be underestimated due to cosmic shear and consistently under- or over-corrected PSF anisotropy \citep[\citetalias{mhb06};][]{hyg04},
 but
 it can also be overestimated due to alignments of galaxies with their extended large-scale environment
 (\citetalias{shh15}).

 Applying an approach introduced by \citetalias{mhb06} to cancel such spurious signal contributions,
 \citetalias{shh15}
obtained
 \mbox{$f_\mathrm{h}=-0.04\pm 0.25$} (\mbox{$f_\mathrm{h}=0.69_{-0.36}^{+0.37}$}) for red (blue) lenses in CFHTLenS \citep{ehm13,hek12,hwm12}. They also analysed mock data that are based on the Millennium Simulation \citep{swj05,hhw09} and incorporate galaxy--halo misalignment models \citep{jsh13}, yielding expected signals  \mbox{$f_\mathrm{h}\simeq 0.285$} for early-type (red) galaxies and \mbox{$f_\mathrm{h}\simeq 0.025$} for late-type (blue) galaxies.
 For red galaxies significantly expanded samples should therefore yield a clear detection of non-zero \mbox{$f_\mathrm{h}$} if halo ellipticities and misalignments are indeed at the expected level.
 This motivated our current study, where we expand from the \citetalias{shh15} analysis by adding observational data from four additional recent weak lensing surveys: RCSLenS \citep{hch16},  NGVSLenS \citep{ferrarese12,raichoor14,parroni17}, CS82 \citep[][]{shan14,hand15,liu15,bls17,lsh17}, and KiDS/KV450 \citep{dejong17,fenechconti17,wright19}.
 With this study we aim to achieve the first clear systematics-corrected weak lensing detection of the signature of
 halo flattening at the mass scale of galaxies.

 Obtaining observational constraints on \mbox{$f_\mathrm{h}$} at the mass scale of galaxies is also of interest
 in the context of large upcoming cosmological weak gravitational lensing surveys \citep[e.g.][]{laureijs11,lsst09}.
 Physical alignments between galaxies and their surrounding large-scale structure
introduce shape--shear correlations  \citep[e.g.][]{his04,jck15},
which  constitute a major source of systematic uncertainty when constraining cosmological parameters with  cosmological weak lensing surveys \citep[e.g.][in this context they are also referred to as `gravitational--intrinsic' (GI) alignments]{scm17,tus18}.
Their impact must therefore be carefully corrected for using theoretical modelling \citep[e.g.][]{brk07} and  calibrations from  simulations and observations \citep[e.g.][]{tmd16,ckj17,hilbert17,pjs18,bcc20} in order to not degrade the  constraining power of these surveys.
This is linked to  \mbox{$f_\mathrm{h}$} measurements in two ways: First, the signature of halo ellipticity itself contributes to  the shape--shear alignments at small scales \citep{bra07}.
Second, misalignment simultaneously reduces  \mbox{$f_\mathrm{h}$}  and  shape--shear correlations.

This paper is organised as follows: We provide an overview of the different data sets employed in our study in Sect.\thinspace\ref{se:data}.
 Sect.\thinspace\ref{se:analysis} describes the analysis and the approach used to extract the anisotropic lensing signal and halo shape signature.
We present our results in  Sect.\thinspace\ref{se:halo_shapes_results}, discuss them in a wider context in
 Sect.\thinspace\ref{se:discussions}, and conclude in
 Sect.\thinspace\ref{se:conclusions}.

In this paper all magnitudes are given in the AB system.
They have
been corrected for Galactic extinction
as detailed in the corresponding survey data release papers.
For the computation of angular diameter distances
(which affect constraints on halo masses but not on $f_\mathrm{h}$\footnote{$f_\mathrm{h}$ is computed from the ratio of the isotropic and the anisotropic lensing signals (see Sect.\thinspace\ref{se:measuring_halo_shapes}), which is why the cosmology dependence cancels out to leading order. Within the fit range the model expectation is only marginally radius dependent (see Fig.\thinspace\ref{fi:stacked_shear}), which is why the cosmology dependence of the radius can be well neglected.}),
we assume
 a standard
$\Lambda$CDM cosmology characterised by \mbox{$\Omega_\mathrm{m}=0.3$},
\mbox{$\Omega_\Lambda=0.7$}, \mbox{$H_0=70 h_{70}$ km/s/Mpc},
and \mbox{$h_{70}=1$},
as approximately consistent with recent CMB constraints \citep[e.g.][]{planck2018cosmology}.

 \section{Data}
\label{se:data}
 In our analysis we incorporate measurements from five different weak lensing surveys (CFHTLenS,
 NGVSLenS, CS82,
 RCSLenS, and KiDS/KV450), which we briefly summarise in the following subsections.
 For all of these surveys PSF-corrected weak lensing galaxy shapes were computed
 using
 \emph{lens}fit \citep{mkh07,kmh08}, employing the version from \citet{fenechconti17} for KiDS/KV450, and the version from \citet{mhk13a} for all other surveys.
 The differences between these versions are discussed in \citet{fenechconti17},
 consisting mostly of differences in the corrections for multiplicative shear measurement bias.
 Multiplicative shear measurement biases do not significantly affect halo shape constraints because these are derived from the ratio of the anisotropic to the isotropic shear signal (see Sect.\thinspace\ref{se:measuring_halo_shapes}), which would suffer the same bias.
We nevertheless apply empirical bias corrections as provided by the surveys
to reduce potential biases in the reported halo masses.
We note that additive shear measurement biases, which represent residuals from the PSF anisotropy correction,  cancel out for measurements of the isotropic shear signal. And while simple estimators of the halo shape signature \citep[e.g.][]{nar00} can be affected by  PSF anisotropy residuals, this is not the case for the systematics-corrected estimator introduced by  \citetalias{mhb06}, which is used in our analysis (see Sect.\thinspace\ref{se:constrain-ani}).

 \subsection{CFHTLenS}
 \label{se:data_cfhtlens}
The
Canada-France-Hawaii Telescope
Lensing Survey (CFHTLenS) is an analysis of data from the
Wide-component of the Canada-France-Hawaii Telescope (CFHT) Legacy Survey, covering an
effective area of 154 deg$^2$.
These images were obtained in the
$ugriz$ broad band filters using MegaCam on CFHT, reaching a
$5\sigma$ limiting magnitude
in the detection $i$-band for $2^{\prime\prime}$ apertures  of $i_\mathrm{AB} \sim 24.5-24.7$ \citep{ehm13}.
Following the image reduction using \textsc{THELI} \citep{ehl09,ehm13},
the CFHTLenS team
estimated  photometric
redshifts (photo-$z$s) using the \textsc{BPZ}
algorithm \citep{ben00,cbs06}
as described in \citet{hek12}.
Their  $i$-band  \emph{lens}fit shape measurements yielded a weighted galaxy source density of 15.1/arcmin$^2$.

\citetalias{shh15} used this now public data set\footnote{CFHTLenS: \url{http://www.cadc-ccda.hia-iha.nrc-cnrc.gc.ca/en/community/CFHTLens}} for their analysis of galaxy halo shapes, from which we expand in the current work.
For their primary results, \citetalias{shh15} limited the analysis to the 129 out of 171 fields passing the systematic tests
implemented by  \citet{hwm12} for cosmic shear measurements. To be consistent with the \citetalias{shh15} analysis we apply the same field selection here, but we note that the applied formalism for  halo shape measurement is fairly insensitive to both multiplicative and additive shear systematics (see Sect.\thinspace\ref{se:measuring_halo_shapes}).
As done by \citetalias{shh15} we use stellar mass estimates to subdivide lens galaxies.
For the CFHTLenS data the stellar mass estimates were derived using \textsc{LePhare} \citep{acm99,iam06} as described in \citet{vuh14} and \citetalias{shh15}.

\subsection{RCSLenS}
\label{se:data:rcslens}
RCSLenS is a public\footnote{RCSLenS: \url{https://www.cadc-ccda.hia-iha.nrc-cnrc.gc.ca/en/community/rcslens}} CFHTLenS-like analysis of the CFHT observations of the Red-sequence Cluster Survey 2 \citep{gilbank11} presented by \citet{hch16}.
RCSLenS covers a total unmasked  area of 571.7deg$^2$ to an $r$-band depth of \mbox{$\sim 24.3$ mag} (for a point source at $7\sigma$),
providing $r$-band galaxy shape measurements with  \emph{lens}fit for a weighted galaxy source density of 5.5/arcmin$^2$.
We limit our analysis to the 383.5 deg$^2$ of the survey that have a uniform coverage in $g$, $r$, $i$, and $z$ band, as needed for  photo-$z$ computation.
In addition to \textsc{BPZ} photo-$z$s the RCSLenS team has also released stellar mass estimates computed using  \textsc{LePhare}  \citep[see][]{hch16,chb16}.

\subsection{NGVSLenS}
The Next Generation Virgo Survey \citep[NGVS,][]{ferrarese12} has covered 104 deg$^2$ using MegaCam on CFHT in the $ugiz$ broad band filters (\mbox{$\sim 30\%$} of the area has additional $r$-band coverage).
Reaching a $5\sigma$ limiting magnitude of 24.4 in $2^{\prime\prime}$ apertures \citep{raichoor14}, the $i$-band imaging  was obtained under superb \mbox{$<0\farcs6$} seeing conditions,
yielding a high effective weak lensing galaxy source density of 24/arcmin$^2$.
In our analysis we also employ photometric redshifts and stellar mass estimates
computed by the
 NGVSLenS team using \textsc{BPZ} \citep[see][]{raichoor14}.

\subsection{CS82}
The CFHT/MegaCam Stripe 82 Survey
\citep[CS82,][]{shan14,hand15,liu15,bls17,lsh17}
obtained excellent-seeing (0\farcs59 median FWHM of the PSF) $i$-band imaging
to \mbox{$i\sim 24.1$} ($5\sigma$ in $2^{\prime\prime}$ apertures) in order to obtain high-resolution  \emph{lens}fit
weak lensing shape measurements (as configured for the CFHTLenS pipeline)
for the area covered by the SDSS equatorial
Stripe 82 $ugriz$ survey \citep{adelman07,annis14}.
CS82
comprises 173 MegaCam pointings, covering 160 deg$^2$ (129.2 deg$^2$ after masking).
Our \emph{lens}fit shape selection (see Sect.\thinspace\ref{se:source_sample}) corresponds
to the one applied by \citet{hand15},
yielding an
 effective weighted source density of 12.3/arcmin$^2$.
In our analysis we also make use of photometric redshifts
computed for the CS82 area based on the
SDSS equatorial
Stripe 82 $ugriz$
data
using
\textsc{EAZY} \citep{brammer08}.

\subsection{KiDS/KV450}
We also include public data\footnote{\url{KV450:http://kids.strw.leidenuniv.nl/cosmicshear2018.php}} from the third data release \citep{dejong17} of the Kilo-Degree
Survey \citep[KiDS,][]{kuijken15}, encompassing 447 deg$^2$ observed in $ugri$ with ESO's VLT Survey Telescope (VST), which were processed using \textsc{THELI} \citep{ehm13} and \textsc{Astro-WISE} \citep{bbb13}.
Good-seeing $r$-band images with a mean PSF FWHM of 0\farcs68 and $5\sigma$ limiting magnitude of
25.0
(computed in $2^{\prime\prime}$ apertures) were used for  \emph{lens}fit galaxy shape measurements \citep{fenechconti17,khm19}, yielding a weighted  galaxy source density of 8.53/arcmin$^2$ \citep{hildebrandt17}.
For the photometric analysis we use an updated \textsc{BPZ} photometric redshift catalogue  which incorporates NIR $ZYJHK_\mathrm{s}$ photometry from VISTA  and provides stellar mass estimates from \textsc{LePhare}
as presented by
\citet[][KV450]{wright19}.

\section{Analysis}
\label{se:analysis}
\subsection{Shape measurements for bright galaxies}
\label{se:shapes_bright}
The \emph{lens}fit algorithm, which was employed for shape measurements in all surveys included in our study,
has been optimised to provide accurate shape estimates for the typically small and distant source
galaxies included in cosmic shear studies.
\citetalias{shh15} found that many of the bright (\mbox{$i\lesssim 20$}) foreground galaxies acting as lens sample (see Sect.\thinspace\ref{se:lenssample})
are excluded by \emph{lens}fit. For example, this can be caused by their large extent, which is not sufficiently covered by the employed postage stamp size.
Galaxies may also be flagged because of the presence of nearby galaxies, whose outer
 isophotes overlap, or the presence of resolved substructure if this is not well described by
the  bulge+disc model employed in \emph{lens}fit \citep[see][]{mhk13a}.
Following \citetalias{shh15} we therefore obtained additional shape measurements using the KSB+ algorithm \citep{ksb95,luk97,hfk98}
for the bright galaxies without successful \emph{lens}fit shape estimates.
This method
is less affected by  nearby galaxies or resolved substructure.
In particular, we employ the KSB+ implementation described in
\citet{hfk98} and \citet{hfk00}, which was tested in the STEP blind challenges
\citep{hwb06,mhb07}.
The bright galaxies in question (those without \emph{lens}fit shape estimates)
are all well resolved and typically have high signal-to-noise ratios \mbox{$S/N=\mathrm{FLUX\_AUTO/FLUXERR\_AUTO}\gtrsim 100$} \citep[defined via the FLUX\_AUTO and FLUXERR\_AUTO parameters from \textsc{SExtractor},][]{bea96}. Therefore, they require only small PSF corrections
and are essentially insensitive to noise-related biases \citep[e.g.][]{mev12,rka12,kzr12}\footnote{We also verified that shape measurements agree well between KSB+ and  \emph{lens}fit for bright galaxies that are not removed by \emph{lens}fit.}.
Given the high signal-to-noise ratios we also employ slightly wider weight functions in the KSB+ moments computation\footnote{\label{fn:weightfct}We employ a weight function with a scale radius that is larger by a factor of $\sqrt{2}$ compared to the Gaussian scale radius that would optimise the signal-to-noise ratio in the case of Gaussian brightness profiles.}, increasing the sensitivity to the outer galaxy light distributions.

\subsection{Lens galaxies}
\label{se:lenssample}
Our selection of lens galaxies closely follows \citetalias{shh15}.
From the \textsc{SExtractor} \citep{bea96} object catalogues provided by the different surveys (see Sect.\thinspace\ref{se:data}),
we pre-select comparably bright objects (\mbox{$i<23.5$}, except for RCSLenS, where we require \mbox{$r<23.5$} due to the non-perfect overlap of the $r$ and $i$-band data) that
are well resolved\footnote{To remove stars we employ the \texttt{SExtractor}
\mbox{$\mathrm{CLASS\_STAR}$} parameter, requiring
\mbox{$\mathrm{CLASS\_STAR}<0.5$}.
To be consistent with  \citetalias{shh15} we additionally require \mbox{$\mathrm{star\_flag}=0$} for CFHTLenS and apply a similar selection \mbox{$\mathrm{SG\_FLAG}=1$} for RCSLenS as recommended by these surveys. We however find that these latter cuts have a completely negligible impact, removing \mbox{$<0.2\%$} of the otherwise selected lens candidates only. We verified through visual inspection that our final lens samples do not suffer from a significant contamination by stars.}
and
have non-zero
shape weights either from \emph{lens}fit or KSB+\footnote{This is the case for \mbox{$\gtrsim 98\%$} of the lens candidates.}.
When constraining the halo shape signature (see Sect.\thinspace\ref{se:measuring_halo_shapes}), the
shear field is stacked with respect to the lens orientation, and
weak lensing contributions are weighted according to the absolute value of the ellipticity of the corresponding lens.
In our analysis we therefore only include lenses with ellipticities in the range \mbox{$0.05<|\epsilon_\mathrm{g}|<0.95$}, for which both the orientation and the absolute value
are well constrained.

\begin{table*}
\caption{Lens galaxy samples.
\label{tab:lenses}}
\begin{center}
\begin{tabular}{cccrc}
\hline
\hline
  Survey & Colour & Stellar mass  [M$_\odot$]&
    \mbox{$N$}
  & $\sigma_\epsilon$\\
\hline
  CFHTLenS & Red&\mbox{$10<\log_{10}M_*<10.5$} & 61569 & 0.35\\
CFHTLenS & Red&\mbox{$10.5<\log_{10}M_*<11$} & 70015 & 0.30\\
CFHTLenS & Red&\mbox{$\log_{10}M_*>11$} & 19624 & 0.23\\
\hline
CFHTLenS & Blue&\mbox{$9.5<\log_{10}M_*<10$} & 125968 & 0.38\\
CFHTLenS & Blue&\mbox{$10<\log_{10}M_*<10.5$} & 69175 & 0.36\\
CFHTLenS & Blue&\mbox{$\log_{10}M_*>10.5$} & 25195 & 0.30\\
\hline
RCSLenS & Red&\mbox{$10<\log_{10}M_*<10.5$} & 84380 & 0.34\\
RCSLenS & Red&\mbox{$10.5<\log_{10}M_*<11$} & 60248 & 0.29\\
RCSLenS & Red&\mbox{$\log_{10}M_*>11$} & 7059 & 0.25\\
 \hline
CS82 & Red&\mbox{$10<\log_{10}M_*<10.5$} & 120640 & 0.39\\
CS82 & Red&\mbox{$10.5<\log_{10}M_*<11$} & 141107 & 0.35\\
CS82 & Red&\mbox{$\log_{10}M_*>11$} & 68231 & 0.29\\
 \hline
CS82 & Blue&\mbox{$9.5<\log_{10}M_*<10$} & 294854 & 0.37\\
CS82 & Blue&\mbox{$10<\log_{10}M_*<10.5$} & 120989 & 0.33\\
CS82 & Blue&\mbox{$\log_{10}M_*>10.5$} & 25528 & 0.27\\
 \hline
KV450 & Red&\mbox{$10<\log_{10}M_*<10.5$} & 178394 & 0.33\\
KV450 & Red&\mbox{$10.5<\log_{10}M_*<11$} & 190479 & 0.30\\
KV450 & Red&\mbox{$\log_{10}M_*>11$} & 32543 & 0.30\\
 \hline
KV450 & Blue&\mbox{$9.5<\log_{10}M_*<10$} & 421945 & 0.37\\
KV450 & Blue&\mbox{$10<\log_{10}M_*<10.5$} & 182514 & 0.31\\
KV450 & Blue&\mbox{$\log_{10}M_*>10.5$} & 53465 & 0.31\\
 \hline
NGVSLenS & Red&\mbox{$10<\log_{10}M_*<10.5$} & 66637 & 0.37\\
NGVSLenS & Red&\mbox{$10.5<\log_{10}M_*<11$} & 121840 & 0.36\\
NGVSLenS & Red&\mbox{$\log_{10}M_*>11$} & 85231 & 0.31\\
 \hline
NGVSLenS & Blue&\mbox{$9.5<\log_{10}M_*<10$} & 277549 & 0.41\\
NGVSLenS & Blue&\mbox{$10<\log_{10}M_*<10.5$} & 262997 & 0.39\\
NGVSLenS & Blue&\mbox{$\log_{10}M_*>10.5$} & 150760 & 0.33\\
\hline
\end{tabular}
\end{center}
\vspace{0.2cm}

{\flushleft
Notes. --- Overview over the sub-sample of lens galaxies used:
{\it Column 2:} Split between red
and blue lenses
using $g-i$ colour for CS82 and the photometric type
$T_\mathrm{BPZ}$ from BPZ for the other surveys.
{\it Column 3:} Stellar mass range.
{\it Column 4:} Number of selected lenses in the redshift
interval \mbox{$0.4\le z_\mathrm{l}< 0.6$} for RCSLenS and \mbox{$0.2\le z_\mathrm{l}< 0.6$} for the other surveys, where the CFHTLenS numbers are based on
the fields that pass the systematics tests from \citet[][see Sect.\thinspace\ref{se:data_cfhtlens}]{hwm12}.
{\it Column 5:} Ellipticity dispersion of the selected lenses with \mbox{$0.05<|\epsilon_\mathrm{g}|<0.95$} combining both ellipticity components.
\\
}
\end{table*}

We also require that lenses have high-quality photometric redshift estimates: As done by \citetalias{shh15}, we require that lenses feature a single-peaked
photometric redshift probability distribution function
\citep[requiring
  \mbox{$\mathrm{ODDS}>0.9$}, which is computed by both \textsc{BPZ} and  \textsc{EAZY}, see][]{hek12} for CFHTLenS, NGVSLenS, RCSLenS, and CS82.
For KV450 we instead follow \citet{wright19}, who show that the selection of galaxies with successful nine-band photometry yields highly accurate redshift estimates in the magnitude range of our lenses.

Following
\citetalias{shh15} we select lenses in the photometric redshift range
\mbox{$0.2<z_\mathrm{b}<0.6$} (split into four thin lens redshift
slices of width \mbox{$\Delta z_\mathrm{b}=0.1$}).
Here $z_\mathrm{b}$ indicates the
best-fit \textsc{BPZ} redshift for all  surveys except for CS82, where it corresponds to the \textsc{EAZY} redshift estimate $z_\mathrm{p}$
(for which the posterior is maximised).

For the surveys with \textsc{BPZ} catalogues we subdivide the lenses into
red lenses (\mbox{$T_\mathrm{BPZ} \le 1.5$}) and blue lenses
(\mbox{$1.5<T_\mathrm{BPZ} < 3.95$}) using the photometric type
$T_\mathrm{BPZ}$ as done by \citetalias{shh15}.
Given the lack of $u$-band data for RCSLenS, our requirement of highly accurate redshifts  (\mbox{$\mathrm{ODDS}>0.9$}) removes the majority of the blue lens candidates over the full redshift range and many red lens candidates at \mbox{$z_\mathrm{b}<0.4$}. For RCSLenS we therefore limit the analysis to red lenses at \mbox{$0.4<z_\mathrm{b}<0.6$}.

Following \citet{mhb06} and \citetalias{shh15} we also subdivide the galaxies
into stellar mass bins (see Table \ref{tab:lenses}), which provides a proxy for halo mass.
This improves the joint weak lensing measurement signal-to-noise ratio
given the mass dependence of the (anisotropic) NFW shear profile (see Sect.\thinspace\ref{se:measuring_halo_shapes}).
We employ stellar mass estimates from  \textsc{LePhare}
for CFHTLenS, RCSLenS, and KV450, and from  \textsc{BPZ} for NGVSLenS
(see Sect.\thinspace\ref{se:data}).
Since
stellar mass estimates and \textsc{BPZ} photometric types  were not available in the CS82 catalogues employed in our analysis\footnote{\label{footnote:bundy}\citet{bundy15} provide a stellar mass catalogue, but we do not employ it in our analysis since it only covers a part of the footprint of our CS82 analysis.}, we
instead applied a selection in photometric redshift, $g-i$ colour, and $i$-band magnitude, which allowed us to approximately recover the lens bin subdivision employed in CFHTLenS (see Appendix \ref{app:CS82lenses} for details).

We do not expect that stellar mass estimates are exactly comparable from survey to survey, given the differences in the input data (e.g.~available bands) and codes used by the different survey teams to compute them. Similarly, our approach for CS82 only approximately reproduces the stellar mass bins used for CFHTLenS (see Appendix \ref{app:CS82lenses}).
        This is one of the reasons why we initially analyse all surveys separately and only combine their constraints in the final step when constraining the signature of halo ellipticity (see Sections \ref{se:measuring_halo_shapes} and \ref{se:halo_shapes_results}).

    In Sect.\thinspace\ref{se:halo_shapes_results}
    we compare our results to predictions derived by
    \citetalias{shh15}
    for central galaxies from mock data based on the Millennium
    Simulation.
    Likewise, our modelling approach assumes that the shear signal surrounding the lens is dominated by the lens itself.
    This is a reasonable assumption for centrals, but may be a poor approximation for satellites, whose surrounding shear field can be significantly influenced by their more massive central host galaxy.
    We therefore aim  to minimise the number of satellites in our lens sample.
    To achieve this, we follow  \citetalias{shh15} and  exclude bins with low stellar mass that have a substantial satellite fraction \citep{vuh14}.
    For  red lenses located in the lowest stellar mass bin that is included in our analysis (\mbox{$10<\log_{10} M_*<10.5$}),  \citet{vuh14} estimate a satellite fraction of $23\pm 2\%$.
To reduce this fraction further we additionally remove\footnote{On average this cut removes about 16\% of the lens candidates in this colour and stellar mass bin.}  galaxies that are flagged by \textsc{SExtractor} to either be  blended with another object, or
    to have their \texttt{MAG\_AUTO} magnitude measurements significantly contaminated by  a nearby neighbour.
    Many  such galaxies are located in the vicinity of a brighter early-type galaxy, indicating that they may be satellites.

\subsection{Source sample}
\label{se:source_sample}
Our parent source sample includes all galaxies with successful
{\it lens}fit shape measurements
that have shape weights \mbox{$w>0$}.
To select background sources we require both \mbox{$z_\mathrm{s,b}>z_\mathrm{l,max}+0.1$} and \mbox{$z_\mathrm{s,lower95}>z_\mathrm{l,max}$}, where $z_\mathrm{l,max}$ is the upper limit of each of the four lens redshift slices, and \mbox{$z_\mathrm{s,lower95}$} indicates the 95\% lower source redshift limit computed by the photometric redshift codes.
This stringent selection reduces the resulting source densities and therefore the statistical constraining power for those surveys (in particular RCSLenS and CS82) that have noisier photo-$z$s, for example due to fewer bands or shallower photometric data.

\citetalias{shh15} removed galaxies with \mbox{$z_\mathrm{s,b}>1.3$} from their source sample as these are likely subject to an increased photometric redshift scatter. Using the CFHTLenS data we find  that the inclusion of these high-$z$ galaxies actually leads to a moderate tightening of the halo ellipticity constraints. While uncertainties in the redshift calibration of these sources may affect the halo mass constraints, these uncertainties do not lead to bias in the  halo ellipticity constraints, which are derived from the ratio of the (equally affected) anisotropic and isotropic shear profiles (see Sect.\thinspace\ref{se:measuring_halo_shapes}).
Therefore, we do not remove these galaxies from our analysis.

\subsection{Extracting the weak lensing halo shape signature}
\label{se:measuring_halo_shapes}

In our analysis we follow the methodology introduced by \citetalias{mhb06}, which was also applied by  \citet{uhs12} and \citetalias{shh15}.
As typically done in weak lensing studies, we
characterise the shape of a galaxy via its complex ellipticity
\begin{equation}
  \epsilon=\epsilon_1+\mathrm{i}\epsilon_2=|\epsilon|\mathrm{e}^{2\mathrm{i}\phi} \,.
\end{equation}
In the case of an idealised galaxy that has co-centric elliptical isophotes with a constant ratio of their major and minor axes $a$ and $b$
 and a constant position angle,
the  absolute value of the ellipticity
is given by
\begin{equation}
  |\epsilon|=(a-b)/(a+b)\,.
  \end{equation}
In this case $\phi$ corresponds to the position angle  of the major axis with respect to the
coordinate $x$-axis.
The ellipticity transforms under a reduced shear
\begin{equation}
  \label{se:reducedshear}
g=\frac{\gamma}{1-\kappa}\,,
\end{equation}
which is a rescaled version of the  anisotropic shear $\gamma$ depending on the convergence $\kappa$,
as
\begin{equation}
\label{eq:eofges}
\epsilon\simeq
  \epsilon_\mathrm{s}+g\simeq  \epsilon_\mathrm{s}+\gamma\,,
\end{equation}
where we assume small distortions (\mbox{$|\gamma |\ll 1$}, \mbox{$|\kappa|\ll 1$}) as adequate for our study \citep[for the general case see][]{seitz97,bas01}.
The intrinsic source ellipticity $\epsilon_\mathrm{s}$ is expected to have a random orientation, yielding expectation  values \mbox{$\mathrm{E}(\epsilon_\mathrm{s})=0$} and \mbox{$\mathrm{E}(\epsilon)=\gamma$}.

In principle, all structures between the source and the observer contribute to the lensing effect.
However, when we constrain the average shear field around the positions
of foreground lens galaxies only structures at the lens redshift contribute coherently to the signal\footnote{Structures in front of the lens do not contribute to the net isotropic galaxy-galaxy lensing signal, but can cause spurious signal for weak lensing constraints on halo ellipticity for simple estimators
  (this is corrected for in our analysis, see Sect.\thinspace\ref{se:constrain-ani}). Structures behind the lens do not cause any bias.}.
The net convergence \mbox{$\kappa=\Sigma/\Sigma_\mathrm{c}$} is given by the
product of
the projected surface mass density $\Sigma$  and the inverse critical surface mass density
\begin{equation}
  \label{eqn:sigmacrit}
  \Sigma_{\mathrm{c}}^{-1} =
  \frac{4\pi G}{c^2}D_{\mathrm{l}} \beta \,,
\end{equation}
which itself depends on the speed of light in vacuum  $c$, the gravitational constant  $G$,
and the  physical angular diameter distances to the source $D_\mathrm{s}$, to the lens $D_\mathrm{l}$, and between lens and source $D_\mathrm{ls}$, given that
the geometric lensing efficiency $\beta$ is defined as
\begin{equation}
\beta=\frac{D_\mathrm{ls}}{D_\mathrm{s}} H(z_\mathrm{s}-z_\mathrm{l}) \, ,
\end{equation}
where $H(x)$ indicates the Heaviside step function.

When stacking the shear field around foreground lens galaxies it is useful to decompose the
shear and the ellipticities of background galaxies into the tangential component
\begin{equation}
\epsilon_\mathrm{t} =  - \epsilon_1 \cos{2 \theta} - \epsilon_2 \sin{2\theta}\,
\end{equation}
and the 45 degrees-rotated cross component
\begin{equation}
  \epsilon_\times = + \epsilon_1 \sin{2\theta} - \epsilon_2 \cos{2 \theta}  \,,
  \label{eq:ex}
\end{equation}
where $\theta$ indicates  the azimuthal angle with respect to the position of the lens when
measured from the $x$-axis.

\subsubsection{Constraining the isotropic shear field}
\label{se:fitisotropic}
The profile of the
azimuthally averaged tangential shear
\mbox{$\langle\gamma_\mathrm{t}\rangle(r)=\Delta \Sigma / \Sigma_{\mathrm{c}}$} directly relates to the differential profile  of the surface mass density
 \mbox{$\Delta \Sigma (r)\equiv \overline{\Sigma}(<r)-\overline{\Sigma}(r)$} \citep[][]{miralda-escude91},
 where
 \mbox{$\overline{\Sigma}(<r)$} and  \mbox{$\overline{\Sigma}(r)$}
 indicate the mean surface mass density  within the projected radius $r$ and at $r$, respectively \citep[this also holds for non-axis-symmetric mass distributions, see e.g.][]{kaiser95a,umetsu20}.
 Phrasing the analysis in terms of $\Delta \Sigma$ rather than $\gamma$ has the advantage of scaling out the redshift dependence of the shear signal.
The differential surface density can directly be estimated from the source galaxy ellipticities as
\begin{equation}
\widehat{\Delta \Sigma}(r)=\frac{\sum_{ij} w_i \Sigma_{\mathrm{c},ij}^{-2}\left(\epsilon_{\mathrm{t},i}\Sigma_{\mathrm{c},ij}\right)}{\sum_{ij} w_i \Sigma_{\mathrm{c},ij}^{-2}} =
\frac{\sum_{ij} w_i \Sigma_{\mathrm{c},ij}^{-1}\epsilon_{\mathrm{t},i}}{\sum_{ij} w_i \Sigma_{\mathrm{c},ij}^{-2}}
\, ,
\label{eq:isoestsimple}
\end{equation}
where the summation is executed over all pairs of lenses $j$ in the corresponding redshift, colour and stellar mass bin,  and sources $i$ located in an annulus around a radius $r$ from the corresponding lens.
Here we multiply the source shape weight \mbox{$w_i\simeq \sigma_{\epsilon,i}^{-2}$} \citep{mhk13a,fenechconti17}
with $\Sigma_{\mathrm{c},ij}^{-2}$ to obtain optimised combined weights with increased sensitivity.
As also done in later subsections we indicate estimators using the $\widehat{\,\,\,\,}$ symbol.

The critical surface mass density $\Sigma_{\mathrm{c},ij}$ depends on the redshifts of the
 lens $z_{\mathrm{l},j}$ and source $z_{\mathrm{s},i}$ (see Eq.\thinspace\ref{eqn:sigmacrit}), where \mbox{$\Sigma_{\mathrm{c},ij}^{-1}=\Sigma_{\mathrm{c},ij}^{-2}=0$} if \mbox{$z_{\mathrm{s},i}\le z_{\mathrm{l},j}$}.
 We approximate $z_{\mathrm{l},j}$ with the centre $z_{\mathrm{l,c}}$ of each of the \mbox{$\Delta z_\mathrm{l}=0.1$} wide thin lens redshift slides for computational efficiency
 as done by \citetalias{shh15}.
 For the computation of  \mbox{$\Sigma_{\mathrm{c},ij}^{-1}$} we employ the best-fit photometric redshift estimates \mbox{$z_\mathrm{s,b}$} as source redshifts $z_{\mathrm{s},i}$ for KV450, NGVSLenS, and CS82.
 For consistency with \citetalias{shh15}, we instead compute an effective \mbox{$\Sigma_{\mathrm{c},ij}^{-1}$} for each source from
 its
 effective geometric lensing
efficiency  (see Eq.\thinspace\ref{eqn:sigmacrit}) \mbox{$\beta^\mathrm{eff}_i=\int
  \beta(z_{\mathrm{l,c}},z_\mathrm{s})p_i(z_\mathrm{s})\mathrm{d}z_\mathrm{s}$} using
the full posterior redshift probability distribution $p_i(z)$ provided by \texttt{BPZ} for the CFHTLenS analysis.
However, tests using \mbox{$z_\mathrm{b}$} instead indicate that this does not have a significant impact on our halo ellipticity results.
 For RCSLenS we follow the recommendation from \citet{hch16} to compute  \mbox{$\Sigma_{\mathrm{c},ij}^{-1}$} (see Eq.\thinspace\ref{eqn:sigmacrit})
 from the $p_i(z)$ provided for the source galaxies, which is more important for this survey given the lack of $u$-band data, leading to larger redshift uncertainties.
In any case it is important to realise that also the use of the full reported posterior redshift probability distribution $p(z)$ can yield biased estimates of the lensing efficiency if it does not accurately reflect the true redshift probability distribution \citep[see e.g.][]{schrabback18},
 introducing systematic biases in halo mass estimates.
 Fortunately, these biases cancel out for the constraints on halo ellipticity,
 which are derived from the ratio of the  (equally affected) isotropic and anisotropic
 shear profiles
 (see Sect.\thinspace\ref{se:getfh}).
 For  halo ellipticity measurements redshift errors only lead to non-optimal weighting (given the inclusion of $\Sigma_{\mathrm{c},ij}^{-2}$ in the effective weights).
 Therefore, we do  not need to obtain highly accurate calibrations of the
 source redshift distribution for our analysis.

 We fit the isotropic part of the measured shear profile for each lens sample  with model predictions  that assume spherical NFW density profiles \citep{nfw96}
as detailed in \citet{wrb00},  applying the concentration--mass relation
from \citet{dsk08}.
This provides estimates for the mass $M_\mathrm{200c}$ located within a sphere with radius
\mbox{$r_{{200}\mathrm{c}}$},
in which the  mean density
equals $200$ times the critical density of the Universe at the lens redshift.
Following \citetalias{shh15} we account for the minor impact the convergence has in
Eq.\thinspace(\ref{se:reducedshear}) when computing isotropic NFW shear profile predictions to obtain more accurate halo mass estimates, but we can safely ignore this in the formalism used to constrain the anisotropic shear signal.

 \subsubsection{Constraining the anisotropic shear field}
\label{se:constrain-ani}
 \citet{nar00} introduced a formalism to constrain  the anisotropic weak lensing shear field around
 lenses, assuming elliptical isothermal mass distributions.
 \citetalias{mhb06} extended this formalism to other density profiles, including elliptical NFW density profiles.
 Importantly, they also introduced a prescription to correct for the main source of systematics in halo ellipticity measurements, which is spurious signal caused by additional alignments of lenses and sources caused by other effects such as cosmic shear or PSF anisotropy residuals.
 This formalism was also used by \citetalias{shh15}, who
 identified an apparent  sign inconsistency in the \citetalias{mhb06} model predictions via their analysis of simplified mock data, and
 also tested the formalism on the Millennium Simulation, which features actual projected halo mass distributions.
Here we employ this formalism assuming elliptical NFW mass distributions,
following the notation from \citetalias{mhb06} with modifications from
\citetalias{shh15}. Below we only summarise the relevant notation, see  \citetalias{mhb06} and \citetalias{shh15} for more detailed derivations and explanations.
We note that \citet{clj16} and \citet{vanuitert17} introduced alternative notations and slightly different estimators to obtain systematics-corrected halo ellipticity estimates, but as shown by \citet{vanuitert17} these are either equivalent to the approach we use or yield  similar results.

In order to stack the anisotropic shear field around the (in projection approximately) elliptical dark matter haloes,
we would ideally like to rotate the coordinates such that the major axes of all haloes align.
Unfortunately, the true orientations of the haloes are unknown.
Thus, we can only align the shear field
according to the orientations of the ellipticities   $\epsilon_\mathrm{g}$ of the observed
lens light distributions.
As a result, our analysis is only sensitive to the average of the component
\begin{equation}
\epsilon_\mathrm{h,a} =  \cos(2\Delta\phi_\mathrm{h,g}) |\epsilon_\mathrm{h}|
\label{eq:eha}
\end{equation}
of the halo ellipticity $\epsilon_\mathrm{h}$ that is aligned
with the galaxy ellipticity.
Here $\Delta\phi_\mathrm{h,g}$ indicates the misalignment angle, which we assume is independent of $|\epsilon_\mathrm{h}|$.
Considering that the asymmetry in the shear field scales in good approximation with the ellipticity of the
mass distribution (see e.g.~Fig.\thinspace 2 in \citetalias{shh15}),
we can then model the
 average tangential
 shear field (scaled via \mbox{$\Sigma_\mathrm{c}$})
 around lens galaxies
 as
\begin{equation}
\Delta \Sigma_\mathrm{model}(r,\Delta\theta)=\Delta \Sigma_\mathrm{iso}(r)\left[1+4 f_\mathrm{rel}(r) \epsilon_\mathrm{h,a} \cos(2\Delta\theta)\right] \,,
\label{eq:sigmamodel_proper}
\end{equation}
where  $\Delta\theta$ indicates the position angle as measured from the
major axis of the lens galaxy. We note that the corresponding equation in \citetalias{mhb06} uses a different prefactor in the anisotropic term, which is due to the different ellipticity definition employed in their work (see \citetalias{shh15}).

Following \citetalias{mhb06} and \citetalias{shh15} we assume that \mbox{$|\epsilon_\mathrm{h}|\propto |\epsilon_\mathrm{g}|$}
(see \citealt{uhs12} for the exploration of  additional schemes),
in which case Eq.\thinspace(\ref{eq:sigmamodel_proper}) can be written as
\begin{equation}
\Delta \Sigma_\mathrm{model}(r,\Delta\theta)=\Delta \Sigma_\mathrm{iso}(r)\left[1+4 f(r) |\epsilon_\mathrm{g}| \cos(2\Delta\theta)\right] \,,
\label{eq:sigmamodel}
\end{equation}
where
\begin{equation}
  f(r)=f_\mathrm{rel}(r) f_\mathrm{h}\, \label{eq:frfsfh}
\end{equation}
relates to
the average ratio
\begin{equation}
  f_\mathrm{h}=
 \langle\epsilon_\mathrm{h,a} / |\epsilon_\mathrm{g}|\rangle=
 \langle\cos(2\Delta\phi_\mathrm{h,g})|\epsilon_\mathrm{h}|/|\epsilon_\mathrm{g}|\rangle
\, \label{eq:fhfromeg}
\end{equation}
between the aligned component of the halo ellipticity and the ellipticity of the light distribution, which is the main quantity we aim to constrain with our analysis.
The quantity
$f_\mathrm{rel}(r)$ in Equations (\ref{eq:sigmamodel_proper}) and (\ref{eq:frfsfh})
depends on the assumed density profile.
Here we employ numerical model predictions for $f_\mathrm{rel}(r)$ obtained by \citetalias{mhb06} for elliptical NFW profiles \citep[for an analytic computation scheme see][]{vanuitert17}.

\citetalias{mhb06} show that estimators
(which are indicated by the $\widehat{\quad\,\quad}$ symbol) for the isotropic and anisotropic components of the scaled shear field in Eq.\thinspace(\ref{eq:sigmamodel})
are given by
\mbox{$\widehat{\Delta \Sigma_\mathrm{iso}}(r)=\widehat{\Delta \Sigma}(r)$} (see Eq.\thinspace\ref{eq:isoestsimple})
and
\begin{equation}
\widehat{f(r) \Delta \Sigma_\mathrm{iso}(r)} =  \frac{\sum_{ij} w_i \Sigma_{\mathrm{c},ij}^{-1}\epsilon_{\mathrm{t},i} |\epsilon_{\mathrm{g},j}|\cos(2\Delta\theta_{ij})}{4 \sum_{ij} w_i \Sigma_{\mathrm{c},ij}^{-2} |\epsilon_{\mathrm{g},j}|^2\cos^2(2\Delta\theta_{ij})}\,,
\label{eq:fdeltasigma_estimate}
\end{equation}
where we again sum over lenses $j$ and  sources $i$ that are  located in a  separation interval around $r$ around the corresponding lens. Here $\Delta\theta_{ij}$ indicates the position angle of source $i$  as measured from the
major axis of lens $j$.

Cosmic shear caused by structures in front of the lens, as well as potential residuals in the PSF anisotropy correction, can align the observed ellipticities of lenses and sources.
This leads to estimates of halo ellipticity that are biased low when constrained via
Eq.\thinspace(\ref{eq:fdeltasigma_estimate})
(\citetalias{mhb06}).
To remove this spurious contribution, \citetalias{mhb06} introduce an additional
estimator
\begin{equation}
\widehat{{f_{45}(r)\Delta \Sigma_\mathrm{iso}(r)}} =  - \frac{\sum_{ij} w_i \Sigma_{\mathrm{c},ij}^{-1}\epsilon_{\times,i} |\epsilon_{\mathrm{g},j}|\sin(2\Delta\theta_{ij})}{4 \sum_{ij} w_i \Sigma_{\mathrm{c},ij}^{-2} |\epsilon_{\mathrm{g},j}|^2\sin^2(2\Delta\theta_{ij})}\,
\label{eq:fdeltasigma_estimate45}
\end{equation}
analogously to Eq.\thinspace(\ref{eq:fdeltasigma_estimate}), but
 based on
 the ellipticity cross component $\epsilon_\times$ (Eq.\thinspace\ref{eq:ex}).
This estimator
 carries an approximately equal spurious signal at the scales relevant for our analysis (see \citetalias{mhb06} and \citetalias{shh15}),
 which is why the modified estimator
 \begin{equation}
   \widehat{\left[f(r)-f_{45}(r)\right]\Delta
     \Sigma_\mathrm{iso}(r)}\equiv \widehat{f(r) \Delta \Sigma_\mathrm{iso}(r)}   -\widehat{f_{45}(r) \Delta \Sigma_\mathrm{iso}(r)}
   \end{equation}probes halo ellipticity without being affected by this systematic contribution.
 The analysis of mock data based on the Millennium Simulation \citep{swj05,hhw09} by \citetalias{shh15} revealed that this estimator also approximately cancels a further systematic signal contribution, which \citetalias{shh15} interpret as the impact of shape-shear correlations \citep[e.g.][]{his04,jsh13} caused by the wider large-scale environment of the lens dark matter halo.

 Similarly to $f(r)$,
   $f_{45}(r)$ also contains some signal from the flattened
 halo, which is why both must be modelled.
Analogously to Eq.\thinspace(\ref{eq:frfsfh}),
 the model for $f_{45}(r)$ is scaled as
\begin{equation}
f_{45}(r)=f_\mathrm{rel,45}(r) f_\mathrm{h}\,.
\label{eq:f45fs45fh}
\end{equation}
\citetalias{mhb06} obtained numerical predictions  for $f_\mathrm{rel}(r)$ and $f_\mathrm{rel,45}(r)$ for elliptical NFW models as a function of the ratio $r/r_\mathrm{s}$.
 These were kindly provided to us in tabulated form, from which we interpolate. To employ these models, we infer the NFW scale radius
\mbox{$r_\mathrm{s}=r_\mathrm{200c}/c_\mathrm{200c}$} from the fit to the isotropic signal (Sect.\thinspace\ref{se:fitisotropic}) and the adopted relation between the mass $M_\mathrm{200c}$ and concentration $c_\mathrm{200c}$.
In addition, we apply the sign correction to  the $f_\mathrm{rel,45}(r)$ model prediction from \citetalias{shh15}.

\subsubsection{Estimating $f_\mathrm{h}$}
\label{se:getfh}
In order to estimate the
aligned
ellipticity ratio $f_\mathrm{h}$
we
define
\begin{eqnarray}
\widehat{y}(r)&=&
\frac{1}{f_\mathrm{rel}(r)-f_\mathrm{rel,45}(r)}\widehat{\left[f(r)-f_{45}(r)\right]\Delta
  \Sigma_\mathrm{iso}(r)}\,,\\
\widehat{x}(r)&=&\widehat{\Delta
  \Sigma_\mathrm{iso}}(r)\,.
\end{eqnarray}
Simply using
\begin{equation}
\widehat{f_\mathrm{h}^\mathrm{biased}}(r)=\frac{\widehat{y}(r)}{\widehat{x}(r)}
\label{eq:fhest_biased}
\end{equation}
would lead to a biased estimate given the noise in $\widehat{x}(r)$.\footnote{We note, however, that the isotropic shear profile and therefore $\widehat{x}(r)$ is typically well constrained (see Fig.\thinspace\ref{fi:shearfitsred_NGVS} and Figs.\thinspace\ref{fi:shearfitsred_KV450} to \ref{fi:shearfitsred_RCS2}), which is why the uncertainties in $\widehat{y}(r)$ have the biggest impact on the uncertainties of the $f_\mathrm{h}$ estimates.}
Therefore, we instead follow  \citetalias{mhb06} and employ an approach introduced by
\citet{bli35a,bli35b} and \citet{fie54}, which aims to constrain a ratio \mbox{$m=y/x$} of two random variables.
Here $m$ corresponds to $f_\mathrm{h}$ and we assume
 that both  $x$
 and $y$ follow Gaussian distributions, which is a plausible approximation in
 the shape-noise-dominated regime of
 galaxy-galaxy lensing.
 The different radial bins provide
 multiple estimates $\widehat{x_k}$ and $\widehat{y_k}$,
 where we can optionally  add estimates from different stellar mass bins and surveys in order to derive joint constraints.
The quantity \mbox{$\widehat{y_k}-m\widehat{x_k}$} is a Gaussian
random variable drawn from a $\mathcal{N}(\mu=0,\sigma^2=\tilde{w}_k^{-1})$ normal distribution for each $k$, with \mbox{$\tilde{w}_k^{-1}=\sigma_{\widehat{y}_k}^2+m^2\sigma_{\widehat{x}_k}^2$}.\footnote{We note that there is a typo in the corresponding equation in \citetalias{shh15}, but their results were computed correctly.}
As a result, the summation
\begin{equation}
\frac{\sum_k\tilde{w}_k(\widehat{y_k}-m\widehat{x_k})}{\sum_k\tilde{w}_k}\sim\mathcal{N}\left(0,\frac{1}{\sum_k\tilde{w}_k}\right)
\end{equation}
over all measurements also constitutes a Gaussian random variable for a given $m$,
 allowing us to identify
frequentist confidence intervals at the $Z\sigma$ level
as
\begin{equation}
\frac{-Z}{\sqrt{\sum_k\tilde{w}_k}}<\frac{\sum_k\tilde{w}_k(\widehat{y_k}-m\widehat{x_k})}{\sum_k\tilde{w}_k} <\frac{Z}{\sqrt{\sum_k\tilde{w}_k}}\,.
\end{equation}
A grid search in $m$ then yields
the
desired estimator as best-fitting
value
\begin{equation}
\widehat{f_\mathrm{h}}=m(Z=0)\,,
\end{equation}
at which we
also compute a reduced $\chi^2$ as
\begin{equation}
\chi^2/\mathrm{d.o.f.}=\frac{\sum_{k=1}^{k=n}\tilde{w}_k(\widehat{y_k}-m(Z=0)\widehat{x_k})^2}{n-1}\,,
\end{equation}
as well as
68 per cent confidence limits  \mbox{$m(Z=\pm 1)$}.
\citetalias{shh15} have shown that off-diagonal covariance elements are sufficiently small that they can  be neglected when estimating halo ellipticity with this approach.

We  apply an alternative Bayesian approach to estimate $f_\mathrm{h}$ in Appendix \ref{app:likelihood}, which yields broadly consistent constraints.
We report the constraints derived using the approach described here as our main results, especially since we compare them to simulation-based predictions from \citetalias{shh15}, which were computed using the same methodology (see Sect.\thinspace\ref{se:discussions}).

\section{Results}
\label{se:halo_shapes_results}

 \begin{figure*}
   \centering
  \includegraphics[width=7.4cm]{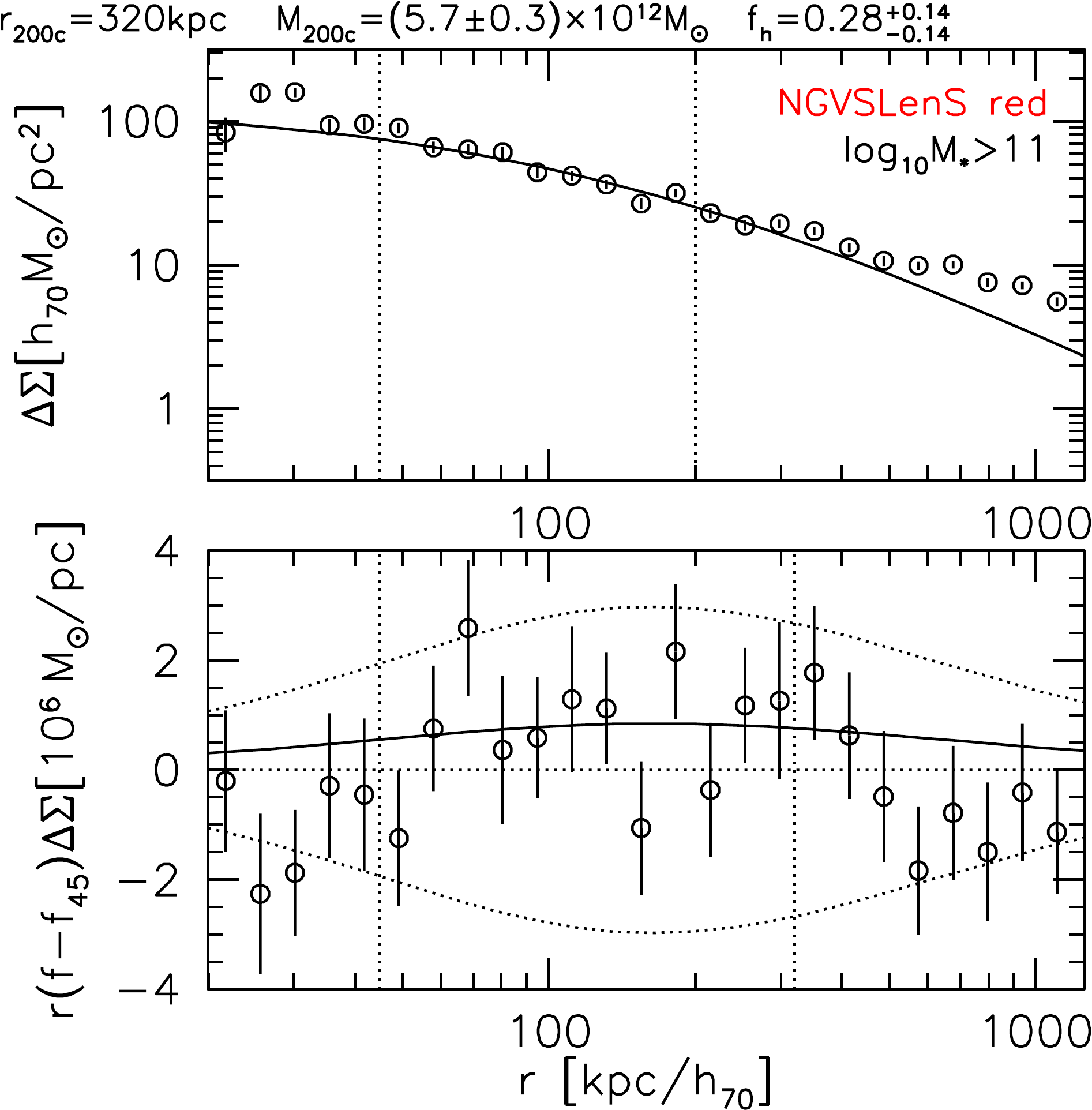}
  \includegraphics[width=7.4cm]{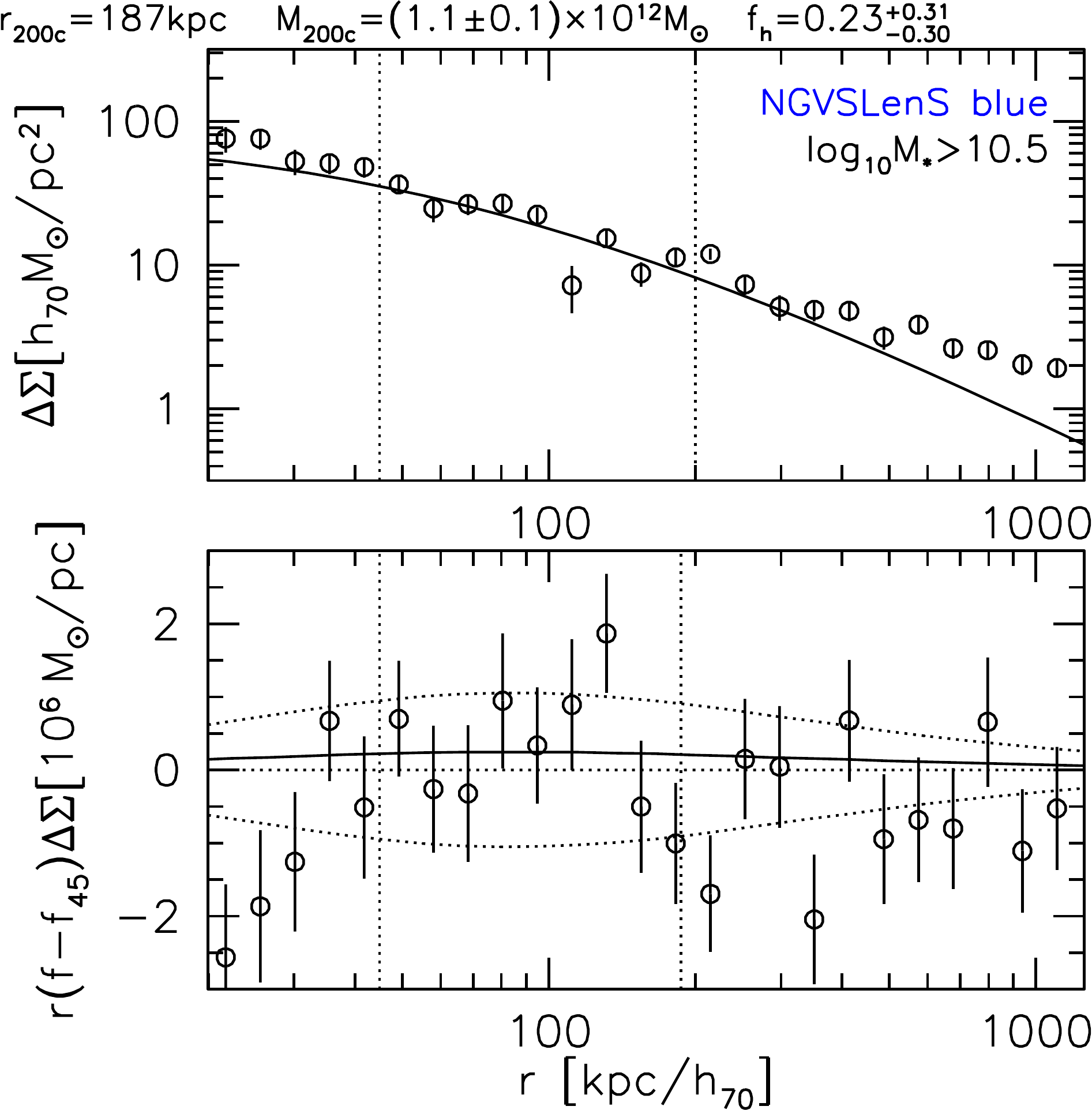}
  \includegraphics[width=7.4cm]{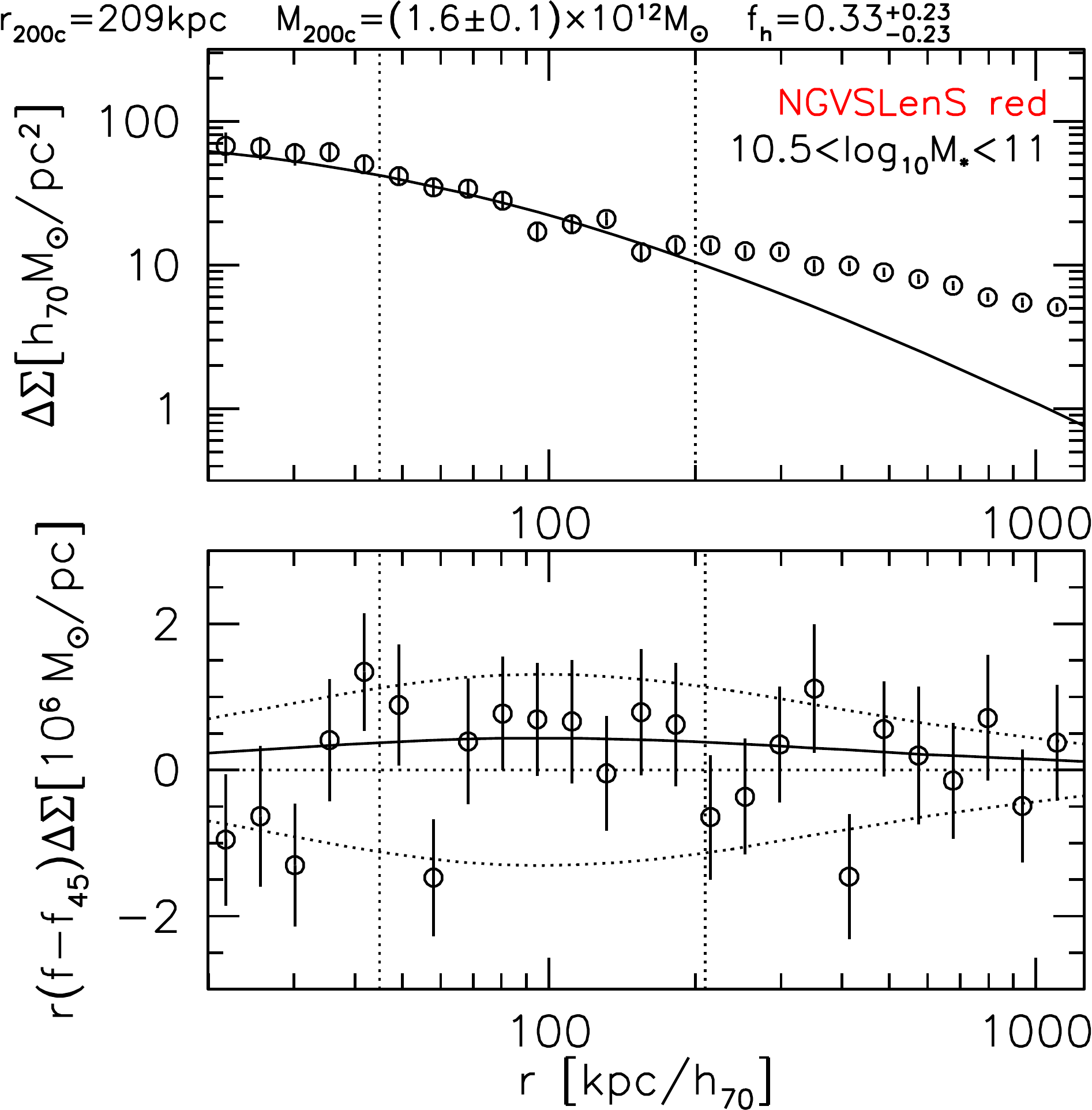}
 \includegraphics[width=7.4cm]{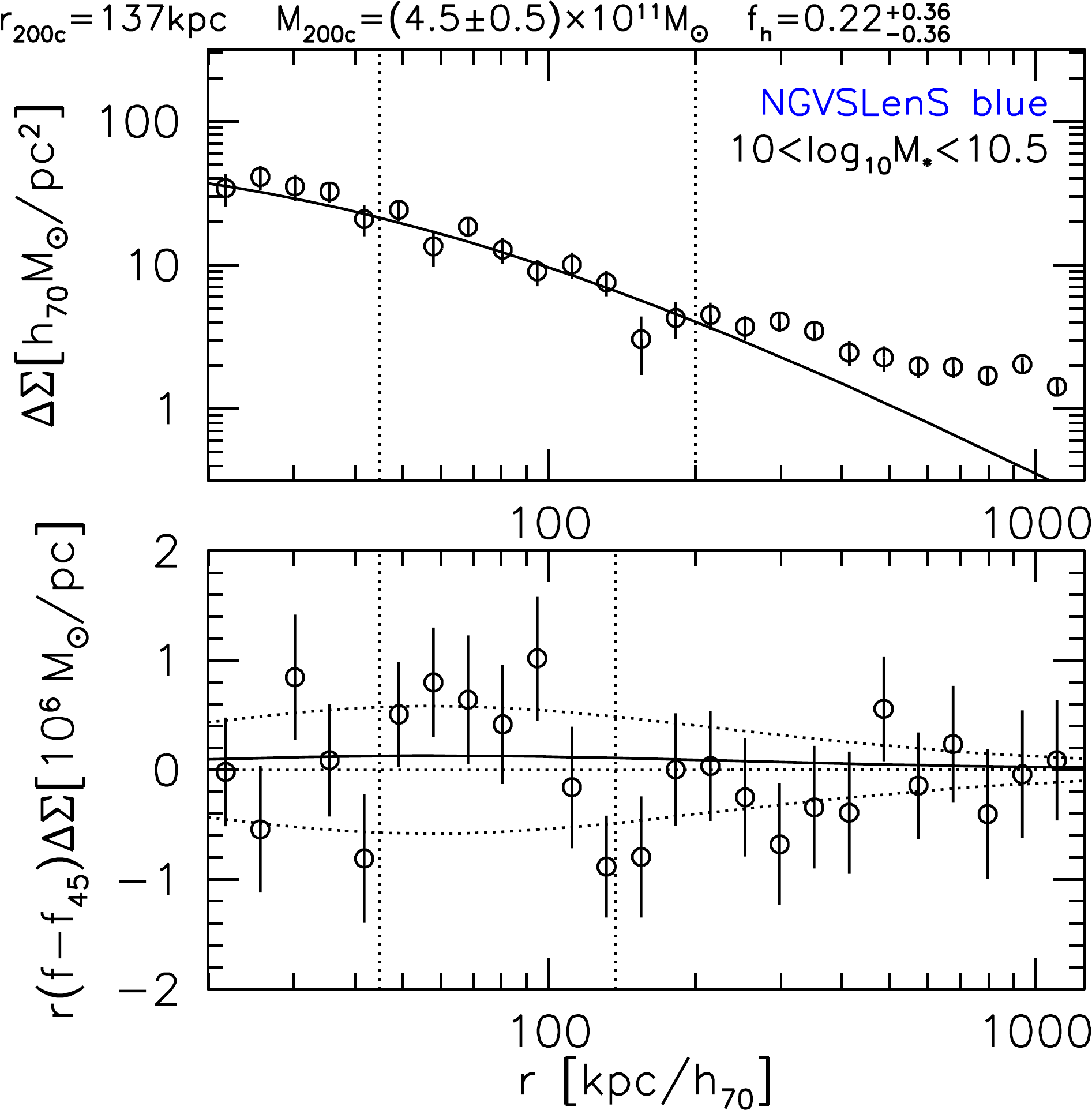}
   \includegraphics[width=7.4cm]{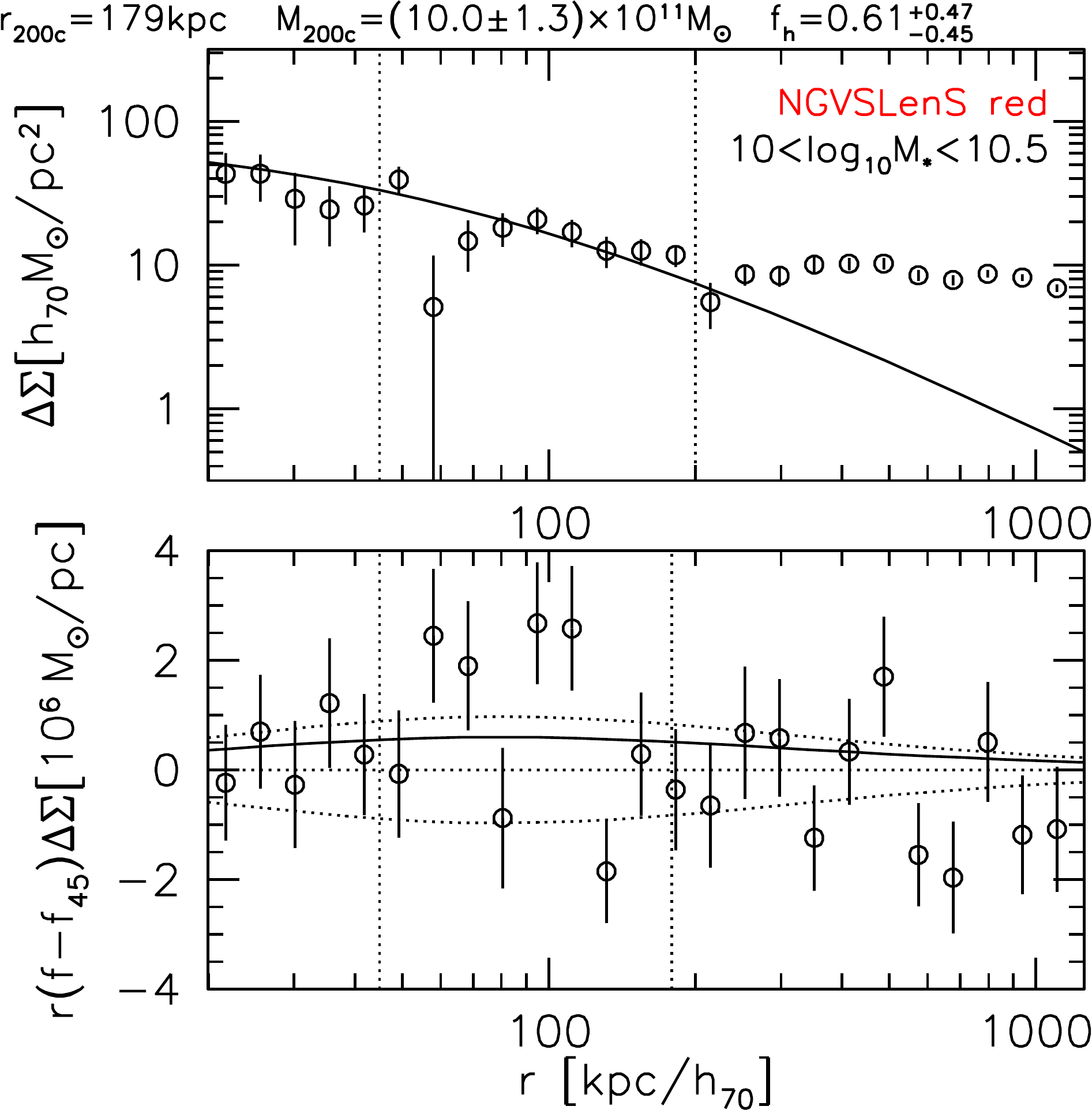}
  \includegraphics[width=7.4cm]{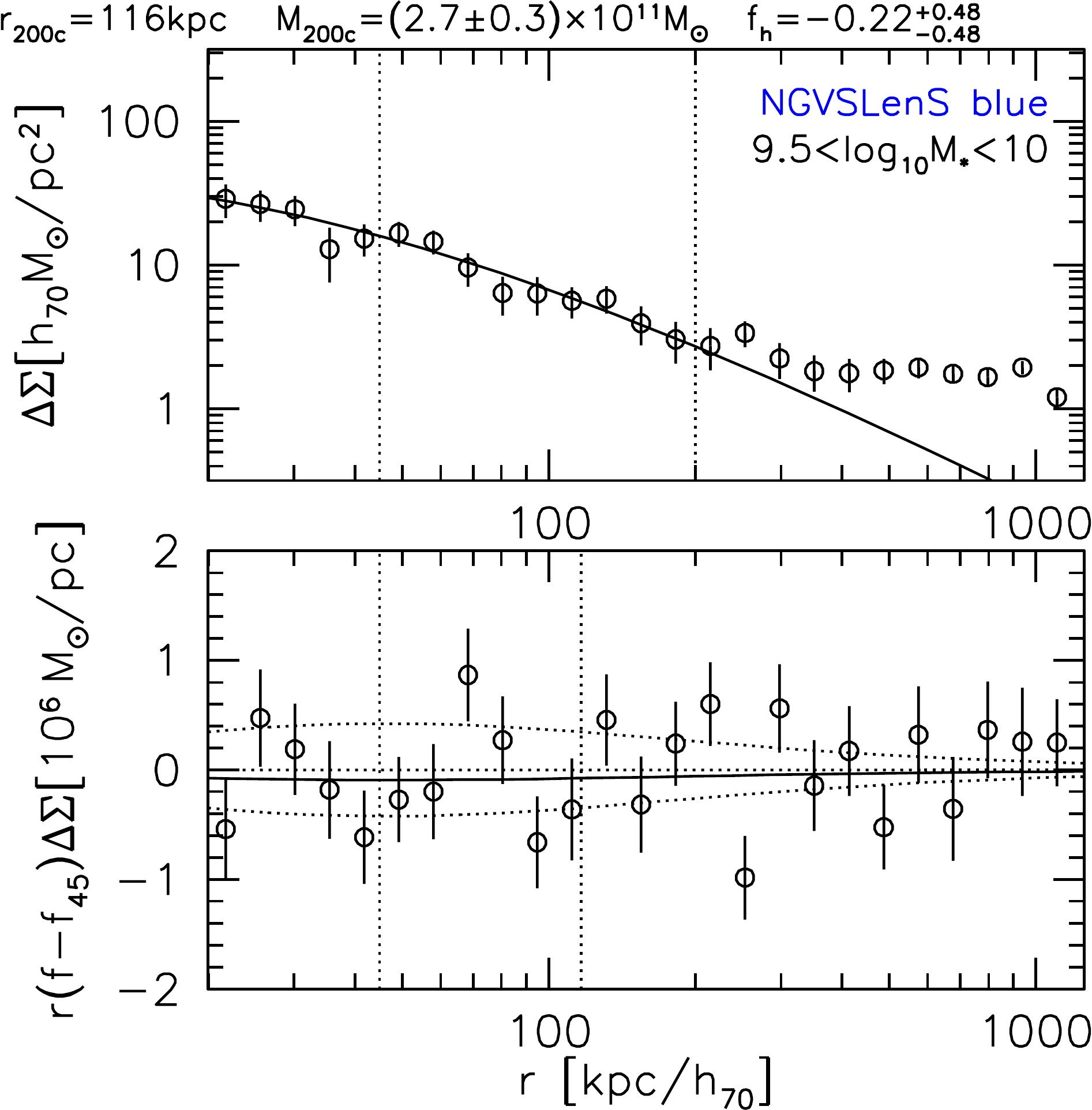}
  \caption{Measured isotropic (top sub-panel in each panel) and anisotropic
(bottom sub-panel in each panel, note the varying $y$-axis scale)
shear signal around red (left) and blue (right) lenses in
the NGVSLenS fields.
From top to bottom the different panels
show decreasing stellar  mass bins as indicated.
For better readability
 the anisotropic signal has been
 scaled by $r$.
 The best-fitting NFW shear profile
 constrained within
 \mbox{$45\thinspace\mathrm{kpc}/h_{70}<r<200\thinspace\mathrm{kpc}/h_{70}$}
is shown by the curve for the isotropic signal.
For \mbox{$(f-f_{45})\Delta\Sigma$}
the curves show
 models computed from  the  best-fit isotropic model and the
  best-fit $f_\mathrm{h}$ for the solid curve, and
  \mbox{$f_\mathrm{h}\in \{+1,0,-1\}$}
for the dotted curves, respectively.
The best-fit  $f_\mathrm{h}$ has been constrained from
\mbox{$(f-f_{45})\Delta\Sigma$} and \mbox{$\Delta\Sigma$}
within
 \mbox{$45\thinspace\mathrm{kpc}/h_{70}<r<r_{200\mathrm{c}}$} (range indicated by
 vertical lines), where
 $r_{200\mathrm{c}}$ has been estimated from the isotropic signal.
 The corresponding figures for the other surveys are shown in Appendix \ref{se:appendix_figures}.
  }
   \label{fi:shearfitsred_NGVS}
    \end{figure*}

\begin{table*}
  \renewcommand{\arraystretch}{1.2}
  \caption{Weak lensing constraints for the different stellar mass and colour bins in the different surveys, as well as joint  \mbox{$f_\mathrm{h}$} constraints.
\label{tab:results}}
\centering
\begin{tabular}{ccccccccc}
\hline
\hline
  Survey &  Colour & Stellar mass & $z_\mathrm{l,midpoint}$ & \mbox{$r_{200\mathrm{c}}$}& \mbox{$M_{200\mathrm{c}}$}& \mbox{$f_\mathrm{h}$}& \mbox{$\chi^2/\mathrm{d.o.f.}$}&  \\

& & [M$_\odot$]& & $[\mathrm{kpc}/h_{70}]$&  $[10^{11}\mathrm{M_\odot}/h_{70}]$& & &\\
  \hline
NGVSLenS& Blue & $ 9.5<\mathrm{log}_{10}M_*<10 $ & 0.4 & 116 & $ 2.7 \pm 0.3 $ & $ -0.22^{+0.48}_{-0.48} $ & $ 8.3 / 5 $ \\
NGVSLenS& Blue & $ 10<\mathrm{log}_{10}M_*<10.5 $ & 0.4 & 137 & $ 4.5 \pm 0.5 $ & $ 0.22^{+0.36}_{-0.36} $ & $ 10.9 / 6 $ \\
NGVSLenS& Blue & $ \mathrm{log}_{10}M_*>10.5 $ & 0.4 & 187 & $ 11.4 \pm 0.9 $ & $ 0.23^{+0.31}_{-0.30} $ & $ 9.0 / 8 $ \\
  \hline
NGVSLenS& Red & $ 10<\mathrm{log}_{10}M_*<10.5 $ & 0.4 & 179 & $ 10.0 \pm 1.3 $ & $ 0.61^{+0.47}_{-0.45} $ & $ 19.2 / 7 $ \\
NGVSLenS& Red & $ 10.5<\mathrm{log}_{10}M_*<11 $ & 0.4 & 209 & $ 15.9 \pm 1.1 $ & $ 0.33^{+0.23}_{-0.23} $ & $ 7.1 / 8 $ \\
NGVSLenS& Red & $ \mathrm{log}_{10}M_*>11 $ & 0.4 & 320 & $ 56.8 \pm 2.8 $ & $ 0.28^{+0.14}_{-0.14} $ & $ 9.3 / 11 $ \\
\hline
KV450& Blue & $ 9.5<\mathrm{log}_{10}M_*<10 $ & 0.4 & 117 & $ 2.8 \pm 0.3 $ & $ 0.34^{+0.59}_{-0.58} $ & $ 2.9 / 5 $ \\
KV450& Blue & $ 10<\mathrm{log}_{10}M_*<10.5 $ & 0.4 & 152 & $ 6.2 \pm 0.7 $ & $ 0.14^{+0.60}_{-0.60} $ & $ 4.8 / 6 $ \\
KV450& Blue & $ \mathrm{log}_{10}M_*>10.5 $ & 0.4 & 195 & $ 12.8 \pm 1.8 $ & $ 0.13^{+0.55}_{-0.55} $ & $ 6.2 / 8 $ \\
  \hline
KV450& Red & $ 10<\mathrm{log}_{10}M_*<10.5 $ & 0.4 & 174 & $ 9.2 \pm 1.0 $ & $ 0.34^{+0.42}_{-0.41} $ & $ 10.9 / 7 $ \\
KV450& Red & $ 10.5<\mathrm{log}_{10}M_*<11 $ & 0.4 & 253 & $ 28.2 \pm 1.5 $ & $ 0.34^{+0.18}_{-0.18} $ & $ 8.4 / 10 $ \\
KV450& Red & $ \mathrm{log}_{10}M_*>11 $ & 0.4 & 368 & $ 86.2 \pm 6.0 $ & $ 0.30^{+0.20}_{-0.20} $ & $ 21.7 / 12 $ \\
\hline
CFHTLenS& Blue & $ 9.5<\mathrm{log}_{10}M_*<10 $ & 0.4 & 128 & $ 3.6 \pm 0.7 $ & $ 0.75^{+0.72}_{-0.68} $ & $ 7.6 / 5 $ \\
CFHTLenS& Blue & $ 10<\mathrm{log}_{10}M_*<10.5 $ & 0.4 & 186 & $ 11.2 \pm 1.6 $ & $ 0.34^{+0.47}_{-0.47} $ & $ 3.5 / 8 $ \\
CFHTLenS& Blue & $ \mathrm{log}_{10}M_*>10.5 $ & 0.4 & 236 & $ 22.9 \pm 3.8 $ & $ 0.69^{+0.56}_{-0.55} $ & $ 8.2 / 9 $ \\
  \hline
CFHTLenS& Red & $ 10<\mathrm{log}_{10}M_*<10.5 $ & 0.4 & 171 & $ 8.7 \pm 1.4 $ & $ -0.39^{+0.67}_{-0.68} $ & $ 7.4 / 7 $ \\
CFHTLenS& Red & $ 10.5<\mathrm{log}_{10}M_*<11 $ & 0.4 & 247 & $ 26.1 \pm 2.3 $ & $ 0.26^{+0.32}_{-0.32} $ & $ 6.3 / 9 $ \\
CFHTLenS& Red & $ \mathrm{log}_{10}M_*>11 $ & 0.4 & 366 & $ 85.0 \pm 8.5 $ & $ 0.09^{+0.35}_{-0.35} $ & $ 8.7 / 12 $ \\
\hline
CS82& Blue & $ 9.5<\mathrm{log}_{10}M_*<10 $ & 0.4 & 107 & $ 2.1 \pm 1.0 $ & $ -2.88^{+5.57}_{-17.12} $ & $ 5.7 / 4 $ \\
CS82& Blue & $ 10<\mathrm{log}_{10}M_*<10.5 $ & 0.4 & 139 & $ 4.6 \pm 2.4 $ & $ 1.01^{+2.79}_{-2.54} $ & $ 2.2 / 6 $ \\
CS82& Blue & $ \mathrm{log}_{10}M_*>10.5 $ & 0.4 & 234 & $ 22.3 \pm 11.8 $ & $ 1.96^{+2.99}_{-1.89} $ & $ 7.2 / 9 $ \\
  \hline
CS82& Red & $ 10<\mathrm{log}_{10}M_*<10.5 $ & 0.4 & 146 & $ 5.4 \pm 2.4 $ & $ -0.66^{+1.16}_{-1.36} $ & $ 7.7 / 6 $ \\
CS82& Red & $ 10.5<\mathrm{log}_{10}M_*<11 $ & 0.4 & 235 & $ 22.5 \pm 5.0 $ & $ 0.25^{+0.75}_{-0.74} $ & $ 11.2 / 9 $ \\
CS82& Red & $ \mathrm{log}_{10}M_*>11 $ & 0.4 & 340 & $ 68.1 \pm 11.7 $ & $ 0.64^{+0.47}_{-0.46} $ & $ 10.4 / 11 $ \\
\hline
RCSLenS& Red & $ 10<\mathrm{log}_{10}M_*<10.5 $ & 0.5 & 199 & $ 15.4 \pm 4.8 $ & $ 1.10^{+1.26}_{-1.17} $ & $ 11.7 / 8 $ \\
RCSLenS& Red & $ 10.5<\mathrm{log}_{10}M_*<11 $ & 0.5 & 242 & $ 27.7 \pm 7.5 $ & $ 0.40^{+0.94}_{-0.92} $ & $ 6.8 / 9 $ \\
  RCSLenS& Red & $ \mathrm{log}_{10}M_*>11 $ & 0.5 & 403 & $ 127.4 \pm 50.0 $ & $ -0.22^{+1.40}_{-1.38} $ & $ 12.7 / 12 $ \\
  \hline
  All & Red & $\mathrm{log}_{10}M_*>10$ & & & &  $0.303^{+0.080}_{-0.079}$& $160.1/152$ \\
 All & Red & $\mathrm{log}_{10}M_*>10.5$ & & & &  $0.304^{+0.083}_{-0.081}$& $104.1/112$  \\
 All & Blue & $\mathrm{log}_{10}M_*>9.5$ & & & &  $0.217^{+0.160}_{-0.159}$& $78.8/90$\\

  \hline
\hline
\end{tabular}
\renewcommand{\arraystretch}{1.0}
\end{table*}

Figure \ref{fi:shearfitsred_NGVS} shows the measured isotropic $\widehat{\Delta \Sigma}(r)$  and anisotropic  $\widehat{\left[f(r)-f_{45}(r)\right]\Delta
  \Sigma_\mathrm{iso}(r)}$ profiles, where the latter is scaled by $r$ for better readability,
estimated in 25 logarithmic
bins of transverse physical separation between 20{\thinspace}kpc/$h_{70}$  and 1.2{\thinspace}Mpc/$h_{70}$
for the different stellar mass and colour bins of the NGVSLenS analysis.
The corresponding figures for the other surveys can be found in Appendix \ref{se:appendix_figures}.

In order to compute the average signal and error-bars shown in Fig.\thinspace\ref{fi:shearfitsred_NGVS} we follow \citetalias{shh15}, first splitting the lens catalogues
into patches covering \mbox{$\sim 1 \mathrm{deg}^2$} (defined via the original survey pointings).
For each patch we compute the profiles for each lens bin, employing larger cutouts from the mosaic source catalogue to ensure good coverage at the edges of a patch.
For each lens colour and stellar mass bin we then compute the combined signal from all patches and thin lens redshift slices, weighting contributions
according to their individual weight sums in the corresponding  radial bin.
Error-bars are then computed from 30,000 bootstrap re-samples of the contributing patches.

To robustly constrain halo shapes we aim to only  include scales in the fit that are dominated by the host dark matter haloes of the lens galaxies.
Following \citetalias{shh15} we therefore include scales  \mbox{$45$\thinspace{kpc}/$h_{70}<r<r_{200\mathrm{c}}$} when constraining $f_\mathrm{h}$,
and scales
\mbox{$45$\thinspace{kpc}/$h_{70}<r<200$\thinspace{kpc}/$h_{70}$} for an initial fit of the isotropic signal that is used to estimate $r_{200\mathrm{c}}$ and  halo masses.
Smaller scales  can carry significant contributions from the baryonic matter distribution (see the small-scale increase in the isotropic signal in the top panels of Fig.\thinspace\ref{fi:shearfitsred_NGVS}), and for the lenses with the highest stellar mass these small scales also suffer from
a dependence of the density of the selected sources on the position angle from the lens major axis (see \citetalias{shh15}).
Larger radii are excluded from the fits as the shear profile may be significantly affected (see the excess isotropic signal at large radii in Fig.\thinspace\ref{fi:shearfitsred_NGVS}) either by neighbouring haloes or the parent halo if the lens is not a central galaxy but a satellite \citep[see e.g.][but we note that we apply cuts to reduce the fraction of satellites as explained in Sect.\thinspace\ref{se:lenssample}]{vuh14}.
Within our fit range the $\widehat{\Delta\Sigma}(r)$ profiles are generally well described by the employed NFW models (see Fig.\thinspace\ref{fi:shearfitsred_NGVS} and Figs.\thinspace\ref{fi:shearfitsred_KV450} to \ref{fi:shearfitsred_RCS2}).

Table \ref{tab:results} summarises the results we obtain for the different surveys and lens colour and stellar mass bins.
Here we notice significant differences in the estimated halo masses between the different surveys for some of the colour and stellar mass bins.
Uncertainties in the shape and redshift calibrations of the source galaxies may contribute to these differences, but we suspect that they are dominantly caused
by differences in the stellar mass definitions of the different surveys (see Sect.\thinspace\ref{se:data}).
This is also hinted at
by the fact that the lens ellipticity dispersion in particular lens colour and stellar mass bin combinations differs significantly between the different surveys
(see Table \ref{tab:lenses}).
Likewise, the fractions of how many lenses fall into the different stellar mass bins vary between the surveys (compare Table \ref{tab:lenses}).
These differences in the stellar mass definitions, as well as the significant depth differences between the surveys  are the main reasons
why  we initially analyse the surveys separately.
However, this does not limit our ability to constrain $f_\mathrm{h}$, given that the stellar masses are only used as a proxy to select approximately similar lenses within each particular lens bin.

\begin{figure*}
   \centering
  \includegraphics[width=9cm]{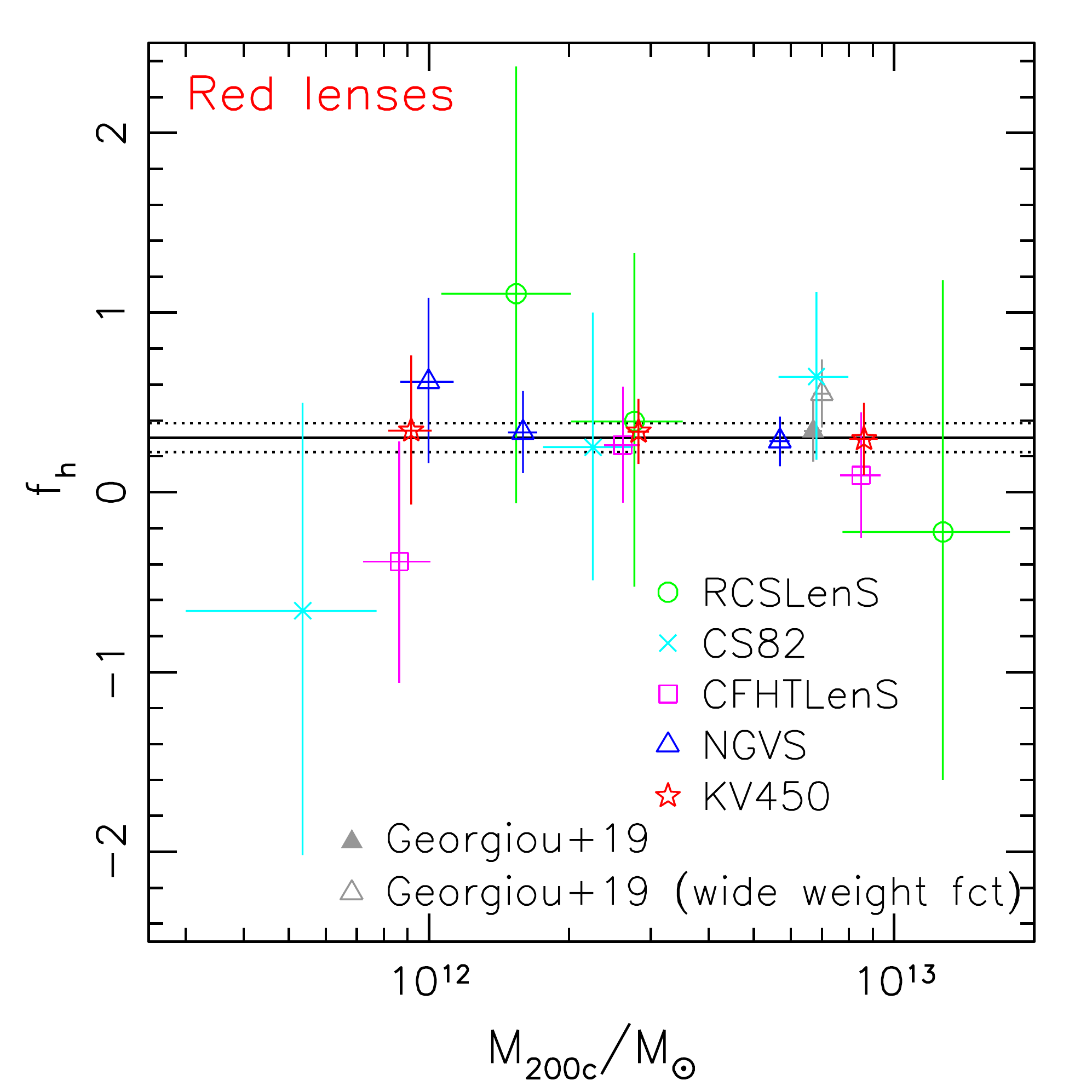}
  \includegraphics[width=9cm]{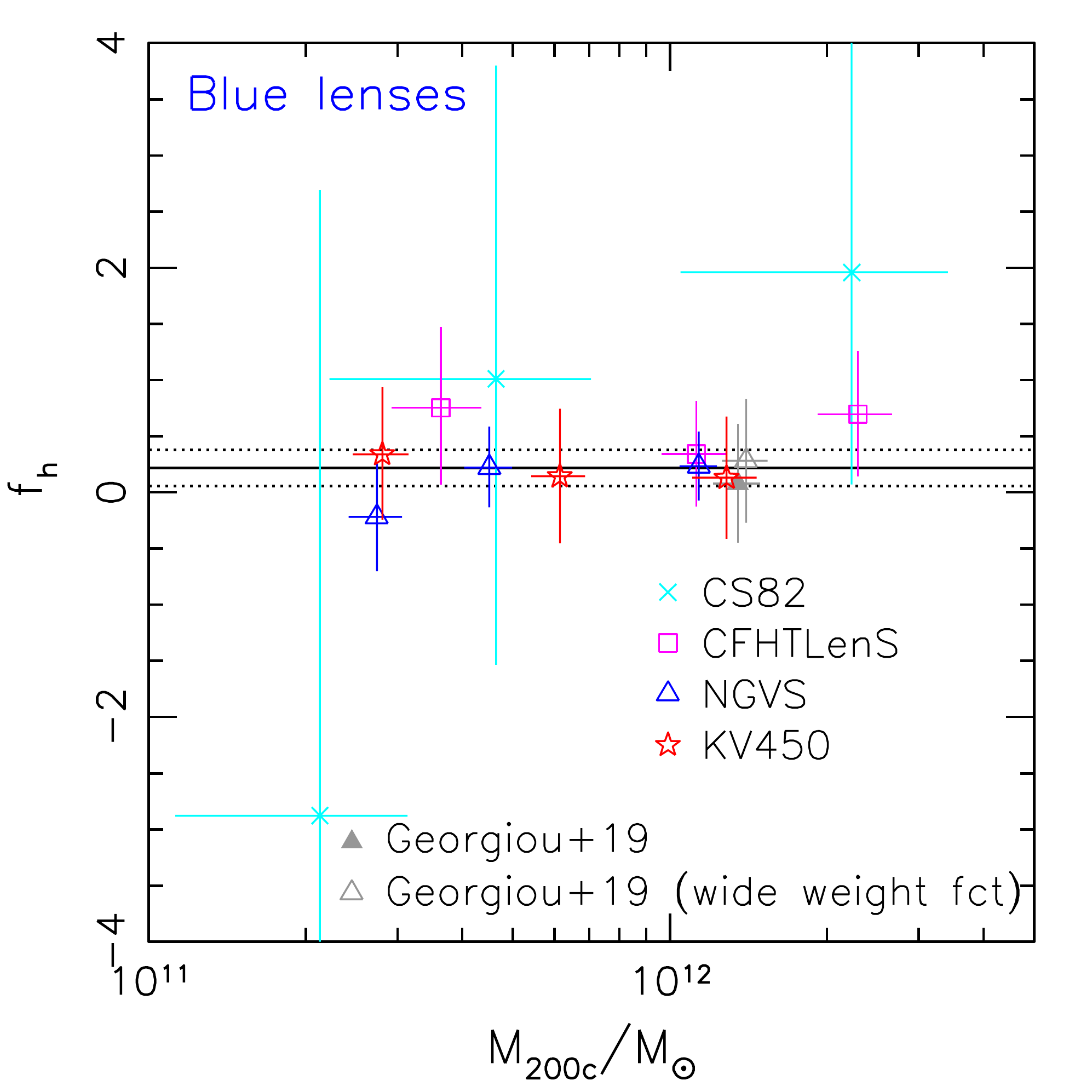}
  \caption{Constraints on \mbox{$f_\mathrm{h}$} and \mbox{$M_\mathrm{200c}$} from the different surveys for red (left) and blue (right) lenses.
    For each survey, the different data points correspond to the different stellar mass bins. The horizontal solid and dotted lines mark the best-fit joint \mbox{$f_\mathrm{h}$} constraints and the \mbox{$\pm 1\sigma$} limits, respectively.
    For comparison we also show results from  the KiDS-1000 analysis of central galaxies from \citet{georgiou19}, including their default constraints and their results achieved using a wider weight function for the shape measurements (shown with an offset in mass for clarity).
}
   \label{fi:join_constraints}
 \end{figure*}

Individual $f_\mathrm{h}$ constraints for the different surveys and lens bins are noisy (see Table \ref{tab:results}), where the most significant (\mbox{$\simeq 1.5$--$2\sigma$}) deviations from zero are found for the more massive red lenses in NGVSLenS and KV450.
As explained in Sect.\thinspace\ref{se:getfh} we then compute joint constraints on $f_\mathrm{h}$
from the $\widehat{\Delta \Sigma}(r)$  and  $\widehat{\left[f(r)-f_{45}(r)\right]\Delta
  \Sigma_\mathrm{iso}(r)}$ profiles of the different surveys.
When including all surveys and stellar mass bins, we obtain \mbox{$f_\mathrm{h}=0.303^{+0.080}_{-0.079}$} for red lenses, and \mbox{$f_\mathrm{h}=0.217^{+0.160}_{-0.159}$} for blue lenses.
These joint constraints are also indicated in Fig.\thinspace\ref{fi:join_constraints},
in which we plot the constraints on \mbox{$f_\mathrm{h}$} versus $M_\mathrm{200c}$ for the individual surveys and stellar mass bins. Our results can also be compared to a recent halo shape analysis of the fourth KiDS data release \citep[][see Fig.\thinspace\ref{fi:join_constraints}]{georgiou19},
which we discuss further in Sect.\thinspace\ref{se:discussions}.
Following  \citetalias{shh15}
we alternatively drop the lowest stellar mass bin for red lenses, since this bin likely includes the highest fraction of satellite galaxies \citep[e.g.][]{vuh14}. In this case the resulting joint constraints \mbox{$f_\mathrm{h}=0.304^{+0.083}_{-0.081}$} are essentially unchanged with only minimally inflated errors, suggesting that they
are robust and dominated by the more massive lenses.
We note that the \mbox{$\chi^2/\mathrm{d.o.f.}$} values for the joint constraints are
fully  consistent with expected statistical fluctuations (see Table \ref{tab:results}), corresponding to $p$-values of 0.31 (0.79) for the analysis combining all red (blue) lenses.

\begin{figure*}
   \centering
  \includegraphics[width=9cm]{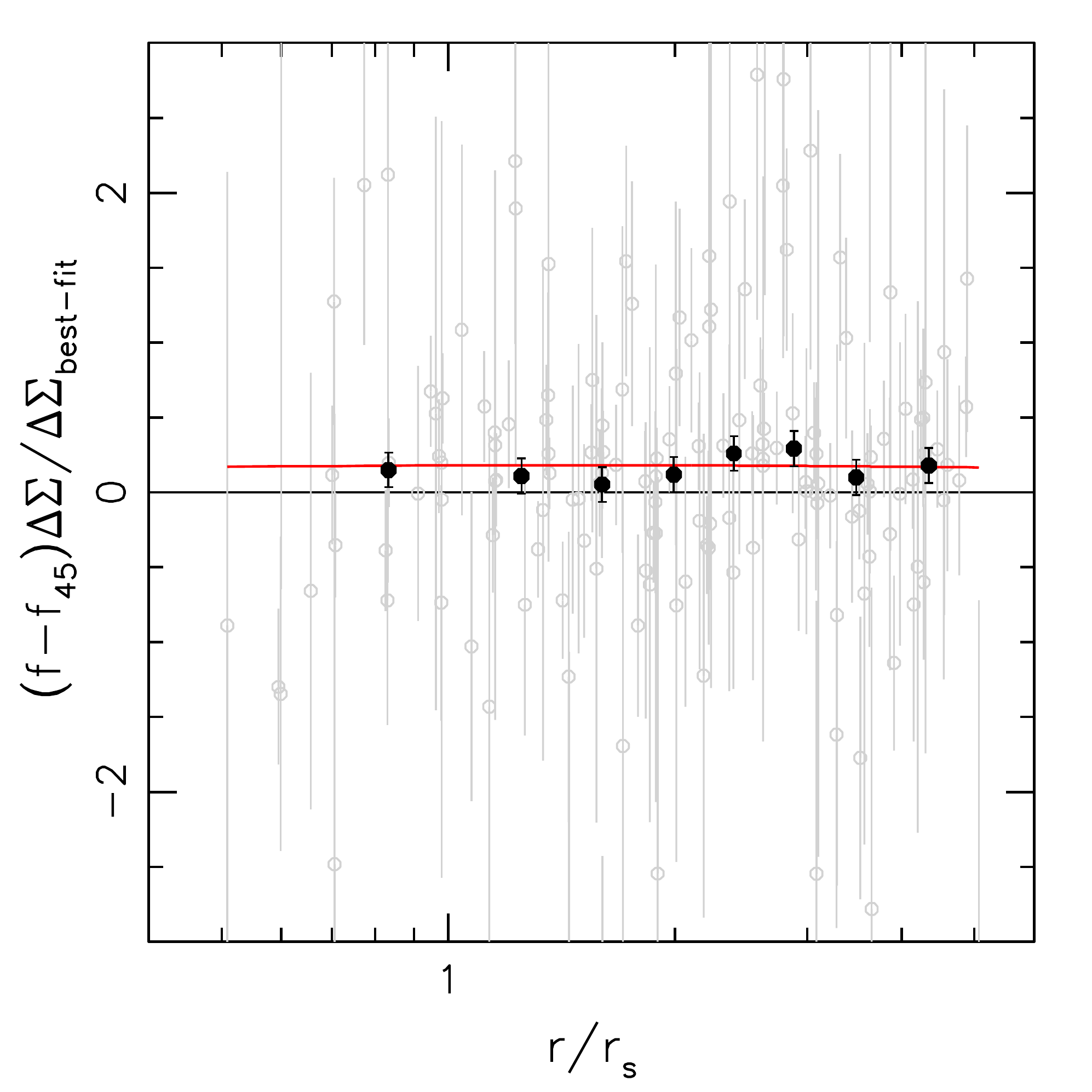}
  \includegraphics[width=9cm]{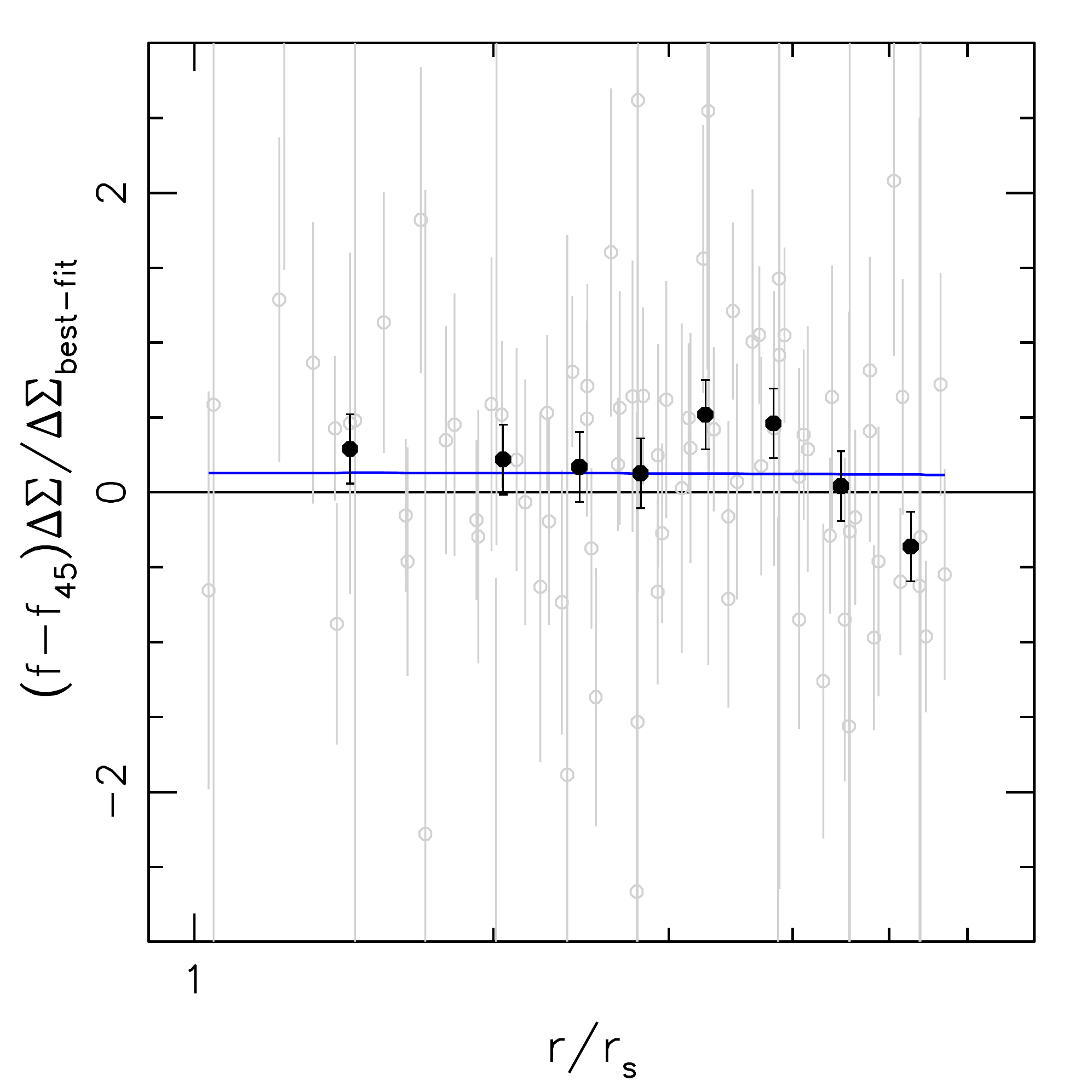}
  \caption{Ratio of the measured \mbox{$(f-f_{45})\Delta\Sigma(r)$} profiles and the best-fit models for the spherical   \mbox{$\Delta\Sigma(r)$}  profiles as a function of the radius in units of the NFW scale radius $r_\mathrm{s}$.
The left panel corresponds to red lenses, while blue lenses are shown on the right.
Each grey point corresponds to one data point
from Figs.\thinspace\ref{fi:shearfitsred_NGVS} and  \ref{fi:shearfitsred_KV450} to \ref{fi:shearfitsred_RCS2} (if it is located within the included fit range, see Sect.\thinspace\ref{se:halo_shapes_results}). Binned averages of these points are shown in black. The curves show model predictions \mbox{$(f_\mathrm{rel}-f_\mathrm{rel,45})f_\mathrm{h}$} for the best-fit joint $f_\mathrm{h}$ estimates for all red lenses (left) and all blue lenses (right).
}
   \label{fi:stacked_shear}
 \end{figure*}

For illustration we also show joint representations for
the anisotropy of the shear field from all surveys and lens bins as a function of $r/r_\mathrm{s}$ in Fig.\thinspace\ref{fi:stacked_shear}.
To achieve this, we divide the measured  \mbox{$(f-f_{45})\Delta\Sigma(r)$} profiles by the best-fit models for the spherical   \mbox{$\Delta\Sigma(r)$}  profiles.
As visible from the binned data points (shown in black) there is
no significant radius-dependent deviation from the model  prediction \mbox{$(f_\mathrm{rel}-f_\mathrm{rel,45})f_\mathrm{h}$}, where $f_\mathrm{h}$ corresponds to the best-fit joint estimate for all red lenses and for all blue lenses, respectively.

\section{Discussion}
\label{se:discussions}
Combining measurements from five different weak lensing surveys, we have been able to
tighten constraints on halo ellipticity substantially compared to the previous CFHTLenS-only analysis conducted by
\citetalias{shh15}, which employed the same methodology.
While all surveys contribute to our new constraints, the most constraining contributions come from NGVSLenS
and KV450 (compare Table \ref{tab:results}), which is thanks to the excellent depth and superb seeing in the case of NGVS, and the excellent data quality and wide area in the case of KV450.

For red lenses we now obtain  \mbox{$f_\mathrm{h}=0.303^{+0.080}_{-0.079}$} (a $3.8\sigma$ detection of non-zero $f_\mathrm{h}$), compared to \mbox{$f_\mathrm{h}=0.217^{+0.160}_{-0.159}$} for blue lenses.
These results are in excellent agreement with
predictions that \citetalias{shh15} derived using mock data based on the Millennium Simulation \citep{swj05,hhw09}.
The lens galaxy shapes for these mock data were initially computed by \citet{jsb13},
employing the scheme from \citet{hwh06}.
This assumes that the shapes of early-type galaxies follow those of their host dark matter haloes, while disc-dominated  late-type galaxies
are initially aligned such that their spin vector is
parallel
to the angular momentum vector of the corresponding host halo.
\citet{jsh13} added misalignment between galaxies and their host dark matter haloes to these mock catalogues.
For early-type galaxies
they
employed a
Gaussian misalignment distribution
with a scatter of 35$^\circ$,
as estimated   from the distribution of satellites around SDSS luminous red galaxies \citep{ojl09}.
This corresponds to a mean absolute misalignment angle of \mbox{$\sim 28^\circ$}, which is broadly consistent with results from hydrodynamical simulations \citep[see][]{tmm14}.
For late-type galaxies \citet{jsh13} used the fitting function from \citet{bet12}, which was derived from simulated haloes that include baryons and incorporate models of galaxy formation physics
\citep{bef10,dmf11,ctd09,oef05}.
Combining lenses in  halo mass ranges that are similar to our analysis, \citetalias{shh15} estimate  \mbox{$f_\mathrm{h}=0.616_{-0.005}^{+0.006}$} (\mbox{$0.095\pm 0.005$}) for perfectly aligned early- (late-) type mock galaxies,
which reduces to \mbox{$f_\mathrm{h}=0.285_{-0.004}^{+0.006}$} (\mbox{$0.025_{-0.004}^{+0.006}$})
when including misalignment,
in excellent agreement with our measurements.
As a caveat we note that the early-type mock galaxies  have noticeably lower intrinsic ellipticity dispersions
compared to the real red lens galaxies. To account for this, \citetalias{shh15} considered different schemes to re-scale their results which lower the  \mbox{$f_\mathrm{h}$} predictions from the simulation.
If we rescale according to the factor \mbox{$\langle|\epsilon_\mathrm{g}|\rangle^\mathrm{surveys}/\langle|\epsilon_\mathrm{g}|\rangle^\mathrm{simulation}=1.53$} (considering lenses with \mbox{$0.05<|\epsilon_\mathrm{g}|<0.95$})
we would expect
\mbox{$f_\mathrm{h}\simeq 0.19$} from the simulation, which is still consistent with our measurement for red lenses at the $1.5\sigma$ level.

The good agreement of our measurements with the current model predictions provides an important consistency check for galaxy formation models.
This is also of direct relevance for cosmic shear measurements \citep[e.g.][]{shj10,hgh13,jth16,troxel18,hamana20,ath20,hildebrandt20}, which can be biased low due to shape--shear alignments \citep[e.g.][]{his04,jck15,tus18}.
These are caused by alignments of galaxies with their surrounding dark matter haloes and large-scale structure environments, which shear the images of background galaxies.
This is linked to our measurements in two ways: First, the flattened halo contributes to the shearing of background galaxies and therefore the shape--shear alignments at small scales \citep{bra07}. Second, the misalignment between the lens shape and its surrounding dark matter distribution reduces the net strength of the shape--shear correlation.
Constraints such as the ones derived from our analysis can therefore be used to update  shape--shear corrections for cosmic shear.
At large scales, these corrections can be calibrated from the alignment of galaxy ellipticities with their surrounding galaxy distributions, but
 at small scales  (\mbox{$\lesssim 200$ kpc}) uncertainties regarding the
galaxy bias prohibit the robust application of this approach \citep{hilbert17}.
Fortunately, these are precisely the scales at which asymmetries in the matter distribution are probed by weak lensing analyses of galaxy halo shapes.
We therefore suggest that future analyses of hydrodynamical simulations, which are used to constrain and calibrate intrinsic alignment models for cosmic shear \citep[see e.g.][]{tmd16,ckj17,hilbert17,pjs18}, also provide direct predictions for $f_\mathrm{h}$, which can then be compared to the observational constraints.

We stress the importance of comparing
observational constraints on \mbox{$f_\mathrm{h}$} to results from consistently analysed mock data, in order to properly account for imperfections in the analysis scheme. For example, the mock analysis from \citetalias{shh15} shows that the systematics-corrected analysis scheme introduced by  \citetalias{mhb06} is capable to approximately, but not perfectly, correct for the impact of cosmic shear and large-scale shape--shear correlations.
  Small residuals should however consistently occur in the real data and the mock analysis, thus allowing for a fair comparison.
  Likewise, our modelling approach assumes a linear scaling between  galaxy ellipticity and halo ellipticity,
  as well as ellipticity-independent misalignment (see Sect.\thinspace\ref{se:constrain-ani}),
  neither of which assumptions is likely met exactly in reality.
  While this likely affects the absolute values of the resulting \mbox{$f_\mathrm{h}$} constraints,
  this is not a concern for a relative comparison
to predictions from simulations,
  as long as the same assumptions are made when analysing the simulated data. In particular,  \citetalias{shh15} make the same assumptions in their analysis of mock data, which is why our results are directly comparable to their predictions.
  Yet, in order to further improve future mock predictions, it will be important to further increase the realism of the galaxy shapes and misalignment models employed when creating the simulations.

Observational constraints on halo ellipticity have been obtained by a number of previous weak lensing studies.
Among the earlier studies aiming to constrain halo ellipticity,
\citet{hyg04} and \citet{phh07} analysed single-band imaging from RCS and CFHTLS without subdividing into red and blue lenses.
Assuming elliptical truncated isothermal sphere models and conducting a maximum likelihood
analysis, \citet{hyg04} obtained \mbox{$f_\mathrm{h}=0.77^{+0.18}_{-0.21}$},
while \citet{phh07} investigated the ratio of the shears averaged in quadrants along
the lens  minor and major axes, yielding a tentative deviation from isotropy (they measured a mean ratio \mbox{$0.76\pm 0.10$} when including scales out to 70$^{\prime\prime}$).
The \mbox{$f_\mathrm{h}$} estimate from \citet{hyg04} is surprisingly high compared to our results.
While the different assumptions regarding the density profiles affect \mbox{$f_\mathrm{h}$} constraints (see \citetalias{mhb06}),
a more important reason for the discrepancy is likely given by the lack of a correction
for the spurious signal caused by other sources of lens--source ellipticity alignments in the \citet{hyg04} analysis.
Such alignments can bias \mbox{$f_\mathrm{h}$} constraints low in the case of contributions from cosmic shear or consistently too large or too small PSF anisotropy corrections  (\citetalias{mhb06}).
Likewise, they can bias them high for contributions from large-scale shape--shear correlations (\citetalias{shh15}).
\citetalias{shh15} argue that the latter effect may have biased the constraints from \citet{hyg04} high given  that their data are relatively shallow, which leads to a stronger
impact of  shape--shear  correlations compared to cosmic shear.

The methodology used in our analysis to correct for spurious signal
was introduced by \citetalias{mhb06}. Their analysis of SDSS data yielded \mbox{$f_\mathrm{h}=0.60 \pm 0.38$}
for red and  \mbox{$f_\mathrm{h}=-1.4^{+1.7}_{-2.0}$} for blue lenses when assuming  elliptical NFW mass profiles. While these constraints are consistent with our results, they are likely affected by a sign error in the $f_{45}$ model prediction as identified by \citetalias{shh15}.
The same is the case for the analysis of  \citet{uhs12}, who
find
\mbox{$f_\mathrm{h}=0.20^{+1.34}_{-1.31}$} for red lenses and \mbox{$f_\mathrm{h}=-2.17^{+1.97}_{-2.03}$}
for blue lenses from RCS2 data when assuming elliptical NFW  profiles and using the same lens ellipticity weighting as employed  in our study.
While we also incorporate RCS2 data into our study (RCSLenS), there are substantial
differences.
In particular, we employ \emph{lens}fit shapes for the sources and fainter lenses \citep[compared to KSB shapes for ][]{uhs12}, restrict the analysis to fields with four-band photometry (see Sect.\thinspace\ref{se:data:rcslens}),
and only employ red lenses at \mbox{$0.4<z_\mathrm{b}<0.6$} due to their better photo-$z$ performance given the lack of $u$-band data (see Sect.\thinspace\ref{se:lenssample}). Most importantly, we apply the sign correction for the  $f_{45}$ model (see Sect.\thinspace\ref{se:constrain-ani}).
Together, this allows us to tighten the joint constraints  for red lenses from RCS2 data noticeably compared to the previous results from  \citet{uhs12} to \mbox{$f_\mathrm{h}=0.59_{-0.66}^{+0.67}$} (see Table \ref{tab:results} for results on the individual stellar mass bins).
Nevertheless, RCSLenS provides the weakest contribution to our overall joint constraints from the five surveys (see Fig.\thinspace\ref{fi:join_constraints}).
This is due to the
comparably shallow source catalogue   and
the lack of $u$-band data, which
leads to noisier photometric redshift estimates
and causes us to only employ lenses in a narrower redshift range (see Sect.\thinspace\ref{se:lenssample}).

Within
the uncertainties our measurements are fully consistent with the
CFHTLenS results from \citetalias{shh15}, who used the same methodology as we do and obtained \mbox{$f_\mathrm{h}=-0.04\pm 0.25$} for red lenses and \mbox{$f_\mathrm{h}=0.69_{-0.36}^{+0.37}$} for blue lenses.
Their and our results are of course not independent given that we include CFHTLenS data in our analysis.
Different to  \citetalias{shh15} we no longer remove \mbox{$z_\mathrm{b}>1.3$} galaxies from our source sample, which leads to a moderate tightening of the CFHTLenS-only constraints
in our analysis\footnote{This is thanks to the
high lensing efficiency weights of the
\mbox{$z_\mathrm{b}>1.3$} sources and their
non-negligible
  number (e.g.~they constitute 29\% of all CFHTLenS sources with \mbox{$z_\mathrm{b}>0.7$}).}
and shifts  best-fitting values
well
within the error-bars to
\mbox{$f_\mathrm{h}=0.09\pm 0.23$} (\mbox{$f_\mathrm{h}=0.54^{+0.33}_{-0.32}$})
for red (blue) lenses  (see Table \ref{tab:results} for results on the individual stellar mass bins).

Very recently, the same methodology was applied by \citet{georgiou19},
who used data from the latest KiDS data release \citep[][KiDS-1000]{kuijken19} to constrain \mbox{$f_\mathrm{h}$} for a highly pure  sample of central galaxies.
In their default analysis they obtain  \mbox{$f_\mathrm{h}=0.34\pm0.17$} for red lenses  and  \mbox{$f_\mathrm{h}=0.08\pm0.53$} for blue lenses, which increases to  \mbox{$f_\mathrm{h}=0.55\pm 0.19$} (\mbox{$f_\mathrm{h}=0.28\pm 0.55$}) for red (blue) lenses when they employ a 1.5 times wider
weight function in the measurement of lens galaxy shapes using \texttt{DEIMOS} \citep{melchior11}.
In our analysis we incorporate data from the previous KiDS KV-450 release, which covers slightly less than half of the area of KiDS-1000.
A direct comparison between their and our study is complicated by  differences in the lens selections, but we can attempt to match samples based
on the estimated halo mass.
Their blue sample is most similar to the blue KV-450 lenses in our highest stellar mass bin, for which we obtain \mbox{$f_\mathrm{h}=0.13\pm 0.55$} in excellent agreement with both of their analyses schemes.
Likewise, their red lenses are best matched by the combination of our two highest stellar mass bins of KV450 red lenses that yield \mbox{$f_\mathrm{h}=0.32\pm 0.14$}, which is in excellent agreement with their default analysis and still consistent with their results obtained using a wider weight function.
Both studies achieve similar statistical uncertainties, where our photometric selection leads to a larger lens sample, which also extends to lower masses (compare Fig.\thinspace\ref{fi:join_constraints}). This is compensated by the larger sky area in their analysis.
The increase  \citet{georgiou19} observe in their \mbox{$f_\mathrm{h}$} constraints when using a wider weight function
is interesting, as it also relates to recent
results that suggest that central galaxies may be more aligned with their satellite distribution if their shapes are measured with more sensitivity to the galaxy outskirts \citep{hmf16,gce19}.
We note that we also employ a slightly widened weight function when computing KSB+ moments for those galaxies without successful \emph{lens}fit shape estimates (see Sect.\thinspace\ref{se:shapes_bright}).
A detailed comparison to this aspect of the \citet{georgiou19}  results is
complicated by differences between the shape measurement algorithms (\emph{lens}fit and KSB+ versus \texttt{DEIMOS}),
which is why we
defer  further investigations of the  weight function-dependence of  \mbox{$f_\mathrm{h}$} constraints to future work.

A number of studies have also constrained halo ellipticity at the mass scale of galaxy groups and clusters.
For example, \citet{vanuitert17} constrain the average halo ellipticity of groups in the GAMA survey using KiDS weak lensing data assuming elliptical NFW density profiles.
They find that the shear signal around the brightest group/cluster member (BCG) shows substantial anisotropy at scales \mbox{$r<250$ kpc}, which yields an average halo ellipticity of \mbox{$\langle|\epsilon_\mathrm{h}|\rangle=0.38\pm 0.12$} when assuming perfect alignment with the BCG orientation.
At larger scales their signal becomes isotropic with respect to the BCG orientation, but still remains anisotropic when compared to the spatial distribution of group members.
\citet{clj16} constrain the anisotropy in the lensing signal around luminous red galaxies (LRGs) that are similar to the BCGs from  \citet{vanuitert17} yielding a smaller value
\mbox{$\langle|\epsilon_\mathrm{h}|\rangle=0.12\pm 0.03$} (in our ellipticity definition).
Building up on this work, \citet{scj18} infer a best-fit average axis ratio of \mbox{$q=0.56\pm 0.09 (\mathrm{stat.}) \pm 0.03
  (\mathrm{sys.})$} in the mass distribution for a  weak lensing analysis of redMaPPer \citep{rykoff14} clusters, corresponding to \mbox{$\langle|\epsilon_\mathrm{h}|\rangle=0.28\pm 0.08$}.
At higher masses,
using weak lensing data of strong lensing clusters
and employing the strong lensing constraints as a proxy for the orientation of the halo, \citet{oguri12} obtain \mbox{$\langle|\epsilon_\mathrm{h}|\rangle=0.31\pm 0.05$}.
Similarly, \citet{ust18} combine weak lensing shear and magnification measurements of 20 massive CLASH clusters to obtain
\mbox{$\langle|\epsilon_\mathrm{h}|\rangle=0.20\pm 0.05$} (when using our ellipticity definition).

 \begin{figure}
   \centering
  \includegraphics[width=9cm]{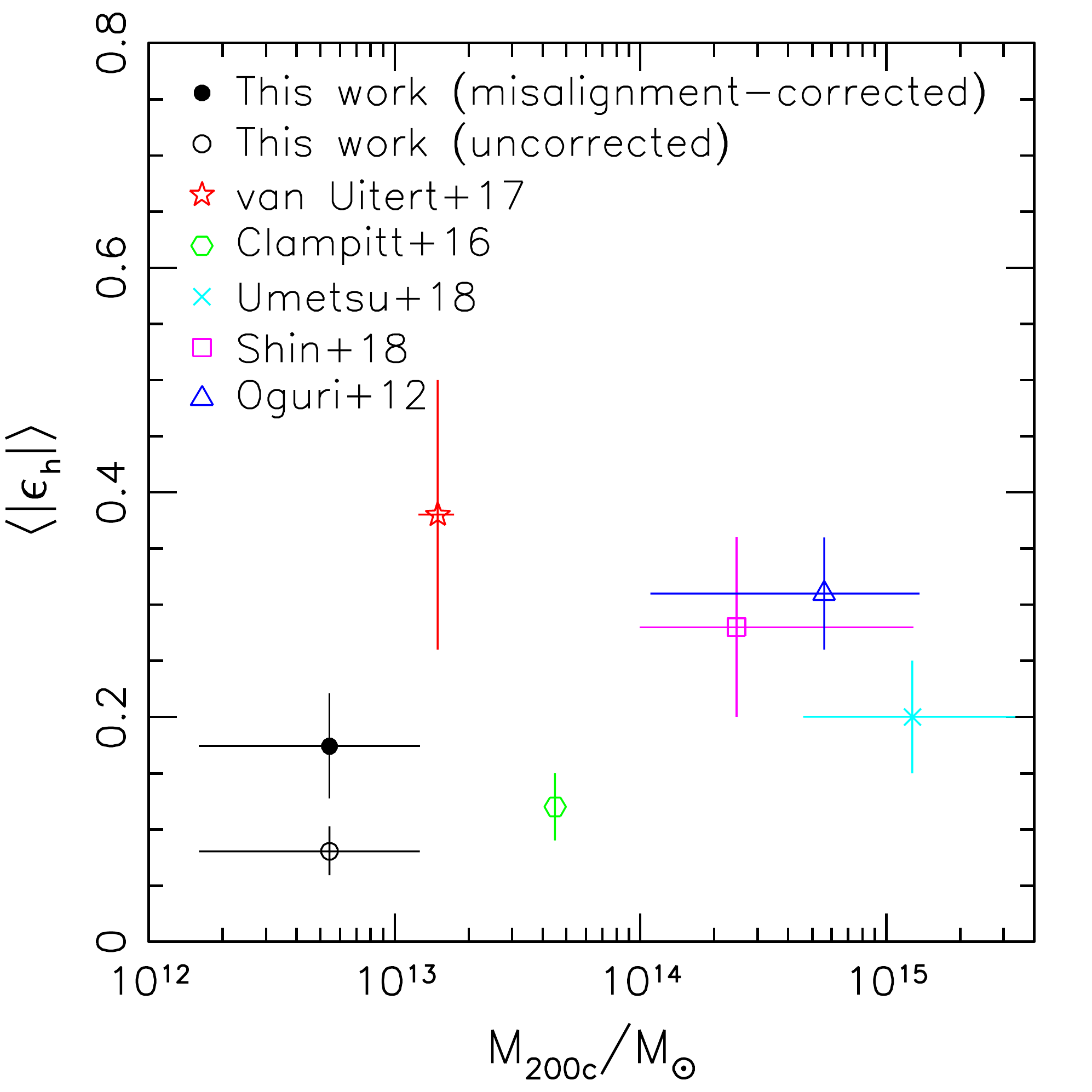}
  \caption{Comparison of constraints on \mbox{$\langle |\epsilon_\mathrm{h}|\rangle$} for red galaxies, galaxy groups, and galaxy clusters from different studies. For our study
we combine the constraints from all red lens samples for our two highest stellar mass bins, showing both the uncorrected estimate (assuming perfect alignment of galaxies and haloes), and the more realistic misalignment-corrected estimate.
The latter assumes that the impact of misalignment is correctly described by the  comparison of the results from the \citetalias{shh15} Millennium Simulation analyses   with and without misalignment (see Sect.\thinspace\ref{se:discussions}).
The horizontal error-bars  indicate the
approximate lens mass range for our results as well as the studies from
\citet{oguri12,scj18}
and \citet{ust18}. For \citet{vanuitert17} the data point is shown at their best-fit mean mass, where the error corresponds to the fit uncertainty. \citet{clj16} only report an approximate mean mass for their sample. We apply mass conversions in case authors employ other definitions than $M_\mathrm{200c}$.
}
   \label{fi:eh_constraints}
 \end{figure}

  It is important to realise that these constraints are obtained at significantly higher mass scales (see Fig.\thinspace\ref{fi:eh_constraints} for an approximate comparison).
For example,   the group sample from
  \citet{vanuitert17} yields an average mass
  \mbox{$M_\mathrm{200c}=1.50^{+0.25}_{-0.24}\times 10^{13}\mathrm{M_\odot}$}, which is still higher by factors of \mbox{$\sim 2$}, \mbox{$\sim 7$}, and  \mbox{$\sim 16$}, respectively, compared to what we typically find for our red lens samples in the three stellar mass bins (compare Table \ref{tab:results}).
  Considering all red lenses  in the five surveys that fall into these stellar mass bins and pass the selections (but without applying the \mbox{$|\epsilon_\mathrm{g}|>0.05$} cut),
  the average
absolute value of the lens ellipticity amounts to
\mbox{$\langle |\epsilon_\mathrm{g}| \rangle=0.265$}.
If galaxies and haloes were perfectly aligned, our estimate of  \mbox{$f_\mathrm{h}=0.303^{+0.080}_{-0.079}$} for these lenses would then imply
\mbox{$\langle|\epsilon_\mathrm{h}|\rangle=0.08\pm 0.02$}.
We can correct this estimate for the impact of misalignment  by considering results from \citetalias{shh15}:
In their analysis of mock data based on the Millennium Simulation, \citetalias{shh15} investigate both cases of  perfectly aligned and misaligned galaxy-halo pairs. Assuming their employed misalignment models \citep[as implemented by][]{jsh13} are correct we can use the ratio of their resulting $f_\mathrm{h}$ estimates for these cases\footnote{Using their combination of early-type lenses at \mbox{$0.2\le z_\mathrm{l}<0.6$} with \mbox{$9.5 <\mathrm{log}_{10}M_*<11$} to best match our halo mass range.}
\mbox{$f_\mathrm{h}(\mathrm{aligned})/f_\mathrm{h}(\mathrm{misaligned})=2.16$} to correct our halo ellipticity estimate for red lenses, yielding
\mbox{$\langle|\epsilon_\mathrm{h}|\rangle=0.174\pm 0.046$}.
This is still
lower compared to the above-mentioned results for group and cluster-scale haloes,
consistent with expectations from numerical simulations, which suggest that less massive haloes are more
spherical  \citep[e.g.][]{bas05,dgt14,vyg17,dgb17}.
The only exception is the constraint for the LRG sample from \citet{clj16}, which is lower, but still consistent with our misalignment-corrected estimate.
However, given that their results do not include a  misalignment correction, the true halo ellipticity is likely larger for their sample.

\section{Conclusions}
\label{se:conclusions}
We have combined weak lensing data from five surveys (CFHTLenS,
 NGVSLenS, CS82,
 RCSLenS, and KiDS/KV450)
 in order to tightly constrain
 \mbox{$f_\mathrm{h}= \langle\cos(2\Delta\phi_\mathrm{h,g})|\epsilon_\mathrm{h}|/|\epsilon_\mathrm{g}|\rangle $}
 the average aligned component of the ellipticity ratio between
 the host dark matter halo and its embedded galaxy for photometrically selected lens galaxies at redshifts \mbox{$0.2<z_\mathrm{b}<0.6$}.
 We obtain \mbox{$f_\mathrm{h}=0.303^{+0.080}_{-0.079}$} for red lenses.
 Similarly clear signals of dark matter halo flattening have previously
 been reported at the mass scale of galaxy groups and clusters \citep[e.g.][]{oguri12,clj16,vanuitert17}, but to our knowledge this presently constitutes  the most significant ($3.8\sigma$) systematics-corrected detection at the mass scale of galaxies.
 We measure \mbox{$f_\mathrm{h}=0.217^{+0.160}_{-0.159}$} for blue galaxies, consistent with a non-detection.
 Both results are in good agreement with theoretical predictions obtained by \citetalias{shh15} using mock data from the Millennium Simulation when including models for galaxy-halo misalignment.
 If we assume that the misalignment models employed in the mock data for early-type galaxies are correct
      \citep[Gaussian misalignment distribution
with a scatter of 35$^\circ$, see][]{ojl09,jsh13}, we can infer an average halo ellipticity of \mbox{$\langle|\epsilon_\mathrm{h}|\rangle=0.174\pm 0.046$} for our red lens galaxies.
      This is
      lower compared to most measurements derived at the scale of galaxy groups and galaxy clusters \citep[][]{evb09,oguri12,clj16,vanuitert17,scj18,ust18}, as consistent with the expectation from simulations that less massive haloes are more spherical  \citep[e.g.][]{bas05,dgt14,vyg17}.

 Observational constraints on \mbox{$f_\mathrm{h}$} sensitively test theoretical models that describe the co-evolution of galaxies and their host dark matter haloes. Improving and
 better calibrating these models is essential in order to reduce systematic uncertainties
 in cosmological weak lensing surveys related to intrinsic galaxy alignments  \citep[see e.g.][]{tmd16,ckj17,hilbert17,pjs18}.
Beyond this, constraints on halo ellipticity can also be used to test non-standard cosmological models.
For example,
interacting dark matter models
predict more spherical dark matter haloes \citep{hellwig13,peter13}.
This is also the case for coupled dark energy models, while the opposite seems to be the case for $f(\mathrm{R})$ theories \citep{LHuillier17}.
The predicted differences of these models compared to standard $\Lambda$CDM expectations are however small, requiring the constraining power of future weak lensing surveys to potentially become useful tests.
This may be different for theories of modified
gravity that aim to remove the need for dark matter, including MOND \citep[Modified Newtonian Dynamics,][]{mil83}, TeVeS
\citep[Scalar--Tensor--Vector theory,][]{bek04}, and MOG/STVG \citep[Modified
Gravity and Scalar--Tensor--Vector Gravity theory,][]{mof06,mot09}.
When considering isolated galaxies in such cosmologies one would naively expect nearly isotropic gravitational signatures  at large separations from the baryons  \citep[e.g.][]{mil01}.
This seems to be at odds with our  $3.8\sigma$ detection of non-zero $f_\mathrm{h}$ for red galaxies.
Unfortunately, detailed forecasts for expected $f_\mathrm{h}$ estimates in such cosmologies
are
not yet available.
If such
predictions are developed in the future,
they would likely need to
incorporate the
potentially contaminating impact of the surrounding large-scale structure,
as included in the  $\Lambda$CDM mock estimates from \citetalias{shh15}, which we compare our measurements to.

Observationally the future is bright for halo ellipticity measurements.
Next generation weak lensing surveys
such as {\it Euclid} \citep[][]{laureijs11} and LSST \citep{lsst09} will provide a tremendous statistical constraining power, providing the prospects to tightly constrain halo ellipticity signatures for different galaxy samples both as a function of mass scale and redshift \citep[simulations predict more spherical haloes at lower redshifts and lower masses, see e.g.][]{tmm14}.
The sensitivity of these surveys may also yield significant detections using estimators that include the ellipticities of lenses and pairs of background sources, providing a route to separate the currently degenerate signatures of halo sphericity and misalignment \citep[e.g.][]{scw05,ssk12,acd15,shy18}.

 \bibliographystyle{aa}
\bibliography{oir,paper1d}

\begin{thebibliography}{148}
\expandafter\ifx\csname natexlab\endcsname\relax\def\natexlab#1{#1}\fi

\bibitem[{{Adelman-McCarthy} {et~al.}(2007){Adelman-McCarthy}, {Ag{\"u}eros},
  {Allam}, {Anderson}, {Anderson}, {Annis}, {Bahcall}, {Bailer-Jones},
  {Baldry}, {Barentine}, {Beers}, {Belokurov}, {Berlind}, {Bernardi},
  {Blanton}, {Bochanski}, {Boroski}, {Bramich}, {Brewington}, {Brinchmann},
  {Brinkmann}, {Brunner}, {Budav{\'a}ri}, {Carey}, {Carliles}, {Carr},
  {Castander}, {Connolly}, {Cool}, {Cunha}, {Csabai}, {Dalcanton}, {Doi},
  {Eisenstein}, {Evans}, {Evans}, {Fan}, {Finkbeiner}, {Friedman}, {Frieman},
  {Fukugita}, {Gillespie}, {Gilmore}, {Glazebrook}, {Gray}, {Grebel}, {Gunn},
  {de Haas}, {Hall}, {Harvanek}, {Hawley}, {Hayes}, {Heckman}, {Hendry},
  {Hennessy}, {Hindsley}, {Hirata}, {Hogan}, {Hogg}, {Holtzman}, {Ichikawa},
  {Ichikawa}, {Ivezi{\'c}}, {Jester}, {Johnston}, {Jorgensen}, {Juri{\'c}},
  {Kauffmann}, {Kent}, {Kleinman}, {Knapp}, {Kniazev}, {Kron}, {Krzesinski},
  {Kuropatkin}, {Lamb}, {Lampeitl}, {Lee}, {Leger}, {Lima}, {Lin}, {Long},
  {Loveday}, {Lupton}, {Mandelbaum}, {Margon}, {Mart{\'\i}nez-Delgado},
  {Matsubara}, {McGehee}, {McKay}, {Meiksin}, {Munn}, {Nakajima}, {Nash},
  {Neilsen}, {Newberg}, {Nichol}, {Nieto-Santisteban}, {Nitta}, {Oyaizu},
  {Okamura}, {Ostriker}, {Padmanabhan}, {Park}, {Peoples}, {Pier}, {Pope},
  {Pourbaix}, {Quinn}, {Raddick}, {Re Fiorentin}, {Richards}, {Richmond},
  {Rix}, {Rockosi}, {Schlegel}, {Schneider}, {Scranton}, {Seljak}, {Sheldon},
  {Shimasaku}, {Silvestri}, {Smith}, {Smol{\v{c}}i{\'c}}, {Snedden},
  {Stebbins}, {Stoughton}, {Strauss}, {SubbaRao}, {Suto}, {Szalay}, {Szapudi},
  {Szkody}, {Tegmark}, {Thakar}, {Tremonti}, {Tucker}, {Uomoto}, {Vanden Berk},
  {Vandenberg}, {Vidrih}, {Vogeley}, {Voges}, {Vogt}, {Weinberg}, {West},
  {White}, {Wilhite}, {Yanny}, {Yocum}, {York}, {Zehavi}, {Zibetti}, \&
  {Zucker}}]{adelman07}
{Adelman-McCarthy}, J.~K., {Ag{\"u}eros}, M.~A., {Allam}, S.~S., {et~al.} 2007,
  \apjs, 172, 634

\bibitem[{{Adhikari} {et~al.}(2015){Adhikari}, {Chue}, \& {Dalal}}]{acd15}
{Adhikari}, S., {Chue}, C.~Y.~R., \& {Dalal}, N. 2015, \jcap, 1, 9

\bibitem[{{Agustsson} \& {Brainerd}(2010)}]{agb10}
{Agustsson}, I. \& {Brainerd}, T.~G. 2010, \apj, 709, 1321

\bibitem[{{Annis} {et~al.}(2014){Annis}, {Soares-Santos}, {Strauss}, {Becker},
  {Dodelson}, {Fan}, {Gunn}, {Hao}, {Ivezi{\'c}}, {Jester}, {Jiang},
  {Johnston}, {Kubo}, {Lampeitl}, {Lin}, {Lupton}, {Miknaitis}, {Seo}, {Simet},
  \& {Yanny}}]{annis14}
{Annis}, J., {Soares-Santos}, M., {Strauss}, M.~A., {et~al.} 2014, \apj, 794,
  120

\bibitem[{{Arnouts} {et~al.}(1999){Arnouts}, {Cristiani}, {Moscardini},
  {Matarrese}, {Lucchin}, {Fontana}, \& {Giallongo}}]{acm99}
{Arnouts}, S., {Cristiani}, S., {Moscardini}, L., {et~al.} 1999, \mnras, 310,
  540

\bibitem[{{Asgari} {et~al.}(2020){Asgari}, {Tr{\"o}ster}, {Heymans},
  {Hildebrandt}, {van den Busch}, {Wright}, {Choi}, {Erben}, {Joachimi},
  {Joudaki}, {Kannawadi}, {Kuijken}, {Lin}, {Schneider}, \& {Zuntz}}]{ath20}
{Asgari}, M., {Tr{\"o}ster}, T., {Heymans}, C., {et~al.} 2020, \aap, 634, A127

\bibitem[{{Bailin} \& {Steinmetz}(2005)}]{bas05}
{Bailin}, J. \& {Steinmetz}, M. 2005, \apj, 627, 647

\bibitem[{{Bartelmann} \& {Schneider}(2001)}]{bas01}
{Bartelmann}, M. \& {Schneider}, P. 2001, \physrep, 340, 291

\bibitem[{{Bate} {et~al.}(2020){Bate}, {Chisari}, {Codis}, {Martin}, {Dubois},
  {Devriendt}, {Pichon}, \& {Slyz}}]{bcc20}
{Bate}, J., {Chisari}, N.~E., {Codis}, S., {et~al.} 2020, \mnras, 491, 4057

\bibitem[{{Begeman} {et~al.}(2013){Begeman}, {Belikov}, {Boxhoorn}, \&
  {Valentijn}}]{bbb13}
{Begeman}, K., {Belikov}, A.~N., {Boxhoorn}, D.~R., \& {Valentijn}, E.~A. 2013,
  Experimental Astronomy, 35, 1

\bibitem[{{Bekenstein}(2004)}]{bek04}
{Bekenstein}, J.~D. 2004, \prd, 70, 083509

\bibitem[{{Ben{\'{\i}}tez}(2000)}]{ben00}
{Ben{\'{\i}}tez}, N. 2000, \apj, 536, 571

\bibitem[{{Bertin} \& {Arnouts}(1996)}]{bea96}
{Bertin}, E. \& {Arnouts}, S. 1996, \aaps, 117, 393

\bibitem[{{Bett}(2012)}]{bet12}
{Bett}, P. 2012, \mnras, 420, 3303

\bibitem[{{Bett} {et~al.}(2010){Bett}, {Eke}, {Frenk}, {Jenkins}, \&
  {Okamoto}}]{bef10}
{Bett}, P., {Eke}, V., {Frenk}, C.~S., {Jenkins}, A., \& {Okamoto}, T. 2010,
  \mnras, 404, 1137

\bibitem[{{Bhowmick} {et~al.}(2020){Bhowmick}, {Chen}, {Tenneti}, {Di Matteo},
  \& {Mandelbaum}}]{bct20}
{Bhowmick}, A.~K., {Chen}, Y., {Tenneti}, A., {Di Matteo}, T., \& {Mandelbaum},
  R. 2020, \mnras, 491, 4116

\bibitem[{{Bliss}(1935{\natexlab{a}})}]{bli35a}
{Bliss}, C.~I. 1935{\natexlab{a}}, Ann.~Appl.~Biol., 22, 134

\bibitem[{{Bliss}(1935{\natexlab{b}})}]{bli35b}
{Bliss}, C.~I. 1935{\natexlab{b}}, Ann.~Appl.~Biol., 22, 307

\bibitem[{{Brainerd} \& {Wright}(2000)}]{brw00}
{Brainerd}, T.~G. \& {Wright}, C.~O. 2000, astro-ph/0006281

\bibitem[{{Brammer} {et~al.}(2008){Brammer}, {van Dokkum}, \&
  {Coppi}}]{brammer08}
{Brammer}, G.~B., {van Dokkum}, P.~G., \& {Coppi}, P. 2008, \apj, 686, 1503

\bibitem[{{Bridle} \& {Abdalla}(2007)}]{bra07}
{Bridle}, S. \& {Abdalla}, F.~B. 2007, \apjl, 655, L1

\bibitem[{{Bridle} \& {King}(2007)}]{brk07}
{Bridle}, S. \& {King}, L. 2007, New Journal of Physics, 9, 444

\bibitem[{{Bruderer} {et~al.}(2016){Bruderer}, {Read}, {Coles}, {Leier},
  {Falco}, {Ferreras}, \& {Saha}}]{brc16}
{Bruderer}, C., {Read}, J.~I., {Coles}, J.~P., {et~al.} 2016, \mnras, 456, 870

\bibitem[{{Bundy} {et~al.}(2015){Bundy}, {Leauthaud}, {Saito}, {Bolton}, {Lin},
  {Maraston}, {Nichol}, {Schneider}, {Thomas}, \& {Wake}}]{bundy15}
{Bundy}, K., {Leauthaud}, A., {Saito}, S., {et~al.} 2015, \apjs, 221, 15

\bibitem[{{Bundy} {et~al.}(2017){Bundy}, {Leauthaud}, {Saito}, {Maraston},
  {Wake}, \& {Thomas}}]{bls17}
{Bundy}, K., {Leauthaud}, A., {Saito}, S., {et~al.} 2017, \apj, 851, 34

\bibitem[{{Caminha} {et~al.}(2016){Caminha}, {Grillo}, {Rosati}, {Balestra},
  {Karman}, {Lombardi}, {Mercurio}, {Nonino}, {Tozzi}, {Zitrin}, {Biviano},
  {Girardi}, {Koekemoer}, {Melchior}, {Meneghetti}, {Munari}, {Suyu}, {Umetsu},
  {Annunziatella}, {Borgani}, {Broadhurst}, {Caputi}, {Coe}, {Delgado-Correal},
  {Ettori}, {Fritz}, {Frye}, {Gobat}, {Maier}, {Monna}, {Postman}, {Sartoris},
  {Seitz}, {Vanzella}, \& {Ziegler}}]{cgr16}
{Caminha}, G.~B., {Grillo}, C., {Rosati}, P., {et~al.} 2016, \aap, 587, A80

\bibitem[{{Caminha} {et~al.}(2019){Caminha}, {Rosati}, {Grillo}, {Rosani},
  {Caputi}, {Meneghetti}, {Mercurio}, {Balestra}, {Bergamini}, {Biviano},
  {Nonino}, {Umetsu}, {Vanzella}, {Annunziatella}, {Broadhurst},
  {Delgado-Correal}, {Demarco}, {Koekemoer}, {Lombardi}, {Maier}, {Verdugo}, \&
  {Zitrin}}]{crg19}
{Caminha}, G.~B., {Rosati}, P., {Grillo}, C., {et~al.} 2019, \aap, 632, A36

\bibitem[{{Chisari} {et~al.}(2017){Chisari}, {Koukoufilippas}, {Jindal},
  {Peirani}, {Beckmann}, {Codis}, {Devriendt}, {Miller}, {Dubois}, {Laigle},
  {Slyz}, \& {Pichon}}]{ckj17}
{Chisari}, N.~E., {Koukoufilippas}, N., {Jindal}, A., {et~al.} 2017, \mnras,
  472, 1163

\bibitem[{{Choi} {et~al.}(2016){Choi}, {Heymans}, {Blake}, {Hildebrandt},
  {Duncan}, {Erben}, {Nakajima}, {Van Waerbeke}, \& {Viola}}]{chb16}
{Choi}, A., {Heymans}, C., {Blake}, C., {et~al.} 2016, \mnras, 463, 3737

\bibitem[{{Chua} {et~al.}(2019){Chua}, {Pillepich}, {Vogelsberger}, \&
  {Hernquist}}]{cpv19}
{Chua}, K. T.~E., {Pillepich}, A., {Vogelsberger}, M., \& {Hernquist}, L. 2019,
  \mnras, 484, 476

\bibitem[{{Clampitt} \& {Jain}(2016)}]{clj16}
{Clampitt}, J. \& {Jain}, B. 2016, \mnras, 457, 4135

\bibitem[{{Codis} {et~al.}(2018){Codis}, {Jindal}, {Chisari}, {Vibert},
  {Dubois}, {Pichon}, \& {Devriendt}}]{cjc18}
{Codis}, S., {Jindal}, A., {Chisari}, N.~E., {et~al.} 2018, \mnras, 481, 4753

\bibitem[{{Coe} {et~al.}(2006){Coe}, {Ben{\'{\i}}tez}, {S{\'a}nchez}, {Jee},
  {Bouwens}, \& {Ford}}]{cbs06}
{Coe}, D., {Ben{\'{\i}}tez}, N., {S{\'a}nchez}, S.~F., {et~al.} 2006, \aj, 132,
  926

\bibitem[{{Crain} {et~al.}(2009){Crain}, {Theuns}, {Dalla Vecchia}, {Eke},
  {Frenk}, {Jenkins}, {Kay}, {Peacock}, {Pearce}, {Schaye}, {Springel},
  {Thomas}, {White}, \& {Wiersma}}]{ctd09}
{Crain}, R.~A., {Theuns}, T., {Dalla Vecchia}, C., {et~al.} 2009, \mnras, 399,
  1773

\bibitem[{{de Jong} {et~al.}(2017){de Jong}, {Verdoes Kleijn}, {Erben},
  {Hildebrandt}, {Kuijken}, {Sikkema}, {Brescia}, {Bilicki}, {Napolitano},
  {Amaro}, {Begeman}, {Boxhoorn}, {Buddelmeijer}, {Cavuoti}, {Getman}, {Grado},
  {Helmich}, {Huang}, {Irisarri}, {La Barbera}, {Longo}, {McFarland},
  {Nakajima}, {Paolillo}, {Puddu}, {Radovich}, {Rifatto}, {Tortora},
  {Valentijn}, {Vellucci}, {Vriend}, {Amon}, {Blake}, {Choi}, {Conti}, {Gwyn},
  {Herbonnet}, {Heymans}, {Hoekstra}, {Klaes}, {Merten}, {Miller}, {Schneider},
  \& {Viola}}]{dejong17}
{de Jong}, J. T.~A., {Verdoes Kleijn}, G.~A., {Erben}, T., {et~al.} 2017, \aap,
  604, A134

\bibitem[{{Deason} {et~al.}(2011){Deason}, {McCarthy}, {Font}, {Evans},
  {Frenk}, {Belokurov}, {Libeskind}, {Crain}, \& {Theuns}}]{dmf11}
{Deason}, A.~J., {McCarthy}, I.~G., {Font}, A.~S., {et~al.} 2011, \mnras, 415,
  2607

\bibitem[{{Debattista} {et~al.}(2015){Debattista}, {van den Bosch}, {Ro{\v
  s}kar}, {Quinn}, {Moore}, \& {Cole}}]{dbr15}
{Debattista}, V.~P., {van den Bosch}, F.~C., {Ro{\v s}kar}, R., {et~al.} 2015,
  \mnras, 452, 4094

\bibitem[{{Despali} {et~al.}(2017){Despali}, {Giocoli}, {Bonamigo}, {Limousin},
  \& {Tormen}}]{dgb17}
{Despali}, G., {Giocoli}, C., {Bonamigo}, M., {Limousin}, M., \& {Tormen}, G.
  2017, \mnras, 466, 181

\bibitem[{{Despali} {et~al.}(2014){Despali}, {Giocoli}, \& {Tormen}}]{dgt14}
{Despali}, G., {Giocoli}, C., \& {Tormen}, G. 2014, \mnras, 443, 3208

\bibitem[{{Diego} {et~al.}(2015){Diego}, {Broadhurst}, {Benitez}, {Lim}, \&
  {Lam}}]{dbb15}
{Diego}, J.~M., {Broadhurst}, T., {Benitez}, N., {Lim}, J., \& {Lam}, D. 2015,
  \mnras, 449, 588

\bibitem[{{Duffy} {et~al.}(2008){Duffy}, {Schaye}, {Kay}, \& {Dalla
  Vecchia}}]{dsk08}
{Duffy}, A.~R., {Schaye}, J., {Kay}, S.~T., \& {Dalla Vecchia}, C. 2008,
  \mnras, 390, L64

\bibitem[{{Dvornik} {et~al.}(2019){Dvornik}, {Zoutendijk}, {Hoekstra}, \&
  {Kuijken}}]{dzh19}
{Dvornik}, A., {Zoutendijk}, S.~L., {Hoekstra}, H., \& {Kuijken}, K. 2019,
  \aap, 627, A74

\bibitem[{{Erben} {et~al.}(2009){Erben}, {Hildebrandt}, {Lerchster}, {Hudelot},
  {Benjamin}, {van Waerbeke}, {Schrabback}, {Brimioulle}, {Cordes}, {Dietrich},
  {Holhjem}, {Schirmer}, \& {Schneider}}]{ehl09}
{Erben}, T., {Hildebrandt}, H., {Lerchster}, M., {et~al.} 2009, \aap, 493, 1197

\bibitem[{{Erben} {et~al.}(2013){Erben}, {Hildebrandt}, {Miller}, {van
  Waerbeke}, {Heymans}, {Hoekstra}, {Kitching}, {Mellier}, {Benjamin}, {Blake},
  {Bonnett}, {Cordes}, {Coupon}, {Fu}, {Gavazzi}, {Gillis}, {Grocutt}, {Gwyn},
  {Holhjem}, {Hudson}, {Kilbinger}, {Kuijken}, {Milkeraitis}, {Rowe},
  {Schrabback}, {Semboloni}, {Simon}, {Smit}, {Toader}, {Vafaei}, {van Uitert},
  \& {Velander}}]{ehm13}
{Erben}, T., {Hildebrandt}, H., {Miller}, L., {et~al.} 2013, \mnras, 433, 2545

\bibitem[{{Evans} \& {Bridle}(2009)}]{evb09}
{Evans}, A.~K.~D. \& {Bridle}, S. 2009, \apj, 695, 1446

\bibitem[{{Fenech Conti} {et~al.}(2017){Fenech Conti}, {Herbonnet}, {Hoekstra},
  {Merten}, {Miller}, \& {Viola}}]{fenechconti17}
{Fenech Conti}, I., {Herbonnet}, R., {Hoekstra}, H., {et~al.} 2017, \mnras,
  467, 1627

\bibitem[{{Ferrarese} {et~al.}(2012){Ferrarese}, {C{\^o}t{\'e}}, {Cuilland re},
  {Gwyn}, {Peng}, {MacArthur}, {Duc}, {Boselli}, {Mei}, {Erben}, {McConnachie},
  {Durrell}, {Mihos}, {Jord{\'a}n}, {Lan{\c{c}}on}, {Puzia}, {Emsellem},
  {Balogh}, {Blakeslee}, {van Waerbeke}, {Gavazzi}, {Vollmer}, {Kavelaars},
  {Woods}, {Ball}, {Boissier}, {Courteau}, {Ferriere}, {Gavazzi},
  {Hildebrandt}, {Hudelot}, {Huertas-Company}, {Liu}, {McLaughlin}, {Mellier},
  {Milkeraitis}, {Schade}, {Balkowski}, {Bournaud}, {Carlberg}, {Chapman},
  {Hoekstra}, {Peng}, {Sawicki}, {Simard}, {Taylor}, {Tully}, {van Driel},
  {Wilson}, {Burdullis}, {Mahoney}, \& {Manset}}]{ferrarese12}
{Ferrarese}, L., {C{\^o}t{\'e}}, P., {Cuilland re}, J.-C., {et~al.} 2012,
  \apjs, 200, 4

\bibitem[{{Fieller}(1954)}]{fie54}
{Fieller}, E.~C. 1954, J.~R.~Stat.~Soc.B, 16, 175

\bibitem[{{Georgiou} {et~al.}(2019{\natexlab{a}}){Georgiou}, {Chisari},
  {Fortuna}, {Hoekstra}, {Kuijken}, {Joachimi}, {Vakili}, {Bilicki}, {Dvornik},
  {Erben}, {Giblin}, {Heymans}, {Napolitano}, \& {Shan}}]{gce19}
{Georgiou}, C., {Chisari}, N.~E., {Fortuna}, M.~C., {et~al.}
  2019{\natexlab{a}}, \aap, 628, A31

\bibitem[{{Georgiou} {et~al.}(2019{\natexlab{b}}){Georgiou}, {Hoekstra},
  {Kuijken}, {Bilicki}, {Dvornik}, {Hildebrandt}, {Schrabback}, {Shan}, \&
  {Wright}}]{georgiou19}
{Georgiou}, C., {Hoekstra}, H., {Kuijken}, K., {et~al.} 2019{\natexlab{b}},
  \aap \,submitted

\bibitem[{{Gilbank} {et~al.}(2011){Gilbank}, {Gladders}, {Yee}, \&
  {Hsieh}}]{gilbank11}
{Gilbank}, D.~G., {Gladders}, M.~D., {Yee}, H.~K.~C., \& {Hsieh}, B.~C. 2011,
  \aj, 141, 94

\bibitem[{{Hamana} {et~al.}(2020){Hamana}, {Shirasaki}, {Miyazaki}, {Hikage},
  {Oguri}, {More}, {Armstrong}, {Leauthaud}, {Mandelbaum}, {Miyatake},
  {Nishizawa}, {Simet}, {Takada}, {Aihara}, {Bosch}, {Komiyama}, {Lupton},
  {Murayama}, {Strauss}, \& {Tanaka}}]{hamana20}
{Hamana}, T., {Shirasaki}, M., {Miyazaki}, S., {et~al.} 2020, \pasj, 72, 16

\bibitem[{{Hand} {et~al.}(2015){Hand}, {Leauthaud}, {Das}, {Sherwin},
  {Addison}, {Bond}, {Calabrese}, {Charbonnier}, {Devlin}, {Dunkley}, {Erben},
  {Hajian}, {Halpern}, {Harnois-D{\'e}raps}, {Heymans}, {Hildebrandt},
  {Hincks}, {Kneib}, {Kosowsky}, {Makler}, {Miller}, {Moodley}, {Moraes},
  {Niemack}, {Page}, {Partridge}, {Sehgal}, {Shan}, {Sievers}, {Spergel},
  {Staggs}, {Switzer}, {Taylor}, {Van Waerbeke}, {Welker}, \&
  {Wollack}}]{hand15}
{Hand}, N., {Leauthaud}, A., {Das}, S., {et~al.} 2015, \prd, 91, 062001

\bibitem[{{Harvey} {et~al.}(2019){Harvey}, {Tam}, {Jauzac}, {Massey}, \&
  {Rhodes}}]{htj19}
{Harvey}, D., {Tam}, S.-I., {Jauzac}, M., {Massey}, R., \& {Rhodes}, J. 2019,
  \mnras\, submitted (also arXiv:1911.06333)

\bibitem[{{Hayashi} \& {Chiba}(2014)}]{hac14}
{Hayashi}, K. \& {Chiba}, M. 2014, \apj, 789, 62

\bibitem[{{Hellwing} {et~al.}(2013){Hellwing}, {Cautun}, {Knebe},
  {Juszkiewicz}, \& {Knollmann}}]{hellwig13}
{Hellwing}, W.~A., {Cautun}, M., {Knebe}, A., {Juszkiewicz}, R., \&
  {Knollmann}, S. 2013, \jcap, 2013, 012

\bibitem[{{Heymans} {et~al.}(2013){Heymans}, {Grocutt}, {Heavens}, {Kilbinger},
  {Kitching}, {Simpson}, {Benjamin}, {Erben}, {Hildebrandt}, {Hoekstra},
  {Mellier}, {Miller}, {Van Waerbeke}, {Brown}, {Coupon}, {Fu},
  {Harnois-D{\'e}raps}, {Hudson}, {Kuijken}, {Rowe}, {Schrabback}, {Semboloni},
  {Vafaei}, \& {Velander}}]{hgh13}
{Heymans}, C., {Grocutt}, E., {Heavens}, A., {et~al.} 2013, \mnras, 432, 2433

\bibitem[{{Heymans} {et~al.}(2006{\natexlab{a}}){Heymans}, {Van Waerbeke},
  {Bacon}, {Berge}, {Bernstein}, {Bertin}, {Bridle}, {Brown}, {Clowe}, {Dahle},
  {Erben}, {Gray}, {Hetterscheidt}, {Hoekstra}, {Hudelot}, {Jarvis}, {Kuijken},
  {Margoniner}, {Massey}, {Mellier}, {Nakajima}, {Refregier}, {Rhodes},
  {Schrabback}, \& {Wittman}}]{hwb06}
{Heymans}, C., {Van Waerbeke}, L., {Bacon}, D., {et~al.} 2006{\natexlab{a}},
  \mnras, 368, 1323

\bibitem[{{Heymans} {et~al.}(2012){Heymans}, {Van Waerbeke}, {Miller}, {Erben},
  {Hildebrandt}, {Hoekstra}, {Kitching}, {Mellier}, {Simon}, {Bonnett},
  {Coupon}, {Fu}, {Harnois D{\'e}raps}, {Hudson}, {Kilbinger}, {Kuijken},
  {Rowe}, {Schrabback}, {Semboloni}, {van Uitert}, {Vafaei}, \&
  {Velander}}]{hwm12}
{Heymans}, C., {Van Waerbeke}, L., {Miller}, L., {et~al.} 2012, \mnras, 427,
  146

\bibitem[{{Heymans} {et~al.}(2006{\natexlab{b}}){Heymans}, {White}, {Heavens},
  {Vale}, \& {van Waerbeke}}]{hwh06}
{Heymans}, C., {White}, M., {Heavens}, A., {Vale}, C., \& {van Waerbeke}, L.
  2006{\natexlab{b}}, \mnras, 371, 750

\bibitem[{{Hilbert} {et~al.}(2009){Hilbert}, {Hartlap}, {White}, \&
  {Schneider}}]{hhw09}
{Hilbert}, S., {Hartlap}, J., {White}, S.~D.~M., \& {Schneider}, P. 2009, \aap,
  499, 31

\bibitem[{{Hilbert} {et~al.}(2017){Hilbert}, {Xu}, {Schneider}, {Springel},
  {Vogelsberger}, \& {Hernquist}}]{hilbert17}
{Hilbert}, S., {Xu}, D., {Schneider}, P., {et~al.} 2017, \mnras, 468, 790

\bibitem[{{Hildebrandt} {et~al.}(2016){Hildebrandt}, {Choi}, {Heymans},
  {Blake}, {Erben}, {Miller}, {Nakajima}, {van Waerbeke}, {Viola},
  {Buddendiek}, {Harnois-D{\'e}raps}, {Hojjati}, {Joachimi}, {Joudaki},
  {Kitching}, {Wolf}, {Gwyn}, {Johnson}, {Kuijken}, {Sheikhbahaee}, {Tudorica},
  \& {Yee}}]{hch16}
{Hildebrandt}, H., {Choi}, A., {Heymans}, C., {et~al.} 2016, \mnras, 463, 635

\bibitem[{{Hildebrandt} {et~al.}(2012){Hildebrandt}, {Erben}, {Kuijken}, {van
  Waerbeke}, {Heymans}, {Coupon}, {Benjamin}, {Bonnett}, {Fu}, {Hoekstra},
  {Kitching}, {Mellier}, {Miller}, {Velander}, {Hudson}, {Rowe}, {Schrabback},
  {Semboloni}, \& {Ben{\'{\i}}tez}}]{hek12}
{Hildebrandt}, H., {Erben}, T., {Kuijken}, K., {et~al.} 2012, \mnras, 421, 2355

\bibitem[{{Hildebrandt} {et~al.}(2020){Hildebrandt}, {K{\"o}hlinger}, {van den
  Busch}, {Joachimi}, {Heymans}, {Kannawadi}, {Wright}, {Asgari}, {Blake},
  {Hoekstra}, {Joudaki}, {Kuijken}, {Miller}, {Morrison}, {Tr{\"o}ster},
  {Amon}, {Archidiacono}, {Brieden}, {Choi}, {de Jong}, {Erben}, {Giblin},
  {Mead}, {Peacock}, {Radovich}, {Schneider}, {Sif{\'o}n}, \&
  {Tewes}}]{hildebrandt20}
{Hildebrandt}, H., {K{\"o}hlinger}, F., {van den Busch}, J.~L., {et~al.} 2020,
  \aap, 633, A69

\bibitem[{{Hildebrandt} {et~al.}(2017){Hildebrandt}, {Viola}, {Heymans},
  {Joudaki}, {Kuijken}, {Blake}, {Erben}, {Joachimi}, {Klaes}, {Miller},
  {Morrison}, {Nakajima}, {Verdoes Kleijn}, {Amon}, {Choi}, {Covone}, {de
  Jong}, {Dvornik}, {Fenech Conti}, {Grado}, {Harnois-D{\'e}raps}, {Herbonnet},
  {Hoekstra}, {K{\"o}hlinger}, {McFarland}, {Mead}, {Merten}, {Napolitano},
  {Peacock}, {Radovich}, {Schneider}, {Simon}, {Valentijn}, {van den Busch},
  {van Uitert}, \& {Van Waerbeke}}]{hildebrandt17}
{Hildebrandt}, H., {Viola}, M., {Heymans}, C., {et~al.} 2017, \mnras, 465, 1454

\bibitem[{{Hirata} \& {Seljak}(2004)}]{his04}
{Hirata}, C.~M. \& {Seljak}, U. 2004, \prd, 70, 063526

\bibitem[{{Hoekstra} {et~al.}(2000){Hoekstra}, {Franx}, \& {Kuijken}}]{hfk00}
{Hoekstra}, H., {Franx}, M., \& {Kuijken}, K. 2000, \apj, 532, 88

\bibitem[{{Hoekstra} {et~al.}(1998){Hoekstra}, {Franx}, {Kuijken}, \&
  {Squires}}]{hfk98}
{Hoekstra}, H., {Franx}, M., {Kuijken}, K., \& {Squires}, G. 1998, \apj, 504,
  636

\bibitem[{{Hoekstra} {et~al.}(2004){Hoekstra}, {Yee}, \& {Gladders}}]{hyg04}
{Hoekstra}, H., {Yee}, H.~K.~C., \& {Gladders}, M.~D. 2004, \apj, 606, 67

\bibitem[{{Huang} {et~al.}(2016){Huang}, {Mandelbaum}, {Freeman}, {Chen},
  {Rozo}, {Rykoff}, \& {Baxter}}]{hmf16}
{Huang}, H.-J., {Mandelbaum}, R., {Freeman}, P.~E., {et~al.} 2016, \mnras, 463,
  222

\bibitem[{{Ilbert} {et~al.}(2006){Ilbert}, {Arnouts}, {McCracken},
  {Bolzonella}, {Bertin}, {Le F{\`e}vre}, {Mellier}, {Zamorani}, {Pell{\`o}},
  {Iovino}, {Tresse}, {Le Brun}, {Bottini}, {Garilli}, {Maccagni}, {Picat},
  {Scaramella}, {Scodeggio}, {Vettolani}, {Zanichelli}, {Adami}, {Bardelli},
  {Cappi}, {Charlot}, {Ciliegi}, {Contini}, {Cucciati}, {Foucaud}, {Franzetti},
  {Gavignaud}, {Guzzo}, {Marano}, {Marinoni}, {Mazure}, {Meneux}, {Merighi},
  {Paltani}, {Pollo}, {Pozzetti}, {Radovich}, {Zucca}, {Bondi}, {Bongiorno},
  {Busarello}, {de La Torre}, {Gregorini}, {Lamareille}, {Mathez}, {Merluzzi},
  {Ripepi}, {Rizzo}, \& {Vergani}}]{iam06}
{Ilbert}, O., {Arnouts}, S., {McCracken}, H.~J., {et~al.} 2006, \aap, 457, 841

\bibitem[{{Jauzac} {et~al.}(2018){Jauzac}, {Harvey}, \& {Massey}}]{jhm18}
{Jauzac}, M., {Harvey}, D., \& {Massey}, R. 2018, \mnras, 477, 4046

\bibitem[{{Jee} {et~al.}(2016){Jee}, {Tyson}, {Hilbert}, {Schneider},
  {Schmidt}, \& {Wittman}}]{jth16}
{Jee}, M.~J., {Tyson}, J.~A., {Hilbert}, S., {et~al.} 2016, \apj, 824, 77

\bibitem[{{Jing} \& {Suto}(2002)}]{jis02}
{Jing}, Y.~P. \& {Suto}, Y. 2002, \apj, 574, 538

\bibitem[{{Joachimi} {et~al.}(2015){Joachimi}, {Cacciato}, {Kitching},
  {Leonard}, {Mandelbaum}, {Sch{\"a}fer}, {Sif{\'o}n}, {Hoekstra}, {Kiessling},
  {Kirk}, \& {Rassat}}]{jck15}
{Joachimi}, B., {Cacciato}, M., {Kitching}, T.~D., {et~al.} 2015, \ssr, 193, 1

\bibitem[{{Joachimi} {et~al.}(2013{\natexlab{a}}){Joachimi}, {Semboloni},
  {Bett}, {Hartlap}, {Hilbert}, {Hoekstra}, {Schneider}, \&
  {Schrabback}}]{jsb13}
{Joachimi}, B., {Semboloni}, E., {Bett}, P.~E., {et~al.} 2013{\natexlab{a}},
  \mnras, 431, 477

\bibitem[{{Joachimi} {et~al.}(2013{\natexlab{b}}){Joachimi}, {Semboloni},
  {Hilbert}, {Bett}, {Hartlap}, {Hoekstra}, \& {Schneider}}]{jsh13}
{Joachimi}, B., {Semboloni}, E., {Hilbert}, S., {et~al.} 2013{\natexlab{b}},
  \mnras, 436, 819

\bibitem[{{Kacprzak} {et~al.}(2012){Kacprzak}, {Zuntz}, {Rowe}, {Bridle},
  {Refregier}, {Amara}, {Voigt}, \& {Hirsch}}]{kzr12}
{Kacprzak}, T., {Zuntz}, J., {Rowe}, B., {et~al.} 2012, \mnras, 427, 2711

\bibitem[{{Kaiser}(1995)}]{kaiser95a}
{Kaiser}, N. 1995, \apjl, 439, L1

\bibitem[{Kaiser {et~al.}(1995)Kaiser, Squires, \& Broadhurst}]{ksb95}
Kaiser, N., Squires, G., \& Broadhurst, T. 1995, ApJ, 449, 460

\bibitem[{{Kannawadi} {et~al.}(2019){Kannawadi}, {Hoekstra}, {Miller}, {Viola},
  {Fenech Conti}, {Herbonnet}, {Erben}, {Heymans}, {Hildebrandt}, {Kuijken},
  {Vakili}, \& {Wright}}]{khm19}
{Kannawadi}, A., {Hoekstra}, H., {Miller}, L., {et~al.} 2019, \aap, 624, A92

\bibitem[{{Khoperskov} {et~al.}(2014){Khoperskov}, {Moiseev}, {Khoperskov}, \&
  {Saburova}}]{kmk14}
{Khoperskov}, S.~A., {Moiseev}, A.~V., {Khoperskov}, A.~V., \& {Saburova},
  A.~S. 2014, \mnras, 441, 2650

\bibitem[{{Kitching} {et~al.}(2008){Kitching}, {Miller}, {Heymans}, {van
  Waerbeke}, \& {Heavens}}]{kmh08}
{Kitching}, T.~D., {Miller}, L., {Heymans}, C.~E., {van Waerbeke}, L., \&
  {Heavens}, A.~F. 2008, \mnras, 390, 149

\bibitem[{{Kuijken} {et~al.}(2019){Kuijken}, {Heymans}, {Dvornik},
  {Hildebrandt}, {de Jong}, {Wright}, {Erben}, {Bilicki}, {Giblin}, {Shan},
  {Getman}, {Grado}, {Hoekstra}, {Miller}, {Napolitano}, {Paolilo}, {Radovich},
  {Schneider}, {Sutherland }, {Tewes}, {Tortora}, {Valentijn}, \& {Verdoes
  Kleijn}}]{kuijken19}
{Kuijken}, K., {Heymans}, C., {Dvornik}, A., {et~al.} 2019, \aap, 625, A2

\bibitem[{{Kuijken} {et~al.}(2015){Kuijken}, {Heymans}, {Hildebrandt},
  {Nakajima}, {Erben}, {de Jong}, {Viola}, {Choi}, {Hoekstra}, {Miller}, {van
  Uitert}, {Amon}, {Blake}, {Brouwer}, {Buddendiek}, {Conti}, {Eriksen},
  {Grado}, {Harnois-D{\'e}raps}, {Helmich}, {Herbonnet}, {Irisarri},
  {Kitching}, {Klaes}, {La Barbera}, {Napolitano}, {Radovich}, {Schneider},
  {Sif{\'o}n}, {Sikkema}, {Simon}, {Tudorica}, {Valentijn}, {Verdoes Kleijn},
  \& {van Waerbeke}}]{kuijken15}
{Kuijken}, K., {Heymans}, C., {Hildebrandt}, H., {et~al.} 2015, \mnras, 454,
  3500

\bibitem[{{Laigle} {et~al.}(2015){Laigle}, {Pichon}, {Codis}, {Dubois}, {Le
  Borgne}, {Pogosyan}, {Devriendt}, {Peirani}, {Prunet}, {Rouberol}, {Slyz}, \&
  {Sousbie}}]{lpc15}
{Laigle}, C., {Pichon}, C., {Codis}, S., {et~al.} 2015, \mnras, 446, 2744

\bibitem[{{Laureijs} {et~al.}(2011){Laureijs}, {Amiaux}, {Arduini},
  {Augu{\`e}res}, {Brinchmann}, {Cole}, {Cropper}, {Dabin}, {Duvet}, {Ealet},
  \& et~al.}]{laureijs11}
{Laureijs}, R., {Amiaux}, J., {Arduini}, S., {et~al.} 2011, arXiv:1110.3193

\bibitem[{{Leauthaud} {et~al.}(2017){Leauthaud}, {Saito}, {Hilbert},
  {Barreira}, {More}, {White}, {Alam}, {Behroozi}, {Bundy}, {Coupon}, {Erben},
  {Heymans}, {Hildebrandt}, {Mandelbaum}, {Miller}, {Moraes}, {Pereira},
  {Rodr{\'\i}guez-Torres}, {Schmidt}, {Shan}, {Viel}, \&
  {Villaescusa-Navarro}}]{lsh17}
{Leauthaud}, A., {Saito}, S., {Hilbert}, S., {et~al.} 2017, \mnras, 467, 3024

\bibitem[{{L'Huillier} {et~al.}(2017){L'Huillier}, {Winther}, {Mota}, {Park},
  \& {Kim}}]{LHuillier17}
{L'Huillier}, B., {Winther}, H.~A., {Mota}, D.~F., {Park}, C., \& {Kim}, J.
  2017, \mnras, 468, 3174

\bibitem[{{Limousin} {et~al.}(2013){Limousin}, {Morandi}, {Sereno},
  {Meneghetti}, {Ettori}, {Bartelmann}, \& {Verdugo}}]{lms13}
{Limousin}, M., {Morandi}, A., {Sereno}, M., {et~al.} 2013, \ssr, 177, 155

\bibitem[{{Liu} {et~al.}(2015){Liu}, {Pan}, {Li}, {Shan}, {Wang}, {Fu}, {Fan},
  {Kneib}, {Leauthaud}, {Van Waerbeke}, {Makler}, {Moraes}, {Erben}, \&
  {Charbonnier}}]{liu15}
{Liu}, X., {Pan}, C., {Li}, R., {et~al.} 2015, \mnras, 450, 2888

\bibitem[{{LSST Science Collaboration} {et~al.}(2009){LSST Science
  Collaboration}, {Abell}, {Allison}, {Anderson}, {Andrew}, {Angel}, {Armus},
  {Arnett}, {Asztalos}, {Axelrod}, \& et~al.}]{lsst09}
{LSST Science Collaboration}, {Abell}, P.~A., {Allison}, J., {et~al.} 2009,
  arXiv:0912.0201

\bibitem[{{Luppino} \& {Kaiser}(1997)}]{luk97}
{Luppino}, G.~A. \& {Kaiser}, N. 1997, \apj, 475, 20

\bibitem[{{Mandelbaum} {et~al.}(2006){Mandelbaum}, {Hirata}, {Broderick},
  {Seljak}, \& {Brinkmann}}]{mhb06}
{Mandelbaum}, R., {Hirata}, C.~M., {Broderick}, T., {Seljak}, U., \&
  {Brinkmann}, J. 2006, \mnras, 370, 1008

\bibitem[{{Massey} {et~al.}(2007){Massey}, {Heymans}, {Berg{\'e}}, {Bernstein},
  {Bridle}, {Clowe}, {Dahle}, {Ellis}, {Erben}, {Hetterscheidt}, {High},
  {Hirata}, {Hoekstra}, {Hudelot}, {Jarvis}, {Johnston}, {Kuijken},
  {Margoniner}, {Mandelbaum}, {Mellier}, {Nakajima}, {Paulin-Henriksson},
  {Peeples}, {Roat}, {Refregier}, {Rhodes}, {Schrabback}, {Schirmer}, {Seljak},
  {Semboloni}, \& {van Waerbeke}}]{mhb07}
{Massey}, R., {Heymans}, C., {Berg{\'e}}, J., {et~al.} 2007, \mnras, 376, 13

\bibitem[{{Melchior} \& {Viola}(2012)}]{mev12}
{Melchior}, P. \& {Viola}, M. 2012, \mnras, 424, 2757

\bibitem[{{Melchior} {et~al.}(2011){Melchior}, {Viola}, {Sch{\"a}fer}, \&
  {Bartelmann}}]{melchior11}
{Melchior}, P., {Viola}, M., {Sch{\"a}fer}, B.~M., \& {Bartelmann}, M. 2011,
  \mnras, 412, 1552

\bibitem[{{Milgrom}(1983)}]{mil83}
{Milgrom}, M. 1983, \apj, 270, 365

\bibitem[{{Milgrom}(2001)}]{mil01}
{Milgrom}, M. 2001, \mnras, 326, 1261

\bibitem[{{Miller} {et~al.}(2013){Miller}, {Heymans}, {Kitching}, {van
  Waerbeke}, {Erben}, {Hildebrandt}, {Hoekstra}, {Mellier}, {Rowe}, {Coupon},
  {Dietrich}, {Fu}, {Harnois-D{\'e}raps}, {Hudson}, {Kilbinger}, {Kuijken},
  {Schrabback}, {Semboloni}, {Vafaei}, \& {Velander}}]{mhk13a}
{Miller}, L., {Heymans}, C., {Kitching}, T.~D., {et~al.} 2013, \mnras, 429,
  2858

\bibitem[{{Miller} {et~al.}(2007){Miller}, {Kitching}, {Heymans}, {Heavens}, \&
  {van Waerbeke}}]{mkh07}
{Miller}, L., {Kitching}, T.~D., {Heymans}, C., {Heavens}, A.~F., \& {van
  Waerbeke}, L. 2007, \mnras, 382, 315

\bibitem[{{Miralda-Escude}(1991)}]{miralda-escude91}
{Miralda-Escude}, J. 1991, \apj, 370, 1

\bibitem[{{Moffat}(2006)}]{mof06}
{Moffat}, J.~W. 2006, \jcap, 3, 4

\bibitem[{{Moffat} \& {Toth}(2009)}]{mot09}
{Moffat}, J.~W. \& {Toth}, V.~T. 2009, \mnras, 397, 1885

\bibitem[{{Natarajan} \& {Refregier}(2000)}]{nar00}
{Natarajan}, P. \& {Refregier}, A. 2000, \apjl, 538, L113

\bibitem[{{Navarro} {et~al.}(1996){Navarro}, {Frenk}, \& {White}}]{nfw96}
{Navarro}, J.~F., {Frenk}, C.~S., \& {White}, S.~D.~M. 1996, \apj, 462, 563

\bibitem[{{Navarro} {et~al.}(1997){Navarro}, {Frenk}, \& {White}}]{nfw97}
{Navarro}, J.~F., {Frenk}, C.~S., \& {White}, S.~D.~M. 1997, \apj, 490, 493

\bibitem[{{Oguri} {et~al.}(2012){Oguri}, {Bayliss}, {Dahle}, {Sharon},
  {Gladders}, {Natarajan}, {Hennawi}, \& {Koester}}]{oguri12}
{Oguri}, M., {Bayliss}, M.~B., {Dahle}, H., {et~al.} 2012, \mnras, 420, 3213

\bibitem[{{Oguri} {et~al.}(2010){Oguri}, {Takada}, {Okabe}, \& {Smith}}]{oto10}
{Oguri}, M., {Takada}, M., {Okabe}, N., \& {Smith}, G.~P. 2010, \mnras, 405,
  2215

\bibitem[{{Okamoto} {et~al.}(2005){Okamoto}, {Eke}, {Frenk}, \&
  {Jenkins}}]{oef05}
{Okamoto}, T., {Eke}, V.~R., {Frenk}, C.~S., \& {Jenkins}, A. 2005, \mnras,
  363, 1299

\bibitem[{{Okumura} {et~al.}(2009){Okumura}, {Jing}, \& {Li}}]{ojl09}
{Okumura}, T., {Jing}, Y.~P., \& {Li}, C. 2009, \apj, 694, 214

\bibitem[{{Parker} {et~al.}(2007){Parker}, {Hoekstra}, {Hudson}, {van
  Waerbeke}, \& {Mellier}}]{phh07}
{Parker}, L.~C., {Hoekstra}, H., {Hudson}, M.~J., {van Waerbeke}, L., \&
  {Mellier}, Y. 2007, \apj, 669, 21

\bibitem[{{Parroni} {et~al.}(2017){Parroni}, {Mei}, {Erben}, {Van Waerbeke},
  {Raichoor}, {Ford}, {Licitra}, {Meneghetti}, {Hildebrand t}, {Miller},
  {C{\^o}t{\'e}}, {Covone}, {Cuillandre}, {Duc}, {Ferrarese}, {Gwyn}, \&
  {Puzia}}]{parroni17}
{Parroni}, C., {Mei}, S., {Erben}, T., {et~al.} 2017, \apj, 848, 114

\bibitem[{{Paterno-Mahler} {et~al.}(2018){Paterno-Mahler}, {Sharon}, {Coe},
  {Mahler}, {Cerny}, {Johnson}, {Schrabback}, {Andrade-Santos}, {Avila},
  {Brada{\v{c}}}, {Bradley}, {Carrasco}, {Czakon}, {Dawson}, {Frye}, {Hoag},
  {Huang}, {Jones}, {Lam}, {Livermore}, {Lovisari}, {Mainali}, {Oesch}, {Ogaz},
  {Past}, {Peterson}, {Ryan}, {Salmon}, {Sendra-Server}, {Stark}, {Umetsu},
  {Vulcani}, \& {Zitrin}}]{psc18}
{Paterno-Mahler}, R., {Sharon}, K., {Coe}, D., {et~al.} 2018, \apj, 863, 154

\bibitem[{{Peter} {et~al.}(2013){Peter}, {Rocha}, {Bullock}, \&
  {Kaplinghat}}]{peter13}
{Peter}, A. H.~G., {Rocha}, M., {Bullock}, J.~S., \& {Kaplinghat}, M. 2013,
  \mnras, 430, 105

\bibitem[{{Peters} {et~al.}(2017){Peters}, {van der Kruit}, {Allen}, \&
  {Freeman}}]{pka17}
{Peters}, S.~P.~C., {van der Kruit}, P.~C., {Allen}, R.~J., \& {Freeman}, K.~C.
  2017, \mnras, 464, 65

\bibitem[{{Piras} {et~al.}(2018){Piras}, {Joachimi}, {Sch{\"a}fer}, {Bonamigo},
  {Hilbert}, \& {van Uitert}}]{pjs18}
{Piras}, D., {Joachimi}, B., {Sch{\"a}fer}, B.~M., {et~al.} 2018, \mnras, 474,
  1165

\bibitem[{{Planck Collaboration} {et~al.}(2020){Planck Collaboration},
  {Aghanim}, {Akrami}, {Ashdown}, {Aumont}, {Baccigalupi}, {Ballardini},
  {Banday}, {Barreiro}, {Bartolo}, {Basak}, {Battye}, {Benabed}, {Bernard},
  {Bersanelli}, {Bielewicz}, {Bock}, {Bond}, {Borrill}, {Bouchet}, {Boulanger},
  {Bucher}, {Burigana}, {Butler}, {Calabrese}, {Cardoso}, {Carron},
  {Challinor}, {Chiang}, {Chluba}, {Colombo}, {Combet}, {Contreras}, {Crill},
  {Cuttaia}, {de Bernardis}, {de Zotti}, {Delabrouille}, {Delouis}, {Di
  Valentino}, {Diego}, {Dor{\'e}}, {Douspis}, {Ducout}, {Dupac}, {Dusini},
  {Efstathiou}, {Elsner}, {En{\ss}lin}, {Eriksen}, {Fantaye}, {Farhang},
  {Fergusson}, {Fernandez-Cobos}, {Finelli}, {Forastieri}, {Frailis},
  {Franceschi}, {Frolov}, {Galeotta}, {Galli}, {Ganga}, {G{\'e}nova-Santos},
  {Gerbino}, {Ghosh}, {Gonz{\'a}lez-Nuevo}, {G{\'o}rski}, {Gratton},
  {Gruppuso}, {Gudmundsson}, {Hamann}, {Hand ley}, {Herranz}, {Hivon}, {Huang},
  {Jaffe}, {Jones}, {Karakci}, {Keih{\"a}nen}, {Keskitalo}, {Kiiveri}, {Kim},
  {Kisner}, {Knox}, {Krachmalnicoff}, {Kunz}, {Kurki-Suonio}, {Lagache},
  {Lamarre}, {Lasenby}, {Lattanzi}, {Lawrence}, {Le Jeune}, {Lemos},
  {Lesgourgues}, {Levrier}, {Lewis}, {Liguori}, {Lilje}, {Lilley}, {Lindholm},
  {L{\'o}pez-Caniego}, {Lubin}, {Ma}, {Mac{\'\i}as-P{\'e}rez}, {Maggio},
  {Maino}, {Mandolesi}, {Mangilli}, {Marcos-Caballero}, {Maris}, {Martin},
  {Martinelli}, {Mart{\'\i}nez-Gonz{\'a}lez}, {Matarrese}, {Mauri}, {McEwen},
  {Meinhold}, {Melchiorri}, {Mennella}, {Migliaccio}, {Millea}, {Mitra},
  {Miville-Desch{\^e}nes}, {Molinari}, {Montier}, {Morgante}, {Moss}, {Natoli},
  {N{\o}rgaard-Nielsen}, {Pagano}, {Paoletti}, {Partridge}, {Patanchon},
  {Peiris}, {Perrotta}, {Pettorino}, {Piacentini}, {Polastri}, {Polenta},
  {Puget}, {Rachen}, {Reinecke}, {Remazeilles}, {Renzi}, {Rocha}, {Rosset},
  {Roudier}, {Rubi{\~n}o-Mart{\'\i}n}, {Ruiz-Granados}, {Salvati}, {Sandri},
  {Savelainen}, {Scott}, {Shellard}, {Sirignano}, {Sirri}, {Spencer},
  {Sunyaev}, {Suur-Uski}, {Tauber}, {Tavagnacco}, {Tenti}, {Toffolatti},
  {Tomasi}, {Trombetti}, {Valenziano}, {Valiviita}, {Van Tent}, {Vibert},
  {Vielva}, {Villa}, {Vittorio}, {Wand elt}, {Wehus}, {White}, {White},
  {Zacchei}, \& {Zonca}}]{planck2018cosmology}
{Planck Collaboration}, {Aghanim}, N., {Akrami}, Y., {et~al.} 2020, \aap, 641,
  A6

\bibitem[{{Raichoor} {et~al.}(2014){Raichoor}, {Mei}, {Erben}, {Hildebrandt},
  {Huertas-Company}, {Ilbert}, {Licitra}, {Ball}, {Boissier}, {Boselli},
  {Chen}, {C{\^o}t{\'e}}, {Cuillandre}, {Duc}, {Durrell}, {Ferrarese},
  {Guhathakurta}, {Gwyn}, {Kavelaars}, {Lan{\c{c}}on}, {Liu}, {MacArthur},
  {Muller}, {Mu{\~n}oz}, {Peng}, {Puzia}, {Sawicki}, {Toloba}, {Van Waerbeke},
  {Woods}, \& {Zhang}}]{raichoor14}
{Raichoor}, A., {Mei}, S., {Erben}, T., {et~al.} 2014, \apj, 797, 102

\bibitem[{{Refregier} {et~al.}(2012){Refregier}, {Kacprzak}, {Amara}, {Bridle},
  \& {Rowe}}]{rka12}
{Refregier}, A., {Kacprzak}, T., {Amara}, A., {Bridle}, S., \& {Rowe}, B. 2012,
  \mnras, 425, 1951

\bibitem[{{Rykoff} {et~al.}(2014){Rykoff}, {Rozo}, {Busha}, {Cunha},
  {Finoguenov}, {Evrard}, {Hao}, {Koester}, {Leauthaud}, {Nord}, {Pierre},
  {Reddick}, {Sadibekova}, {Sheldon}, \& {Wechsler}}]{rykoff14}
{Rykoff}, E.~S., {Rozo}, E., {Busha}, M.~T., {et~al.} 2014, \apj, 785, 104

\bibitem[{{Sch{\"a}fer} \& {Merkel}(2017)}]{scm17}
{Sch{\"a}fer}, B.~M. \& {Merkel}, P.~M. 2017, \mnras, 470, 3453

\bibitem[{Schneider {et~al.}(2006)Schneider, Kochanek, \& Wambsganss}]{skw06}
Schneider, P., Kochanek, C., \& Wambsganss, J. 2006, {{\sl Gravitational
  Lensing: Strong, Weak \& Micro}, Saas-Fee Advanced Course 33, Swiss Society
  for Astrophysics and Astronomy, G. Meylan, P. Jetzer \& P. North (Eds.),
  Springer-Verlag: Berlin}

\bibitem[{{Schneider} \& {Watts}(2005)}]{scw05}
{Schneider}, P. \& {Watts}, P. 2005, \aap, 432, 783

\bibitem[{{Schrabback} {et~al.}(2018){Schrabback}, {Applegate}, {Dietrich},
  {Hoekstra}, {Bocquet}, {Gonzalez}, {von der Linden}, {McDonald}, {Morrison},
  {Raihan}, {Allen}, {Bayliss}, {Benson}, {Bleem}, {Chiu}, {Desai}, {Foley},
  {de Haan}, {High}, {Hilbert}, {Mantz}, {Massey}, {Mohr}, {Reichardt}, {Saro},
  {Simon}, {Stern}, {Stubbs}, \& {Zenteno}}]{schrabback18}
{Schrabback}, T., {Applegate}, D., {Dietrich}, J.~P., {et~al.} 2018, \mnras,
  474, 2635

\bibitem[{{Schrabback} {et~al.}(2010){Schrabback}, {Hartlap}, {Joachimi},
  {Kilbinger}, {Simon}, {Benabed}, {Brada{\v c}}, {Eifler}, {Erben},
  {Fassnacht}, {High}, {Hilbert}, {Hildebrandt}, {Hoekstra}, {Kuijken},
  {Marshall}, {Mellier}, {Morganson}, {Schneider}, {Semboloni}, {van Waerbeke},
  \& {Velander}}]{shj10}
{Schrabback}, T., {Hartlap}, J., {Joachimi}, B., {et~al.} 2010, \aap, 516, A63

\bibitem[{{Schrabback} {et~al.}(2015){Schrabback}, {Hilbert}, {Hoekstra},
  {Simon}, {van Uitert}, {Erben}, {Heymans}, {Hildebrandt}, {Kitching},
  {Mellier}, {Miller}, {Van Waerbeke}, {Bett}, {Coupon}, {Fu}, {Hudson},
  {Joachimi}, {Kilbinger}, \& {Kuijken}}]{shh15}
{Schrabback}, T., {Hilbert}, S., {Hoekstra}, H., {et~al.} 2015, \mnras, 454,
  1432

\bibitem[{{Seitz} \& {Schneider}(1997)}]{seitz97}
{Seitz}, C. \& {Schneider}, P. 1997, \aap, 318, 687

\bibitem[{{Shan} {et~al.}(2014){Shan}, {Kneib}, {Comparat}, {Jullo},
  {Charbonnier}, {Erben}, {Makler}, {Moraes}, {Van Waerbeke}, {Courbin},
  {Meylan}, {Tao}, \& {Taylor}}]{shan14}
{Shan}, H.~Y., {Kneib}, J.-P., {Comparat}, J., {et~al.} 2014, \mnras, 442, 2534

\bibitem[{{Shin} {et~al.}(2018){Shin}, {Clampitt}, {Jain}, {Bernstein}, {Neil},
  {Rozo}, \& {Rykoff}}]{scj18}
{Shin}, T.-h., {Clampitt}, J., {Jain}, B., {et~al.} 2018, \mnras, 475, 2421

\bibitem[{{Shirasaki} \& {Yoshida}(2018)}]{shy18}
{Shirasaki}, M. \& {Yoshida}, N. 2018, \mnras, 475, 1665

\bibitem[{{Simon} {et~al.}(2012){Simon}, {Schneider}, \& {K{\"u}bler}}]{ssk12}
{Simon}, P., {Schneider}, P., \& {K{\"u}bler}, D. 2012, \aap, 548, A102

\bibitem[{{Springel} {et~al.}(2005){Springel}, {White}, {Jenkins}, {Frenk},
  {Yoshida}, {Gao}, {Navarro}, {Thacker}, {Croton}, {Helly}, {Peacock}, {Cole},
  {Thomas}, {Couchman}, {Evrard}, {Colberg}, \& {Pearce}}]{swj05}
{Springel}, V., {White}, S.~D.~M., {Jenkins}, A., {et~al.} 2005, \nat, 435, 629

\bibitem[{{Suyu} {et~al.}(2012){Suyu}, {Hensel}, {McKean}, {Fassnacht}, {Treu},
  {Halkola}, {Norbury}, {Jackson}, {Schneider}, {Thompson}, {Auger},
  {Koopmans}, \& {Matthews}}]{shk12}
{Suyu}, S.~H., {Hensel}, S.~W., {McKean}, J.~P., {et~al.} 2012, \apj, 750, 10

\bibitem[{{Tenneti} {et~al.}(2016){Tenneti}, {Mandelbaum}, \& {Di
  Matteo}}]{tmd16}
{Tenneti}, A., {Mandelbaum}, R., \& {Di Matteo}, T. 2016, \mnras, 462, 2668

\bibitem[{{Tenneti} {et~al.}(2014){Tenneti}, {Mandelbaum}, {Di Matteo}, {Feng},
  \& {Khandai}}]{tmm14}
{Tenneti}, A., {Mandelbaum}, R., {Di Matteo}, T., {Feng}, Y., \& {Khandai}, N.
  2014, \mnras, 441, 470

\bibitem[{{Troxel} {et~al.}(2018){Troxel}, {MacCrann}, {Zuntz}, {Eifler},
  {Krause}, {Dodelson}, {Gruen}, {Blazek}, {Friedrich}, {Samuroff}, {Prat},
  {Secco}, {Davis}, {Fert{\'e}}, {DeRose}, {Alarcon}, {Amara}, {Baxter},
  {Becker}, {Bernstein}, {Bridle}, {Cawthon}, {Chang}, {Choi}, {De Vicente},
  {Drlica-Wagner}, {Elvin-Poole}, {Frieman}, {Gatti}, {Hartley}, {Honscheid},
  {Hoyle}, {Huff}, {Huterer}, {Jain}, {Jarvis}, {Kacprzak}, {Kirk}, {Kokron},
  {Krawiec}, {Lahav}, {Liddle}, {Peacock}, {Rau}, {Refregier}, {Rollins},
  {Rozo}, {Rykoff}, {S{\'a}nchez}, {Sevilla-Noarbe}, {Sheldon}, {Stebbins},
  {Varga}, {Vielzeuf}, {Wang}, {Wechsler}, {Yanny}, {Abbott}, {Abdalla},
  {Allam}, {Annis}, {Bechtol}, {Benoit-L{\'e}vy}, {Bertin}, {Brooks},
  {Buckley-Geer}, {Burke}, {Carnero Rosell}, {Carrasco Kind}, {Carretero},
  {Castander}, {Crocce}, {Cunha}, {D'Andrea}, {da Costa}, {DePoy}, {Desai},
  {Diehl}, {Dietrich}, {Doel}, {Fernandez}, {Flaugher}, {Fosalba},
  {Garc{\'\i}a-Bellido}, {Gaztanaga}, {Gerdes}, {Giannantonio}, {Goldstein},
  {Gruendl}, {Gschwend}, {Gutierrez}, {James}, {Jeltema}, {Johnson}, {Johnson},
  {Kent}, {Kuehn}, {Kuhlmann}, {Kuropatkin}, {Li}, {Lima}, {Lin}, {Maia},
  {March}, {Marshall}, {Martini}, {Melchior}, {Menanteau}, {Miquel}, {Mohr},
  {Neilsen}, {Nichol}, {Nord}, {Petravick}, {Plazas}, {Romer}, {Roodman},
  {Sako}, {Sanchez}, {Scarpine}, {Schindler}, {Schubnell}, {Smith}, {Smith},
  {Soares-Santos}, {Sobreira}, {Suchyta}, {Swanson}, {Tarle}, {Thomas},
  {Tucker}, {Vikram}, {Walker}, {Weller}, {Zhang}, \& {DES
  Collaboration}}]{troxel18}
{Troxel}, M.~A., {MacCrann}, N., {Zuntz}, J., {et~al.} 2018, \prd, 98, 043528

\bibitem[{{Tugendhat} \& {Sch{\"a}fer}(2018)}]{tus18}
{Tugendhat}, T.~M. \& {Sch{\"a}fer}, B.~M. 2018, \mnras, 476, 3460

\bibitem[{{Umetsu}(2020)}]{umetsu20}
{Umetsu}, K. 2020, \aapr\, submitted (also arXiv:2007.00506)

\bibitem[{{Umetsu} {et~al.}(2018){Umetsu}, {Sereno}, {Tam}, {Chiu}, {Fan},
  {Ettori}, {Gruen}, {Okumura}, {Medezinski}, {Donahue}, {Meneghetti}, {Frye},
  {Koekemoer}, {Broadhurst}, {Zitrin}, {Balestra}, {Ben{\'\i}tez}, {Higuchi},
  {Melchior}, {Mercurio}, {Merten}, {Molino}, {Nonino}, {Postman}, {Rosati},
  {Sayers}, \& {Seitz}}]{ust18}
{Umetsu}, K., {Sereno}, M., {Tam}, S.-I., {et~al.} 2018, \apj, 860, 104

\bibitem[{{van Uitert} {et~al.}(2017){van Uitert}, {Hoekstra}, {Joachimi},
  {Schneider}, {Bland-Hawthorn}, {Choi}, {Erben}, {Heymans}, {Hildebrandt},
  {Hopkins}, {Klaes}, {Kuijken}, {Nakajima}, {Napolitano}, {Schrabback},
  {Valentijn}, \& {Viola}}]{vanuitert17}
{van Uitert}, E., {Hoekstra}, H., {Joachimi}, B., {et~al.} 2017, \mnras, 467,
  4131

\bibitem[{{van Uitert} {et~al.}(2012){van Uitert}, {Hoekstra}, {Schrabback},
  {Gilbank}, {Gladders}, \& {Yee}}]{uhs12}
{van Uitert}, E., {Hoekstra}, H., {Schrabback}, T., {et~al.} 2012, \aap, 545,
  A71

\bibitem[{{Vega-Ferrero} {et~al.}(2017){Vega-Ferrero}, {Yepes}, \&
  {Gottl{\"o}ber}}]{vyg17}
{Vega-Ferrero}, J., {Yepes}, G., \& {Gottl{\"o}ber}, S. 2017, \mnras, 467, 3226

\bibitem[{{Velander} {et~al.}(2014){Velander}, {van Uitert}, {Hoekstra},
  {Coupon}, {Erben}, {Heymans}, {Hildebrandt}, {Kitching}, {Mellier}, {Miller},
  {Van Waerbeke}, {Bonnett}, {Fu}, {Giodini}, {Hudson}, {Kuijken}, {Rowe},
  {Schrabback}, \& {Semboloni}}]{vuh14}
{Velander}, M., {van Uitert}, E., {Hoekstra}, H., {et~al.} 2014, \mnras, 437,
  2111

\bibitem[{{Velliscig} {et~al.}(2015){Velliscig}, {Cacciato}, {Schaye}, {Crain},
  {Bower}, {van Daalen}, {Vecchia}, {Frenk}, {Furlong}, {McCarthy}, {Schaller},
  \& {Theuns}}]{vcs15}
{Velliscig}, M., {Cacciato}, M., {Schaye}, J., {et~al.} 2015, \mnras, 453, 721

\bibitem[{{Wright} {et~al.}(2019){Wright}, {Hildebrandt}, {Kuijken}, {Erben},
  {Blake}, {Buddelmeijer}, {Choi}, {Cross}, {de Jong}, {Edge},
  {Gonzalez-Fernandez}, {Gonz{\'a}lez Solares}, {Grado}, {Heymans}, {Irwin},
  {Kupcu Yoldas}, {Lewis}, {Mann}, {Napolitano}, {Radovich}, {Schneider},
  {Sif{\'o}n}, {Sutherland}, {Sutorius}, \& {Verdoes Kleijn}}]{wright19}
{Wright}, A.~H., {Hildebrandt}, H., {Kuijken}, K., {et~al.} 2019, \aap, 632,
  A34

\bibitem[{{Wright} \& {Brainerd}(2000)}]{wrb00}
{Wright}, C.~O. \& {Brainerd}, T.~G. 2000, \apj, 534, 34

\end{thebibliography}

\section*{Acknowledgements}

We thank Rachel Mandelbaum for providing the tabulated model predictions for the anisotropic shear signal.
We also thank the anonymous referee, whose comments have helped to improve the paper significantly.

This work is based on observations obtained with MegaPrime/MegaCam, a
joint project of CFHT and CEA/DAPNIA, at the Canada-France-Hawaii
Telescope (CFHT) which is operated by the National Research Council
(NRC) of Canada, the Institut National des Sciences de l'Univers of
the Centre National de la Recherche Scientifique (CNRS) of France, and
the University of Hawaii. This research used the facilities of the
Canadian Astronomy Data Centre operated by the National Research
Council of Canada with the support of the Canadian Space Agency.
RCSLenS data processing was made possible thanks to significant
computing support from the NSERC Research Tools and Instruments grant
program.
CS82 is a joint project using Canadian, French and Brazilian time on CFHT (programmes 10BB009, 10BF023 and 10BC022). The Brazilian partnership on CFHT was managed by the Laborat\'orio Nacional de Astrof\'isica (LNA).
Based on data products from observations made with ESO Telescopes at the La Silla Paranal Observatory under programme IDs 177.A-3016, 177.A-3017, 177.A-3018, 179.A-2004, 298.A-5015, and on data products produced by the KiDS consortium.

HH and AK acknowledge support from Vici grant 639.043.512 financed by the Netherlands Organization for Scientic Research.
EvU acknowledges support from an STFC Ernest Rutherford Research Grant, grant reference ST/L00285X/1.
LvW is funded by NSERC and CIfAR.
CH and MA acknowledge support from the European Research Council under grant number 647112.
CH also acknowledges support from the Max Planck Society and the Alexander von Humboldt Foundation in the framework of the Max Planck-Humboldt Research Award endowed by the Federal Ministry of Education and Research.
HHi is supported by a Heisenberg grant of the Deutsche Forschungsgemeinschaft (Hi 1495/5-1) as well as an ERC Consolidator Grant (No. 770935).
LM acknowledges support from STFC grant ST/N000919/1.

{\it Author Contributions:} All authors contributed to the development
and writing of this paper. The authorship list is given in two groups:
the lead authors (TS, HH, LVW, EvU, CG), followed by
an alphabetical
group, which covers those who have either made a significant contribution to the data products or to the scientific analysis.

\appendix

\section{Lens sub-samples in CS82}
\label{app:CS82lenses}

 \begin{figure*}
   \centering
  \includegraphics[width=8cm]{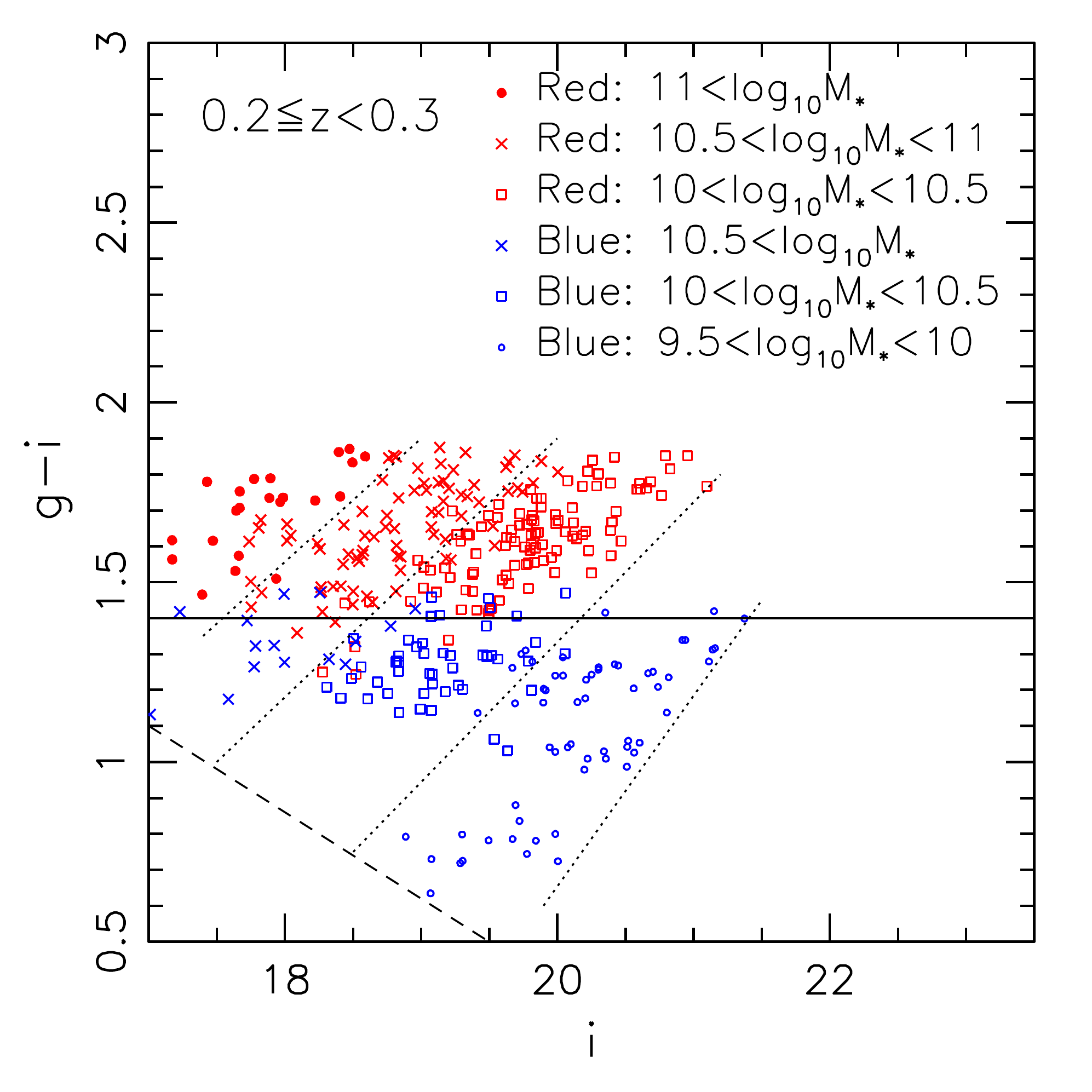}
  \includegraphics[width=8cm]{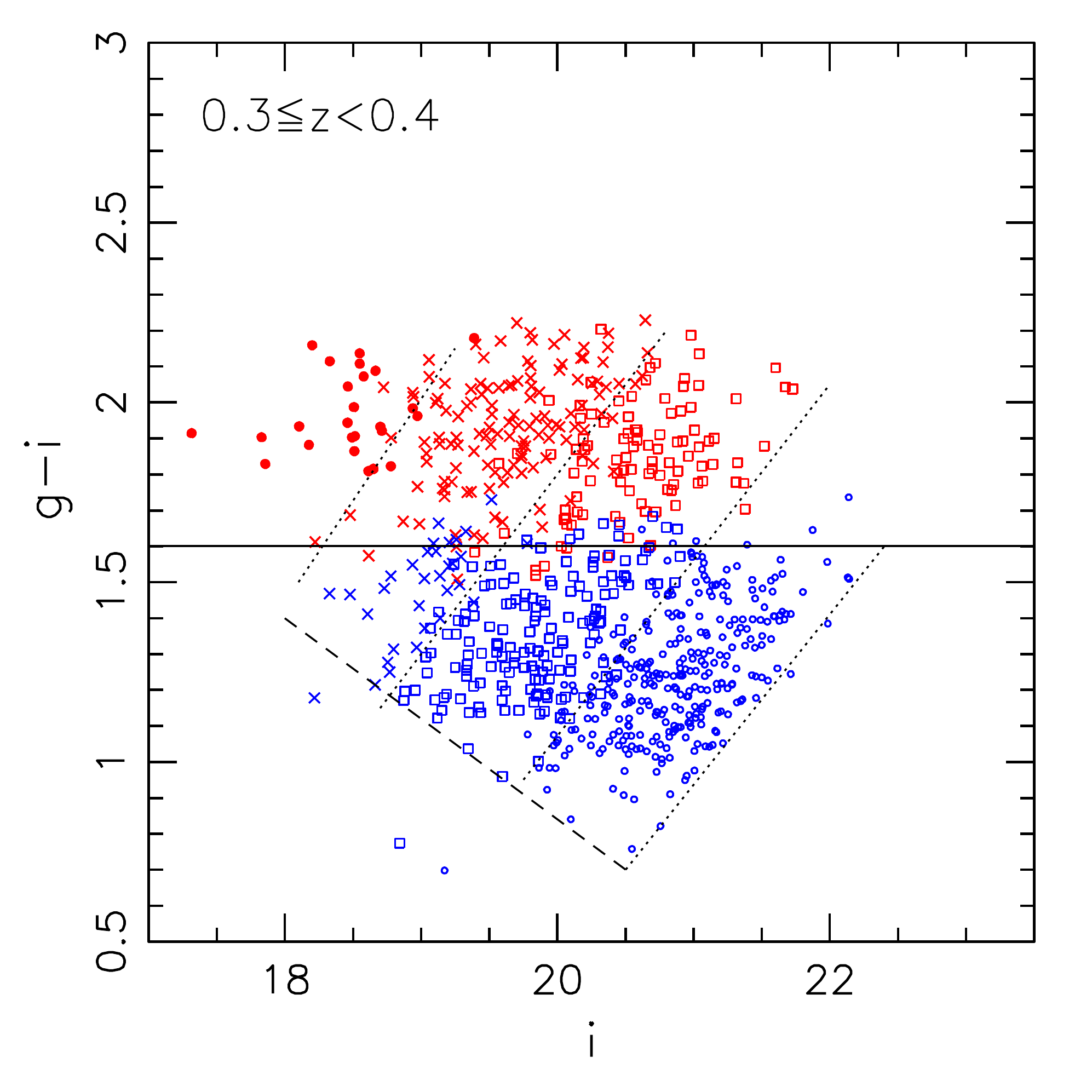}
\includegraphics[width=8cm]{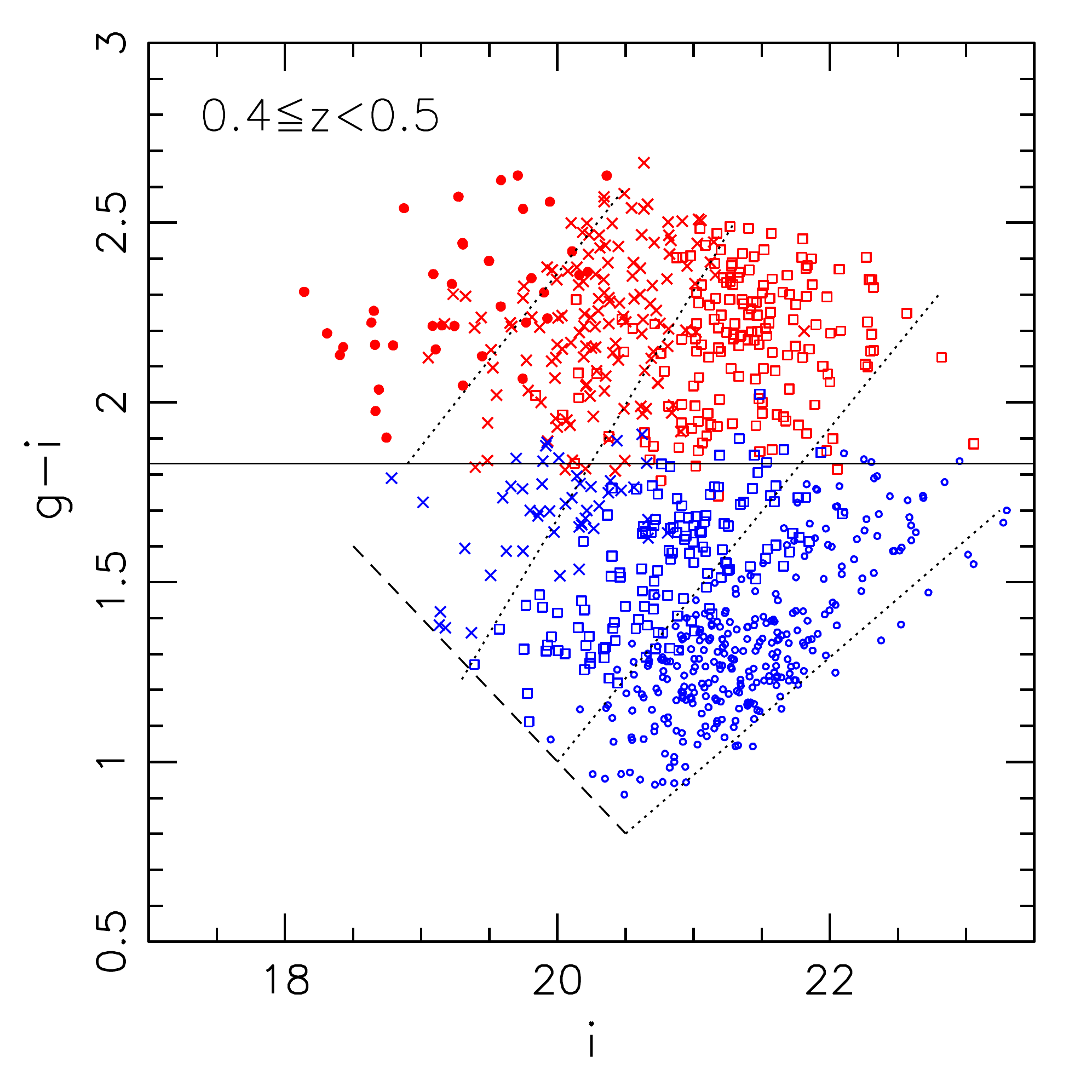}
 \includegraphics[width=8cm]{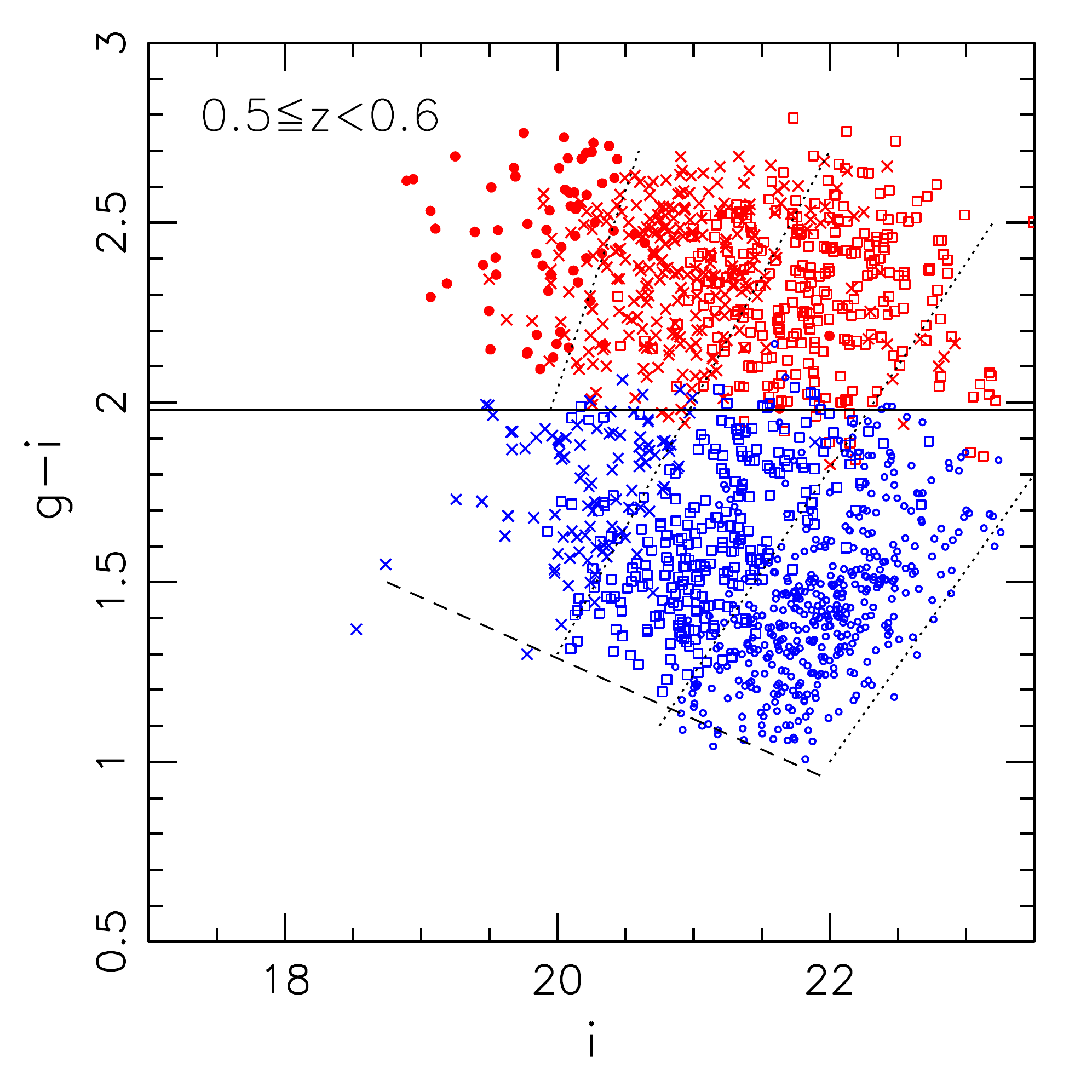}
 \caption{Distribution of red galaxies (\mbox{$T_\mathrm{BPZ} \le 1.5$}) versus blue galaxies  (\mbox{$1.5<T_\mathrm{BPZ} < 3.95$}) selected according to the photometric type from \textsc{BPZ} and split into different stellar mass bins in $g-i$ versus $i$ colour-magnitude space, as found in the CFHTLenS field W1m0m1.
   The solid and dotted lines illustrate how we split the lenses into red versus blue bins and approximate stellar mass bins, respectively, in the analysis of CS82 data. The dashed lines additionally indicate $g-i$ limits used to exclude very blue star forming galaxies in the CS82 analysis.
 }
   \label{fi:gi}
    \end{figure*}

 Since
 stellar mass estimates and \textsc{BPZ} photometric types  were not available in the CS82 catalogues employed in our analysis\textsuperscript{\ref{footnote:bundy}},
 we applied a different approach to split the lenses into
sub-samples for these data.
Using the CFHTLenS data as a template,
we find that the observed $g-i$ colour
provides a good proxy to split the galaxies into the blue and red sub-samples (see Sect.\thinspace\ref{se:lenssample})
after the separation into the thin redshift slices has been applied (see Fig.\thinspace\ref{fi:gi}).
For CS82 we therefore define red versus blue samples by splitting lens candidates at observed (extinction-corrected) colours \mbox{$g-i=(1.4, 1.6, 1.83, 1.98)$} for lenses in the photometric redshifts intervals \mbox{$[0.2,0.3),[0.3,0.4),[0.4,0.5)$}, and  \mbox{$[0.5,0.6)$}, respectively.
        Likewise, the comparison to CFHTLenS reveals that a subdivision into stellar mass bins can approximately be reproduced by selecting different regions in $g-i$ versus $i$ colour--magnitude space after applying the redshift selection (see Fig.\thinspace\ref{fi:gi}), which we then employ for the CS82 analysis.
        However, for the other surveys we still employ the actual
         stellar mass estimates, which are expected to provide
         a more accurate proxy for halo mass given that their computation makes use of additional information (especially the NIR data for KV450).

\section{Bayesian estimation of $f_\mathrm{h}$}
\label{app:likelihood}

 \begin{figure*}
   \centering
  \includegraphics[width=7.4cm]{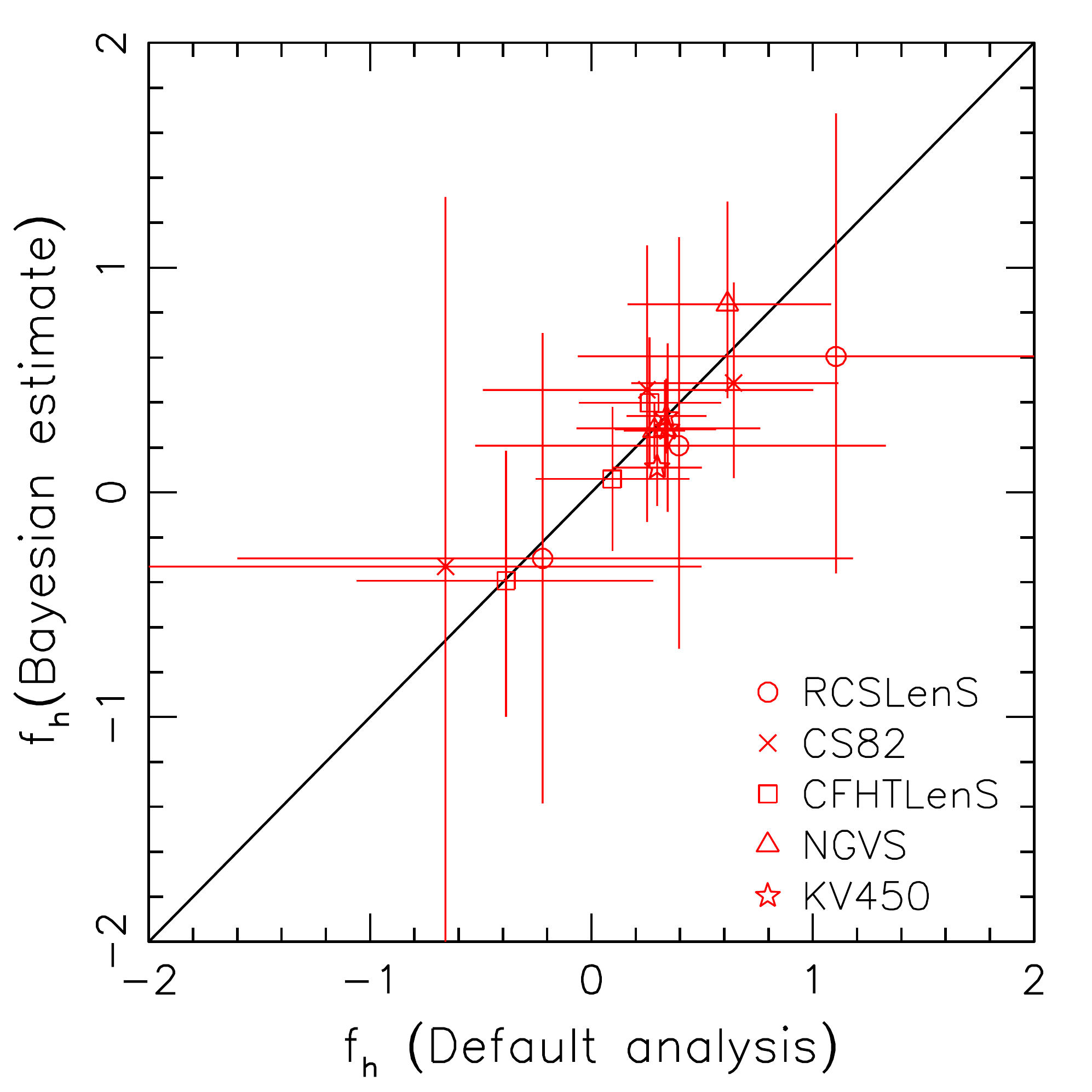}
  \includegraphics[width=7.4cm]{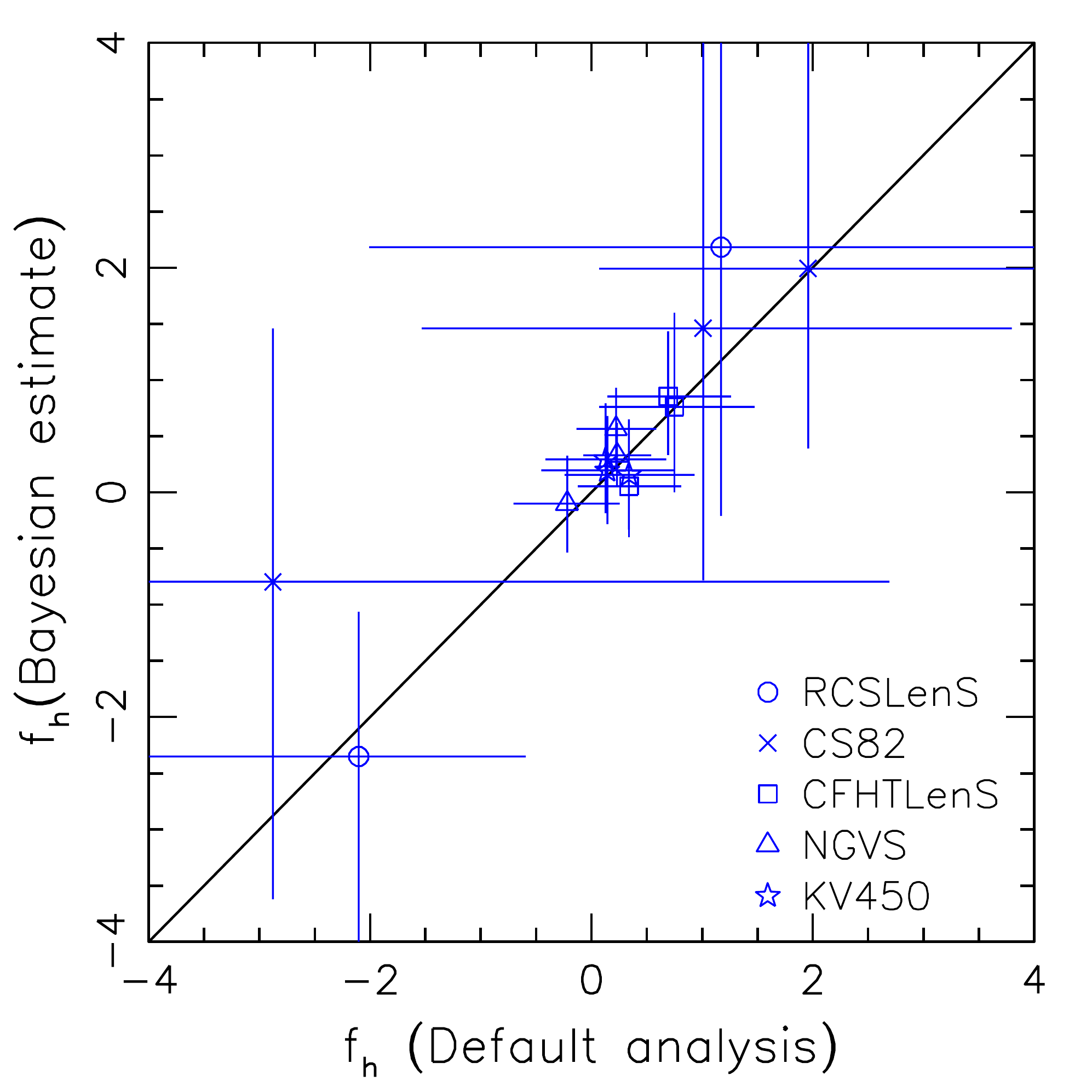}
  \caption{Comparison of $f_\mathrm{h}$ estimates for the different surveys and lens stellar mass bins from our default analysis scheme (see Sect.\thinspace\ref{se:getfh}) and the Bayesian estimation described in Appendix \ref{app:likelihood}. The left and right panels show red and blue lens samples, respectively.
  }
   \label{fi:likelihood_fhfh}
    \end{figure*}

 \begin{figure}
   \centering
  \includegraphics[width=7.4cm]{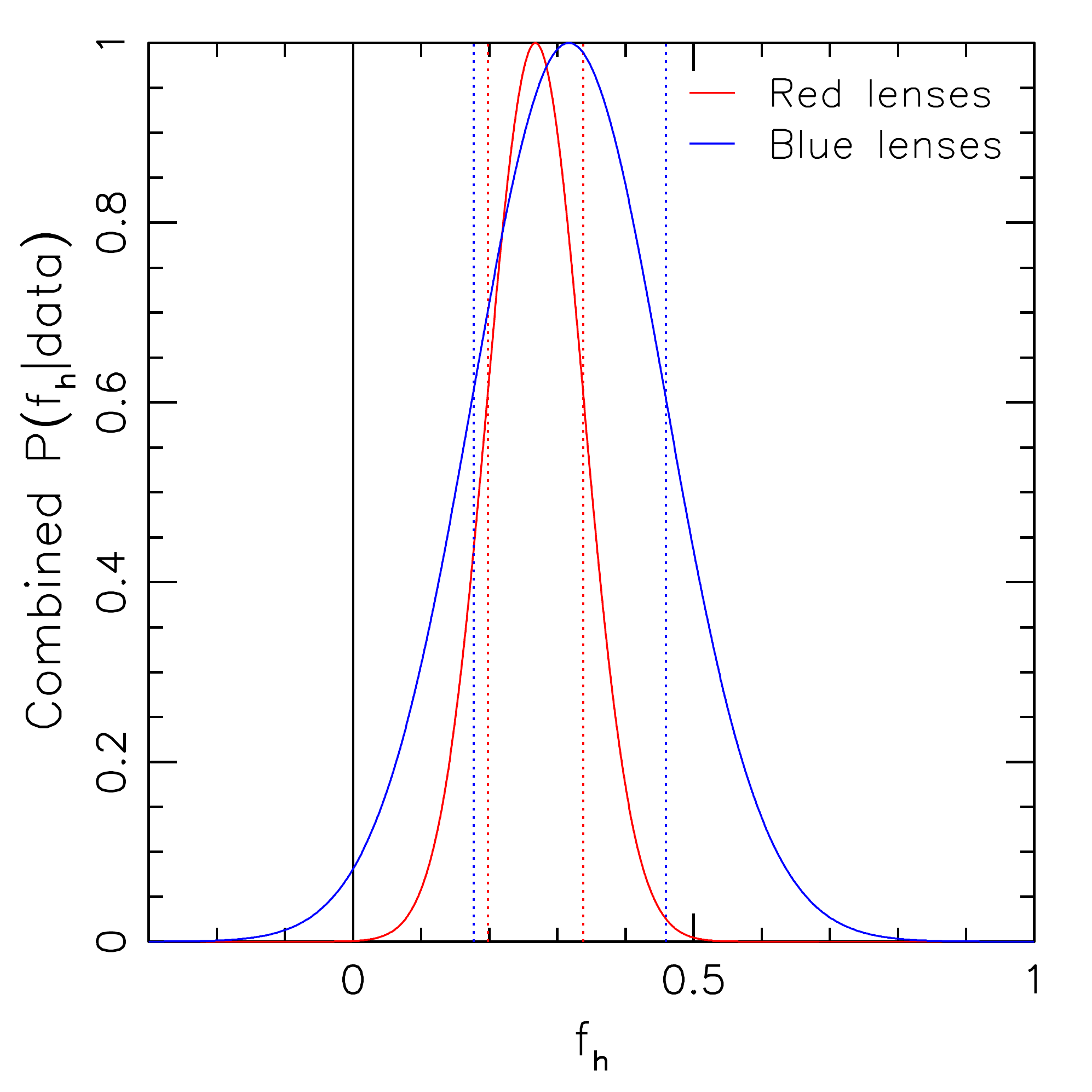}
  \caption{Posterior probability distribution (curves, normalised based on the peak value)
of
    $f_\mathrm{h}$ for red and blue lenses, combining all surveys and stellar mass bins. The
    68\% confidence limits are indicated by the vertical
    dotted lines.
  }
   \label{fi:likelihood_p_of_fh}
    \end{figure}

 In Sect.\thinspace\ref{se:getfh} we described our default analysis approach,
 where we follow \citetalias{mhb06} to estimate $f_\mathrm{h}$ from the noisy $\widehat{\Delta \Sigma_\mathrm{iso}}(r)$ and
 $\widehat{\left[f-f_{45}\right]\Delta  \Sigma_\mathrm{iso}(r)}$\footnote{In Sect.\thinspace\ref{se:measuring_halo_shapes} we denoted these estimates as $\widehat{\left[f(r)-f_{45}(r)\right]\Delta  \Sigma_\mathrm{iso}(r)}$ to stress that the model expectations for
$f$ and $f_{45}$
are radius dependent. However, within our fit range the radius dependence is weak for $f(r)-f_{45}(r)$, as indicated by the almost flat curves in Fig.\thinspace\ref{fi:stacked_shear}. Thus, and to improve the readability we employ this shortened notation here and in the $y$-axis labels of Figs.\thinspace\ref{fi:shearfitsred_NGVS}, \ref{fi:stacked_shear}, and \ref{fi:shearfitsred_KV450} to \ref{fi:shearfitsred_RCS2}.}
 estimates in individual radial bins.
 In this Appendix we explore an alternative
 Bayesian inference approach.
Here we first employ the $\widehat{\Delta \Sigma_\mathrm{iso}}(r)$ profile
to constrain the
posterior probability distribution for the effective mean halo mass $M_\mathrm{200c}$ of the particular lens sample
 \begin{equation}
   \begin{split}
     &
     P\left(M_\mathrm{200c}|\,\widehat{\Delta \Sigma_\mathrm{iso}}(r)\right) \propto  P_0(M_\mathrm{200c})
   P\left(\widehat{\Delta \Sigma_\mathrm{iso}}(r)|M_\mathrm{200c}\right) \\
   & \propto  P_0(M_\mathrm{200c})
     \exp{\left[-\frac{1}{2}\sum_i\left(\frac{\widehat{\Delta \Sigma}_\mathrm{iso}(r_i)-\Delta \Sigma_\mathrm{iso}^\mathrm{model}(r_i|M_\mathrm{200c})}{\sigma_{\widehat{\Delta \Sigma}_\mathrm{iso}}(r_i)}\right)^2\right]}\, ,
  \end{split}
  \label{eq:pofm}
  \end{equation}
  where
  we assume a Gaussian likelihood, employ a flat prior in halo mass $P_0(M_\mathrm{200c})$, and ignore the small and noisy off-diagonal elements of the shape-noise-dominated covariance matrix as done in the equations below\footnote{Confirming results from \citetalias{shh15} we find that the off-diagonal entries in the correlation matrix are generally small.
For example,
    for the  KV-450 data within our fit range the off-diagonal entries in the correlation matrix
    (including the cross-correlation between $\widehat{\Delta \Sigma_\mathrm{iso}}(r)$ and $\widehat{\left[f
            -f_{45}
          \right]\Delta
          \Sigma_\mathrm{iso}(r)}$)
    have an r.m.s. of 0.05 (likely limited by noise from the covariance estimation via  the field-wise bootstrapping) and
have absolute values that
    are always $< 0.15$.}.
In Eq.\thinspace(\ref{eq:pofm}) the sum runs over all radial bins $i$ with $45\thinspace\mathrm{kpc}/h_{70}<r_i<200\thinspace\mathrm{kpc}/h_{70}$, matching the fit range used in Sect.\thinspace\ref{se:halo_shapes_results} to constrain halo masses.

We then compute the posterior probability distribution for $f_\mathrm{h}$ given the data for this lens bin
\begin{equation}
  \begin{split}
  &  P\left(f_\mathrm{h}|\,\mathrm{data}\right)
    \propto P_0(f_\mathrm{h})\, P\left(\mathrm{data}|\, f_\mathrm{h}\right)\propto \\
    &
    P_0(f_\mathrm{h}) \int \mathrm{d} M_\mathrm{200c}
    P\left(M_\mathrm{200c}|\widehat{\Delta \Sigma_\mathrm{iso}}(r)\right)
     P\left( \widehat{\left[f
           -f_{45}
         \right]\Delta
             \Sigma_\mathrm{iso}(r)} | f_\mathrm{h}, M_\mathrm{200c}\right)    \\
    &\propto P_0(f_\mathrm{h}) \int \mathrm{d} M_\mathrm{200c}
    P\left(M_\mathrm{200c}|\widehat{\Delta \Sigma_\mathrm{iso}}(r)\right)
     \exp{\left[-\frac{1}{2}
         \sum_j
\chi^2_j \label{eq:poffh}
       \right]}\, ,
   \end{split}
 \end{equation}
 where
 \begin{equation}
   \chi_j=
     \frac{
       \widehat{\left[f
           -f_{45}
         \right]\Delta
             \Sigma_\mathrm{iso}(r_j)}
           -
            \left[f_\mathrm{rel}(r_j)-f_\mathrm{rel,45}(r_j)\right]
           f_\mathrm{h} \Delta \Sigma_\mathrm{iso}^\mathrm{model}(r_j|M_\mathrm{200c})
     }
     {
       \sigma_{\widehat{\left[f
             -f_{45}
           \right]\Delta
           \Sigma_\mathrm{iso}}}(r_j)
     }\,. \label{eq:chi_bayesian}
\end{equation}
In Eq.\thinspace(\ref{eq:poffh}) we marginalise over halo mass and assume a flat prior $P_0(f_\mathrm{h})$.
The factor $\left[f_\mathrm{rel}(r_j)-f_\mathrm{rel,45}(r_j)\right]$
in Eq.\thinspace(\ref{eq:chi_bayesian}) also depends on halo mass given its dependence on $r/r_\mathrm{s}(M_\mathrm{200c})$ (see Sect.\thinspace\ref{se:constrain-ani}).
However, since this dependence is weak (as indicated by the almost flat curves in Fig.\thinspace\ref{fi:stacked_shear}), and given that the halo masses are well constrained for all of our
lens samples (when combining the different lens redshift slices),
we can safely employ the  $\left[f_\mathrm{rel}(r_j)-f_\mathrm{rel,45}(r_j)\right]$ model prediction computed at the best-fitting halo mass.
Likewise, we keep the fit range  \mbox{$45$\thinspace{kpc}/$h_{70}<r_j<r_{200\mathrm{c}}^\mathrm{best-fit}$} and therefore the  contributing bins $j$ in Eq.\thinspace(\ref{eq:chi_bayesian}) fixed, as defined by the best-fitting halo mass for this lens sample.

Fig.\thinspace\ref{fi:likelihood_fhfh} compares the Bayesian estimates of $f_\mathrm{h}$ to the constraints derived from our default analysis for the different lens samples and surveys.
We generally find very good agreement between the estimates, confirming the previously employed approach.
We note that we do not expect perfect agreement for multiple reasons.
For example, our Bayesian approach assumes that the isotropic signal is perfectly modelled by the reduced shear profile  of a  spherical NFW halo. This is not
strictly required  for the approach described in Sect.\thinspace\ref{se:getfh},
since it directly uses the   $\widehat{\Delta \Sigma_\mathrm{iso}}(r)$ and
$\widehat{\left[f-f_{45}\right]\Delta  \Sigma_\mathrm{iso}(r)}$ estimates in the same radial bins without explicitly assuming a mass model  (of course, both approaches rely on accurate predictions for $\left[f_\mathrm{rel}(r)-f_\mathrm{rel,45}(r)\right]$, which would change for different density profiles).
Likewise, contributions from individual radial bins are weighted differently and change because of the differences between the  $\widehat{\Delta \Sigma_\mathrm{iso}}(r_i)$ and  $\Delta \Sigma_\mathrm{iso}^\mathrm{model}(r_i)$.
Finally, the $\Delta \Sigma_\mathrm{iso}^\mathrm{model}(r)$ model is typically constrained over a fit range with a different upper limit compared to the range
used to constrain $f_\mathrm{h}$.

Multiplying the  $P\left(f_\mathrm{h}|\,\mathrm{data}\right)$
posterior probability distributions computed for the individual surveys and different lens bins we can directly compute joint constraints, which are shown in
Fig.\thinspace\ref{fi:likelihood_p_of_fh} for all red and all blue lens samples, respectively.
From this we
estimate the mode of the distribution
and 68\% confidence intervals of
\mbox{$f_\mathrm{h}=0.268\pm 0.070$} for all red lenses, which is in good agreement
with the \mbox{$f_\mathrm{h}=0.303^{+0.080}_{-0.079}$} constraint derived using our default analysis (see Table \ref{tab:results}).
We suspect that the slightly smaller error-bars found in the Bayesian analysis may result from the stronger assumptions made in this approach (see above). We note that the significance of the `detection' of a non-zero  \mbox{$f_\mathrm{h}$} is basically identical for both approaches ($3.8\sigma$).
For the blue lens samples the  Bayesian approach yields
a joint constraint
\mbox{$f_\mathrm{h}=0.317^{+0.142}_{-0.140}$}, which  again has  moderately smaller error-bars than the constraint derived using our standard approach (\mbox{$f_\mathrm{h}=0.217^{+0.160}_{-0.159}$}).
For the  Bayesian analysis the mode estimate is also  noticeably higher than the best-fitting estimate of the default analysis (by $0.6\sigma$ using the error of the default analysis),
which we suspect may be caused by a combination of the effects  discussed above.
We suggest that future studies of larger samples could also employ both analysis approaches to investigate if similar shifts are found at a
higher  significance, which could hint at relevant systematic differences between the constraints derived from both approaches.

\section{Detailed results for the different surveys}
\label{se:appendix_figures}

\begin{figure*}
   \centering
  \includegraphics[width=7.4cm]{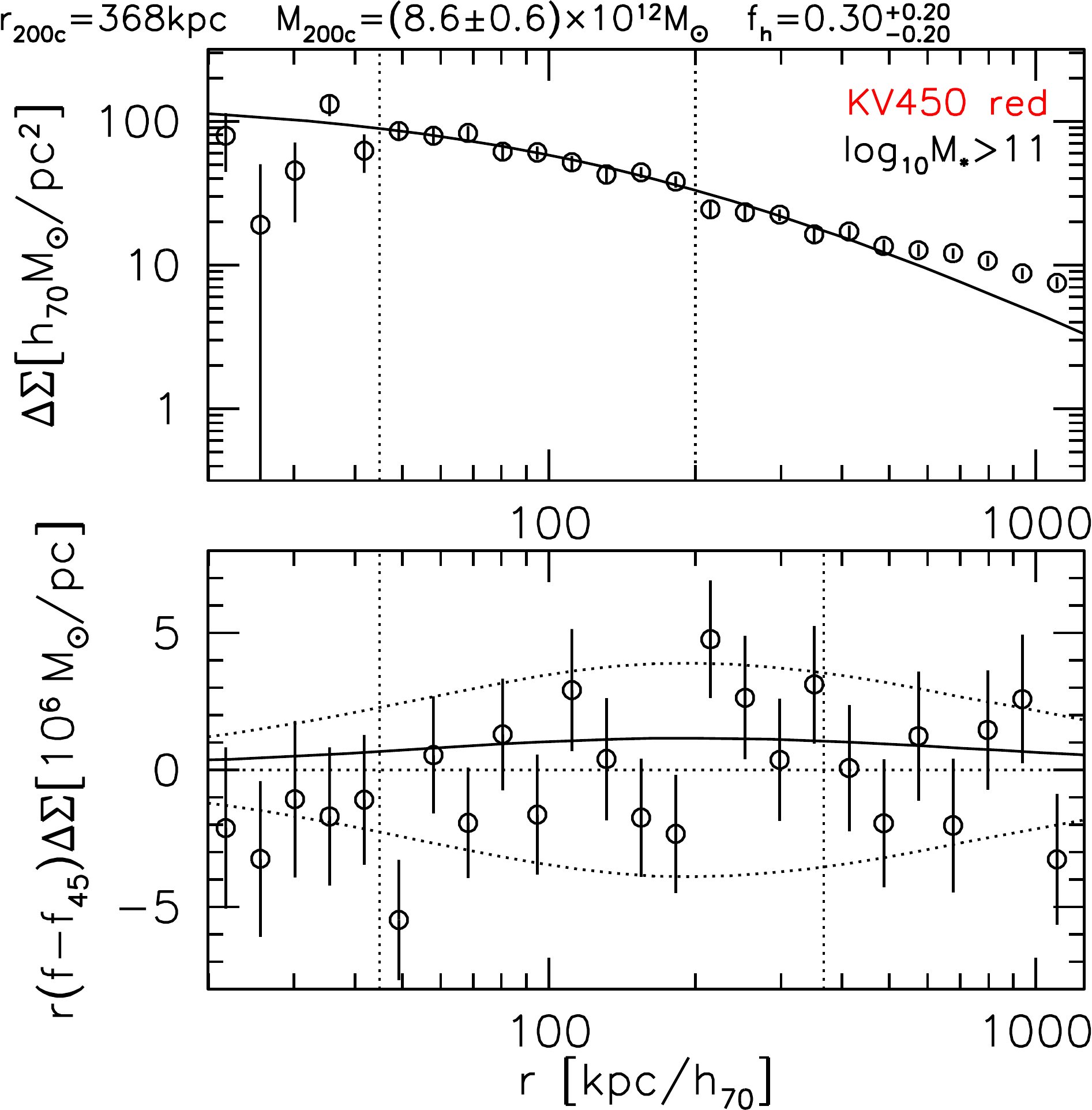}
   \includegraphics[width=7.4cm]{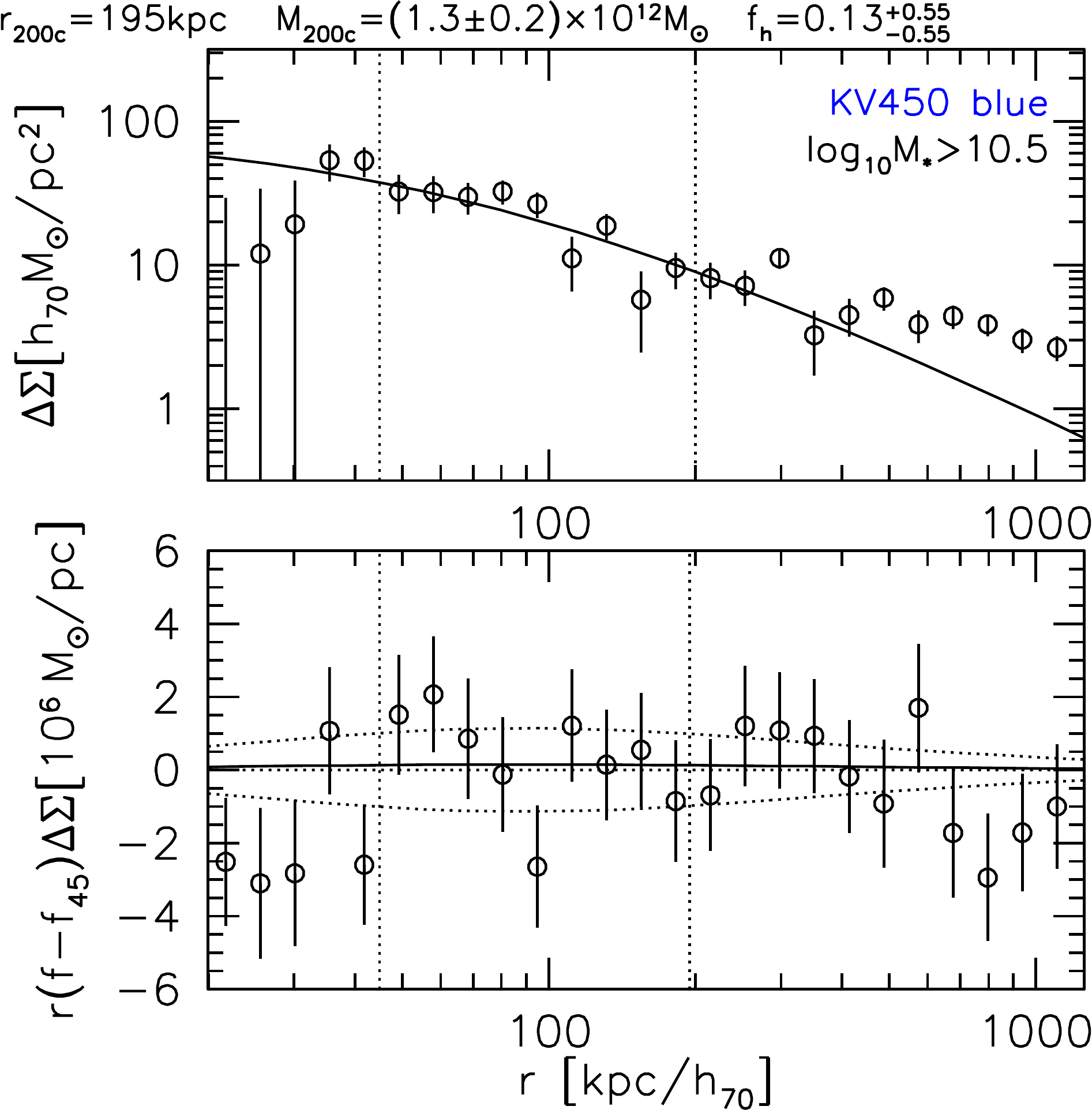}
 \includegraphics[width=7.4cm]{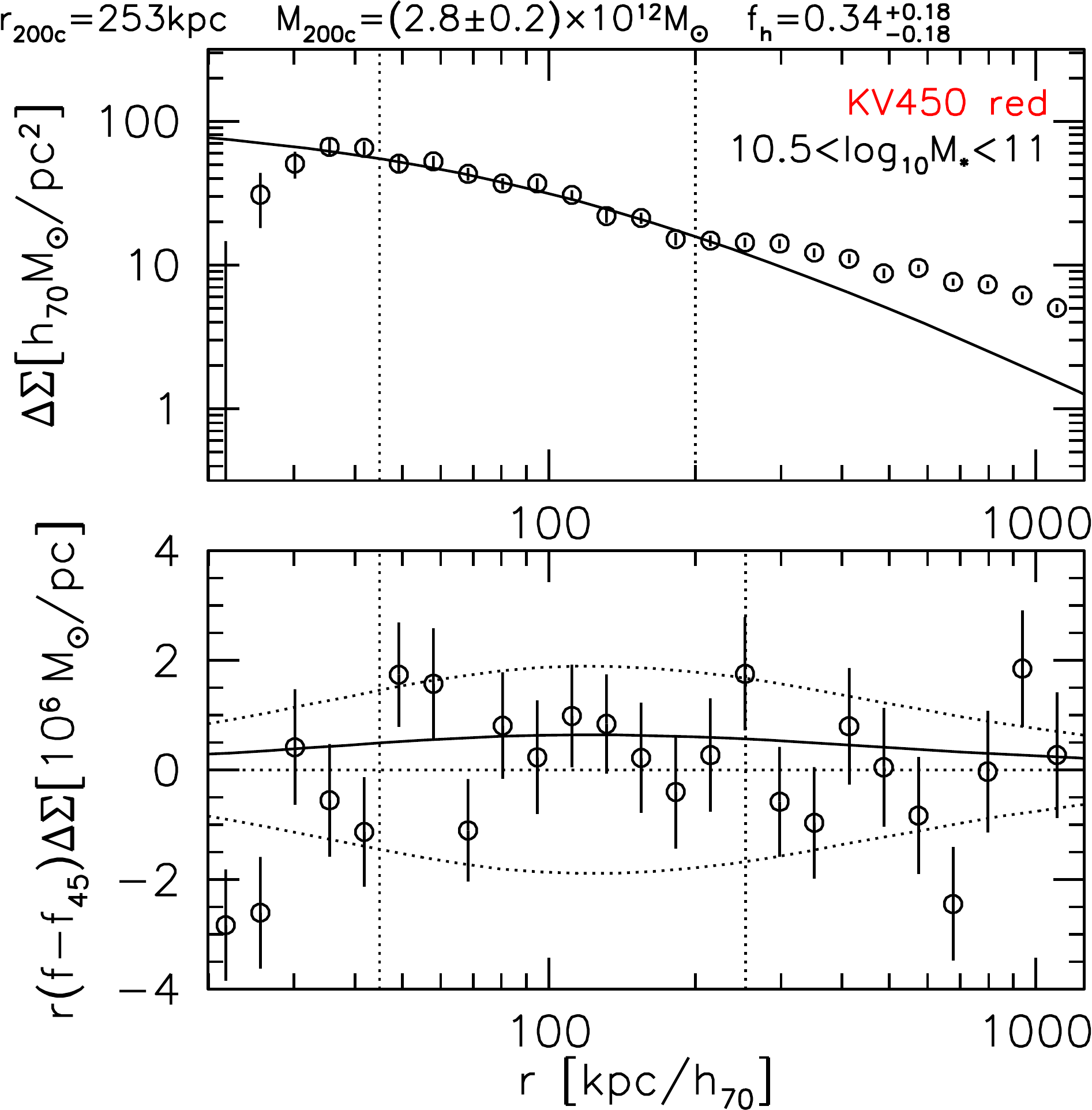}
   \includegraphics[width=7.4cm]{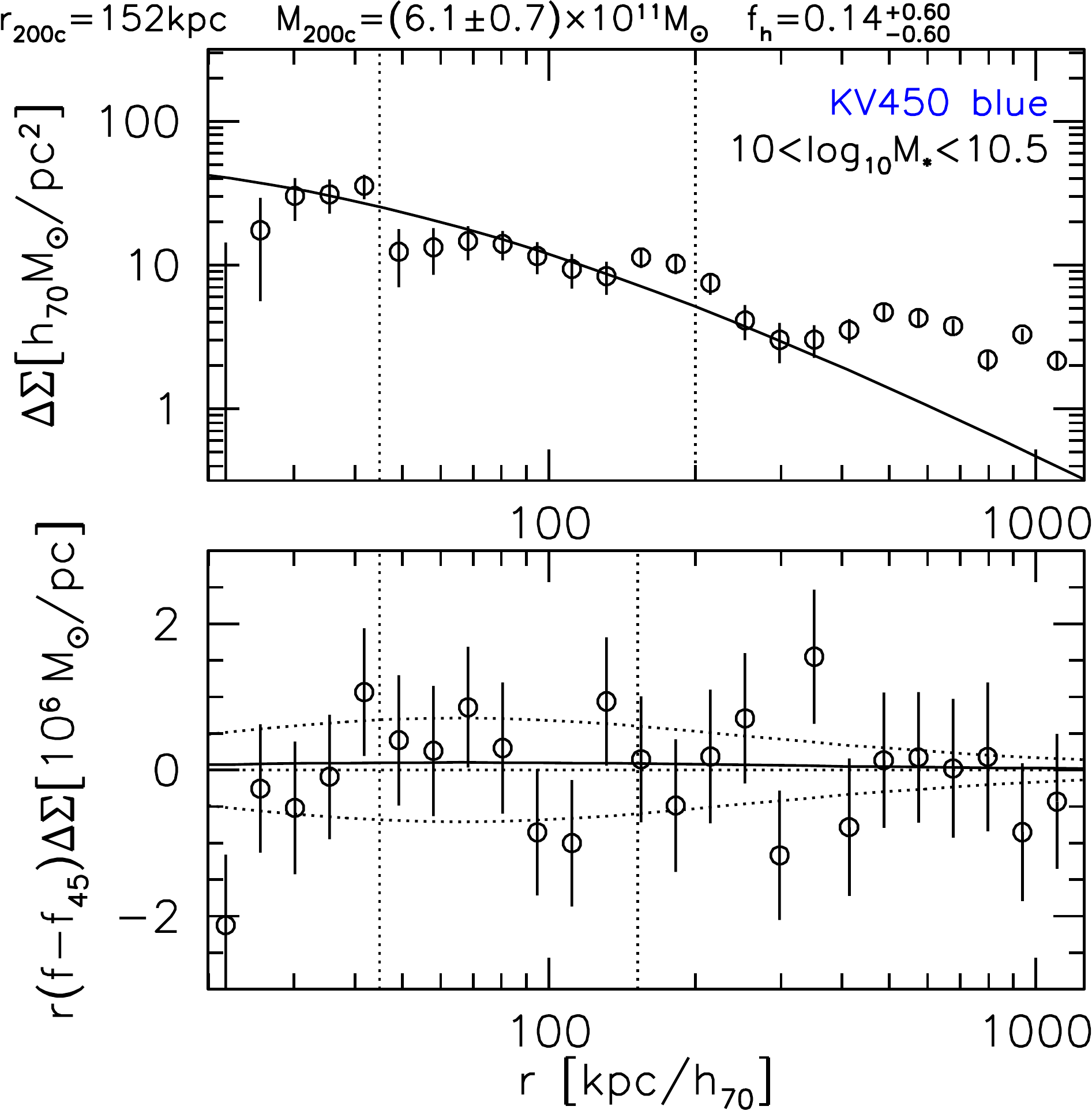}
 \includegraphics[width=7.4cm]{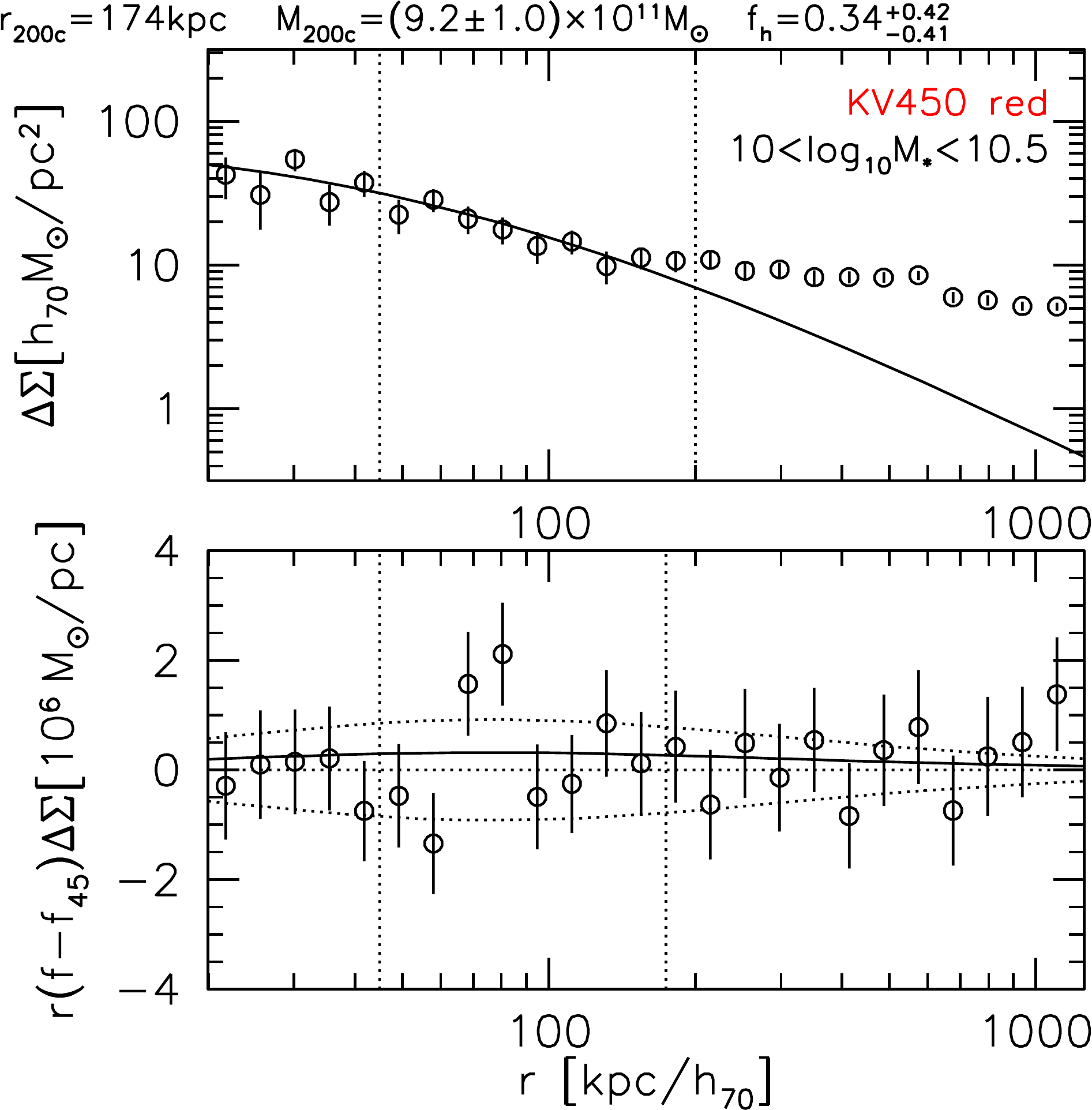}
  \includegraphics[width=7.4cm]{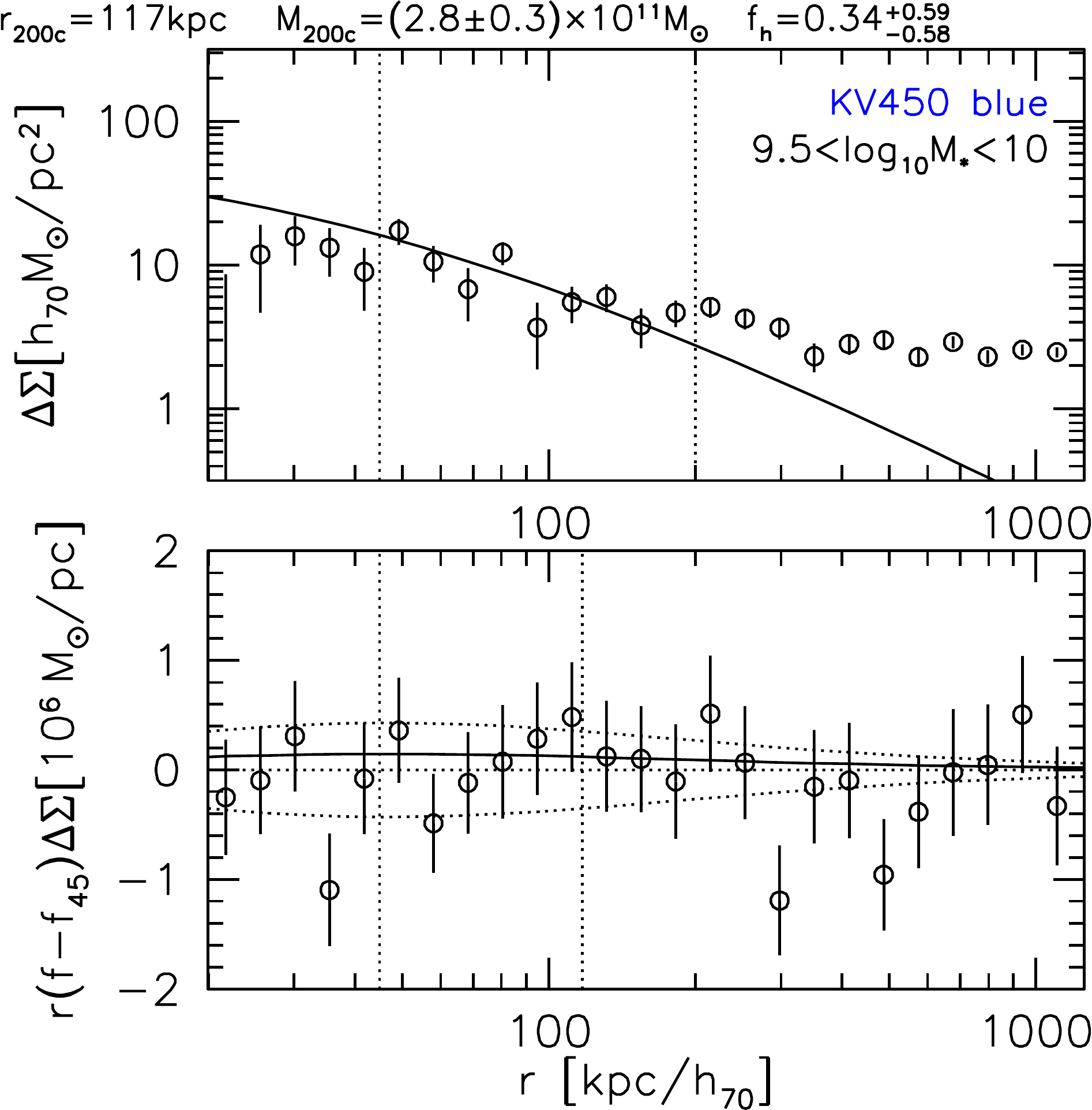}
  \caption{As Fig.\thinspace\ref{fi:shearfitsred_NGVS}, but computed from the KV450 data.
  }
   \label{fi:shearfitsred_KV450}
    \end{figure*}

\begin{figure*}
   \centering
  \includegraphics[width=7.4cm]{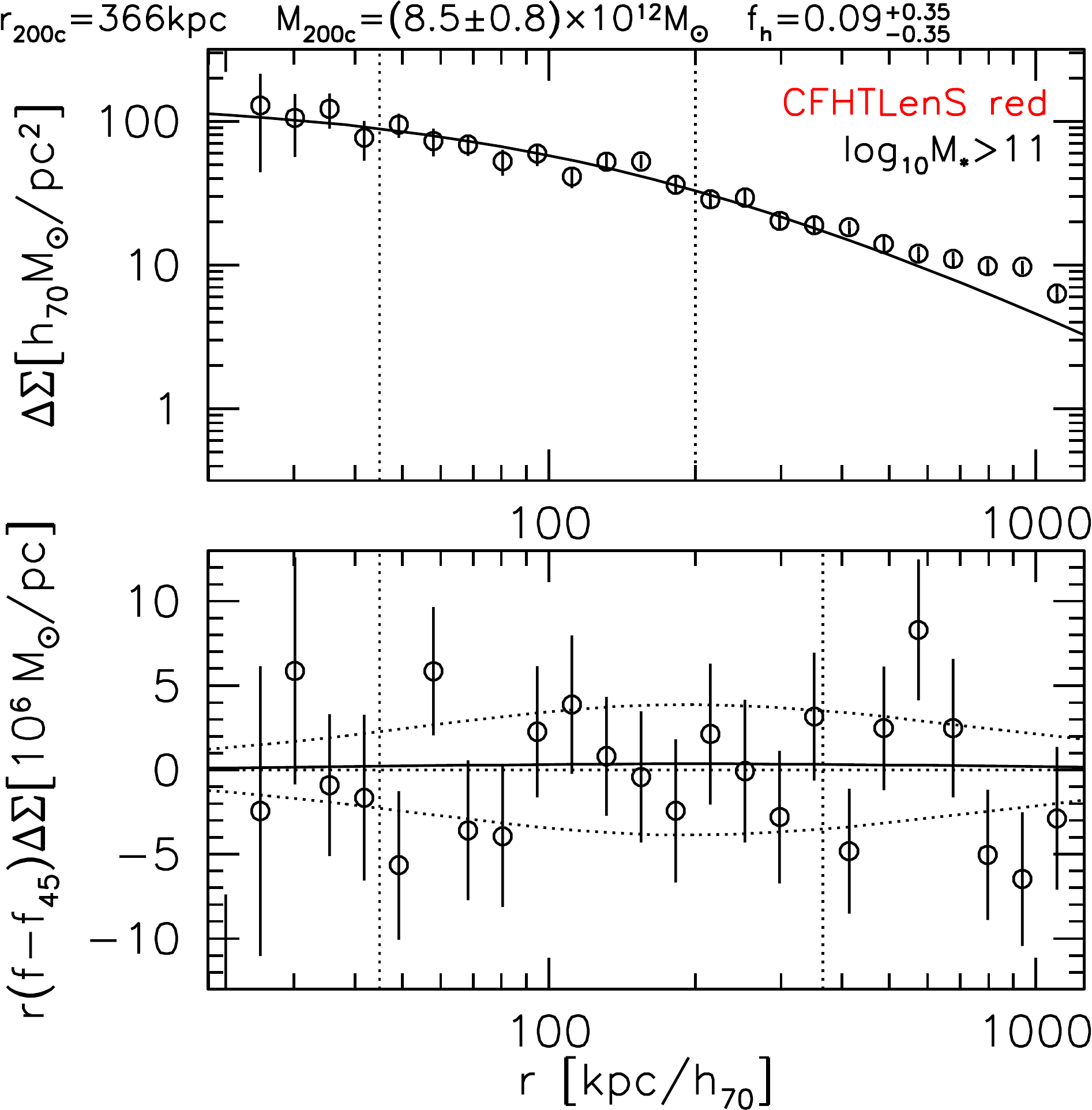}
  \includegraphics[width=7.4cm]{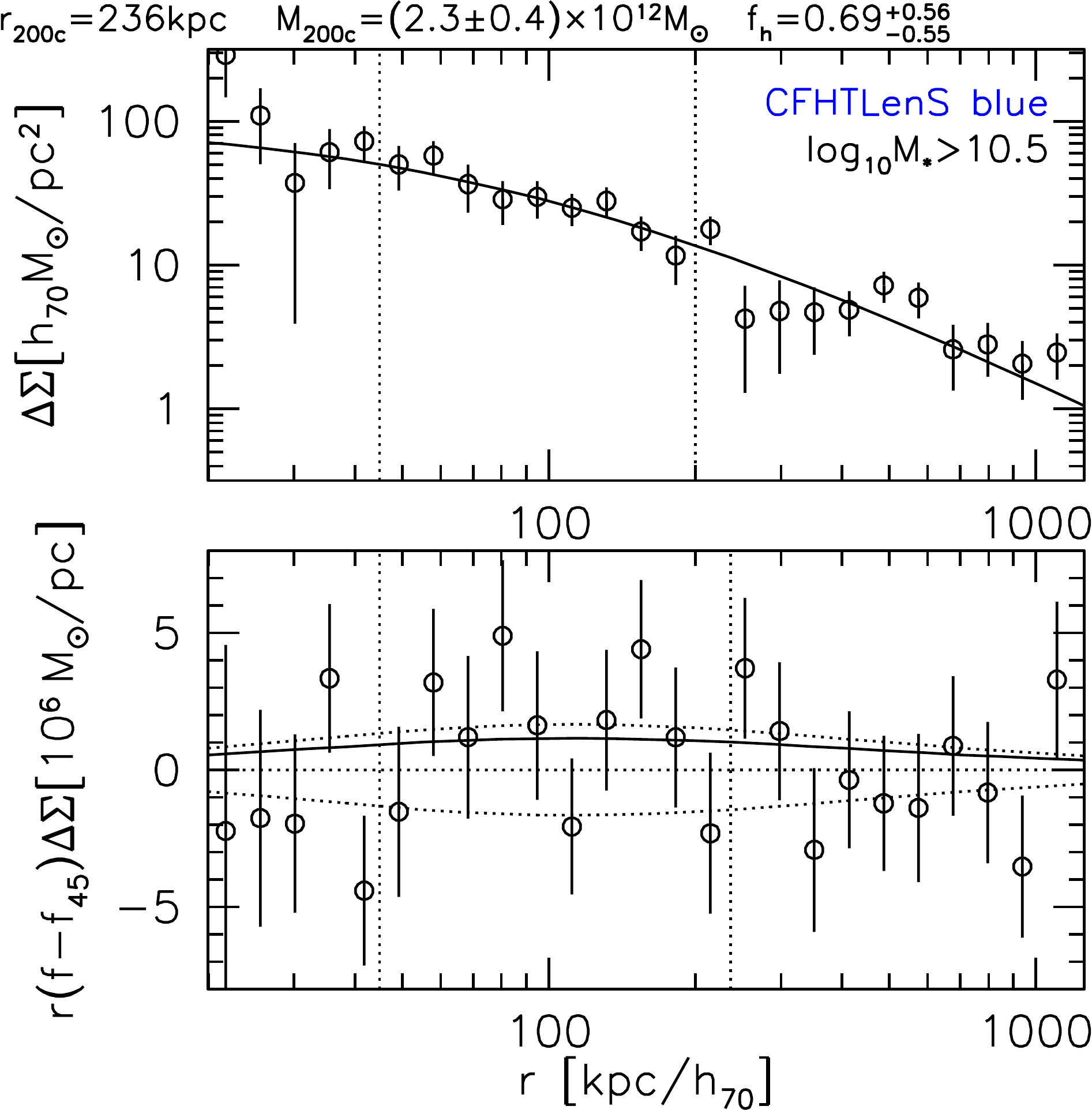}
  \includegraphics[width=7.4cm]{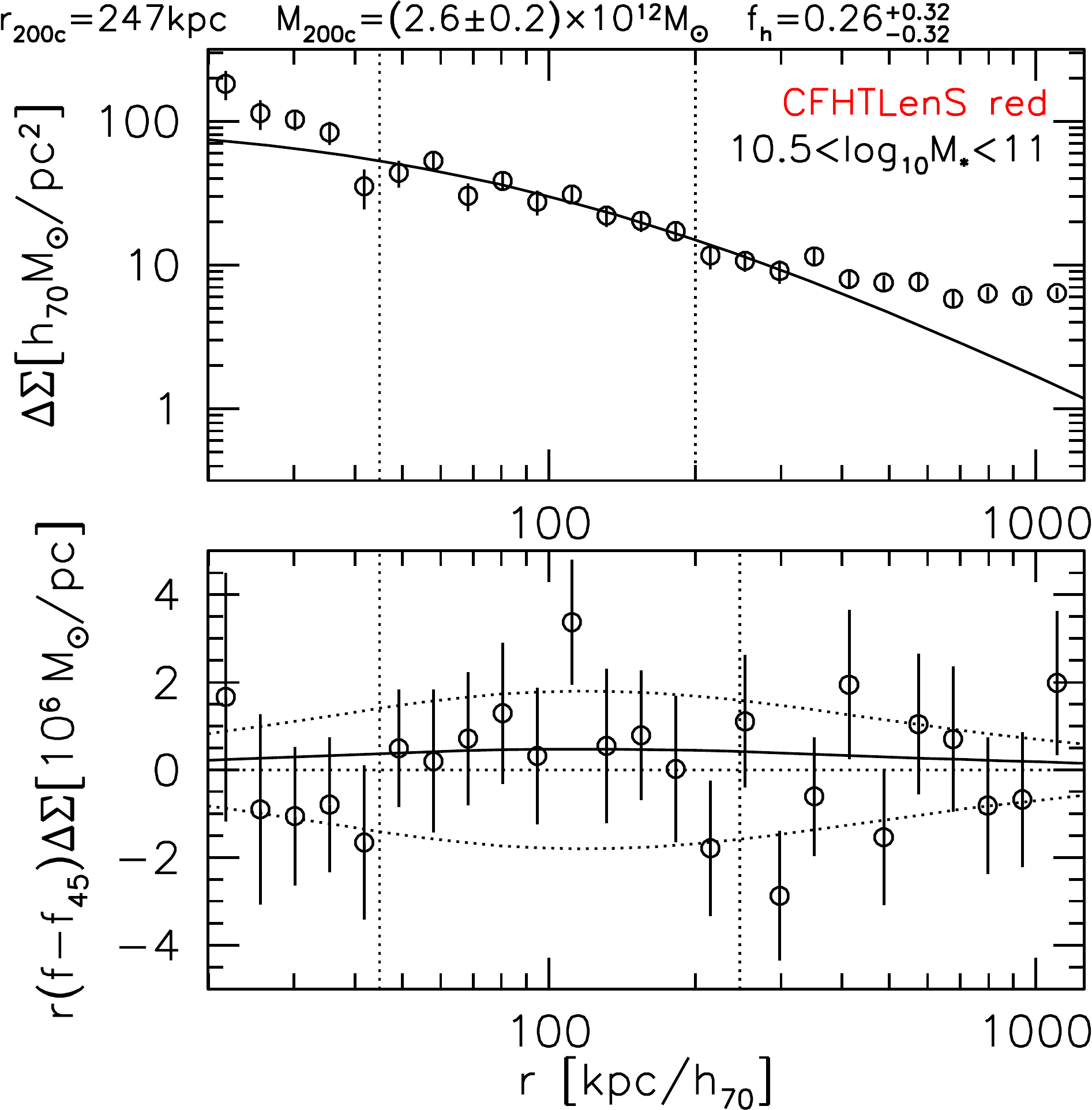}
  \includegraphics[width=7.4cm]{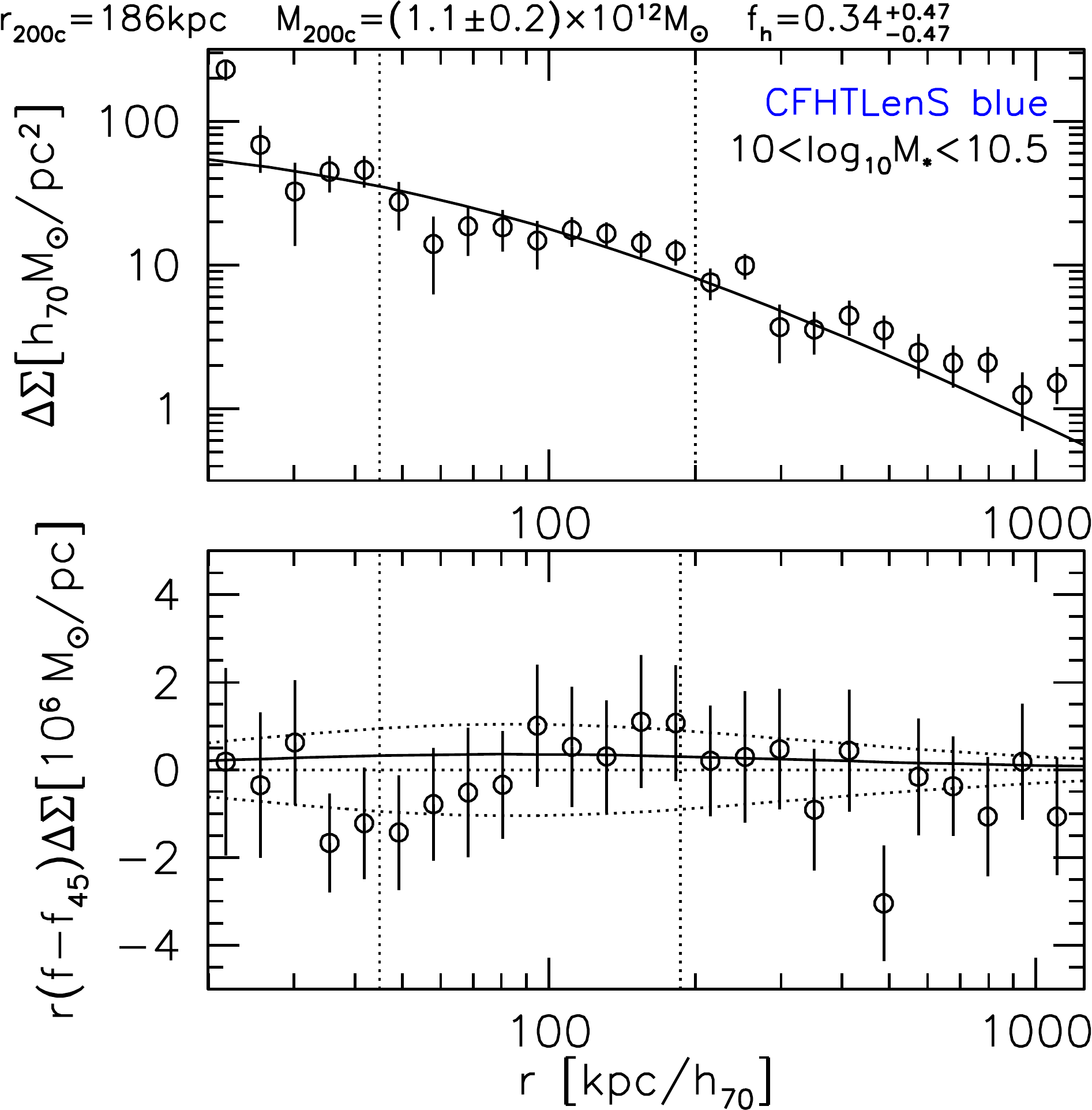}
  \includegraphics[width=7.4cm]{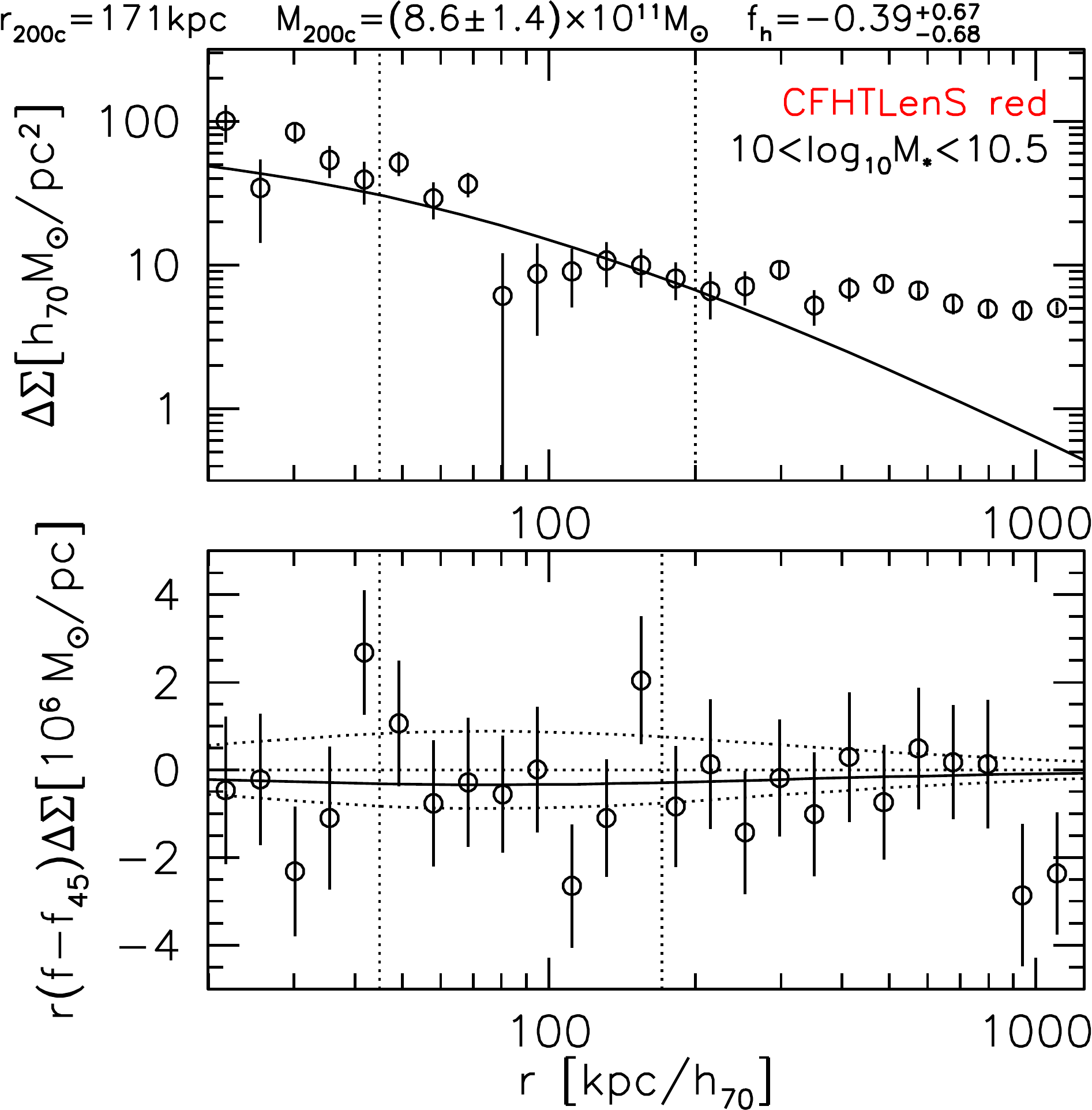}
  \includegraphics[width=7.4cm]{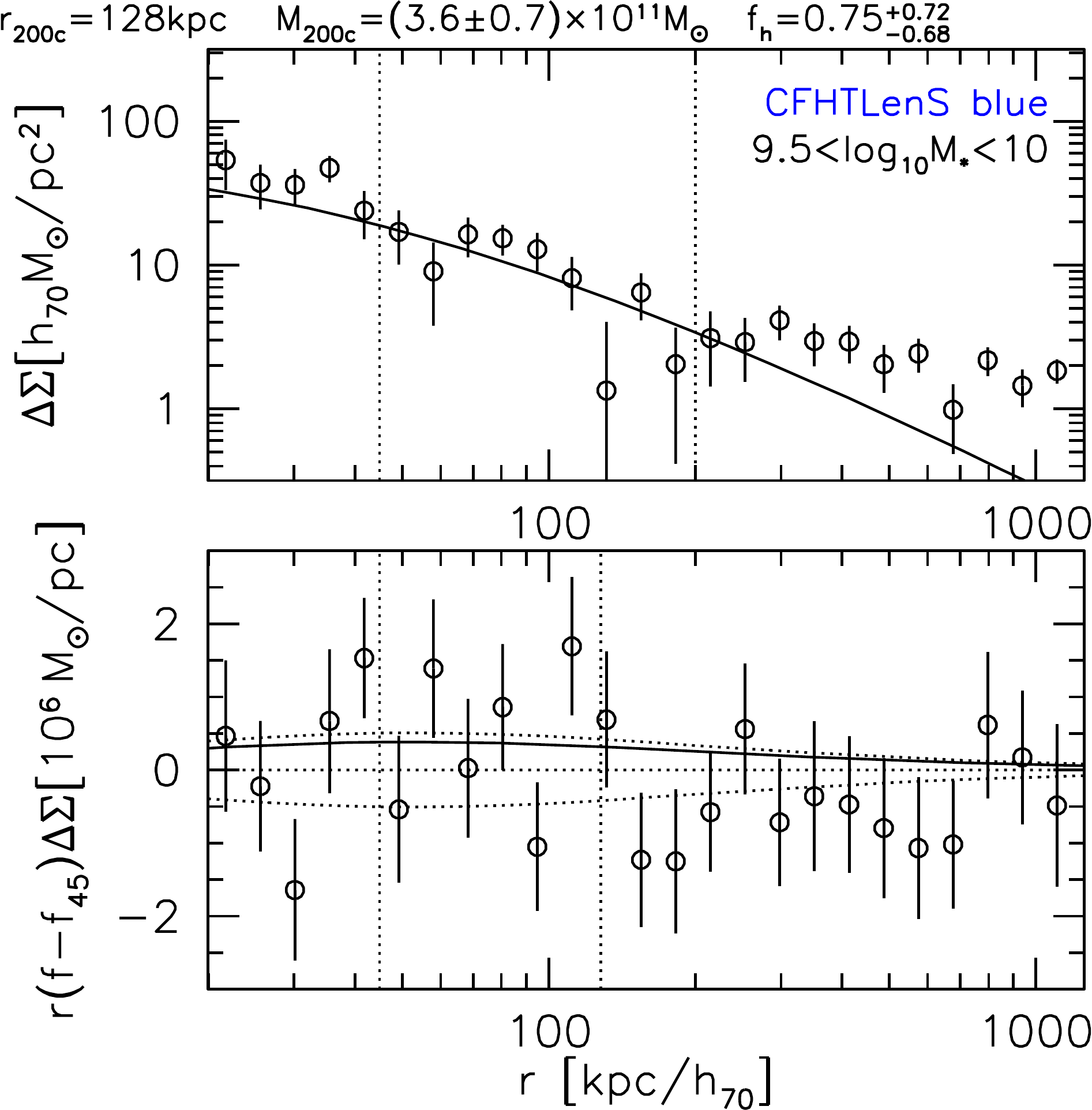}
  \caption{As Fig.\thinspace\ref{fi:shearfitsred_NGVS}, but computed from the CFHTLenS data.
  }
   \label{fi:shearfitsred_cfhtlens}
    \end{figure*}

\begin{figure*}
   \centering
  \includegraphics[width=7.4cm]{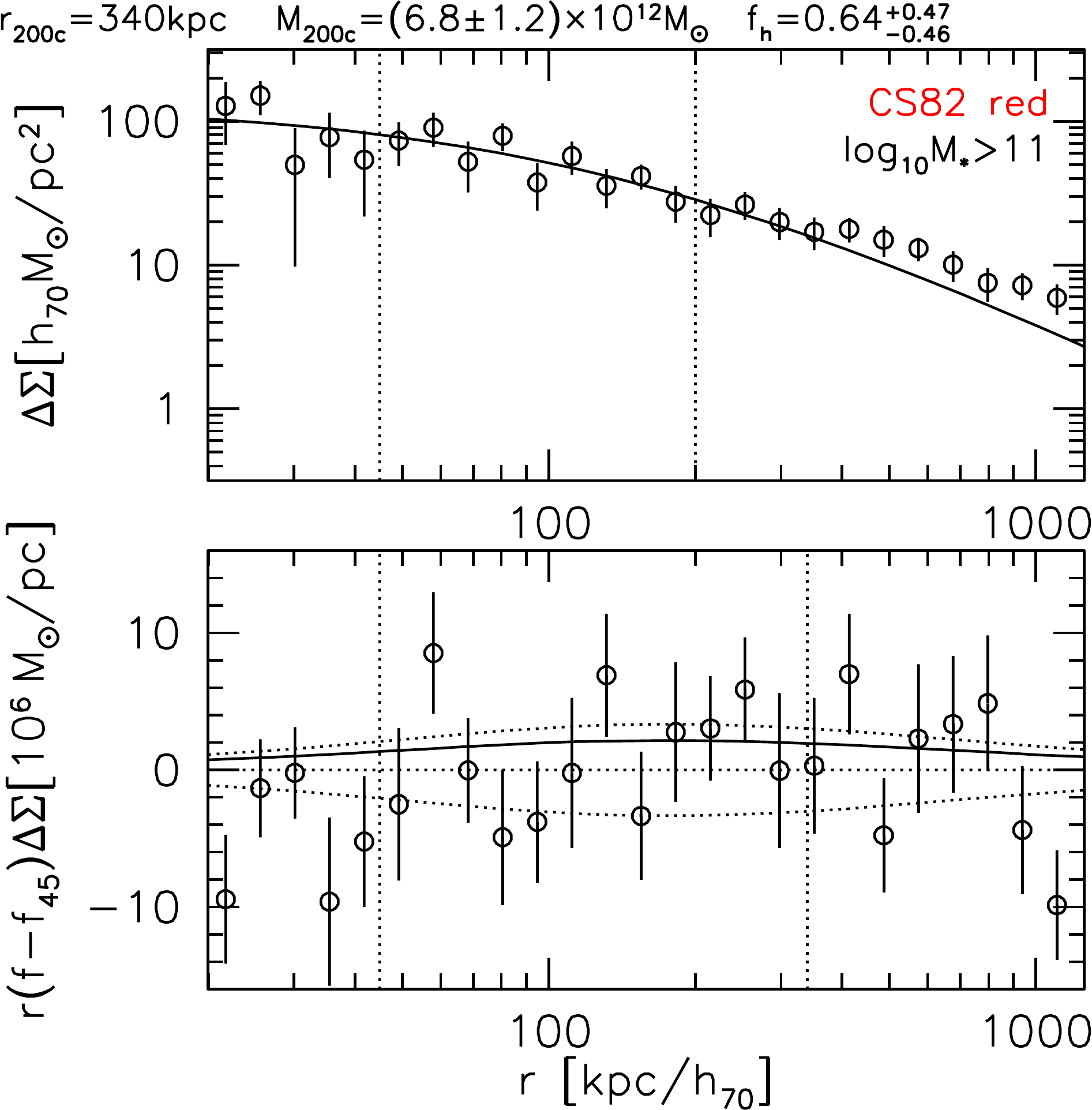}
  \includegraphics[width=7.4cm]{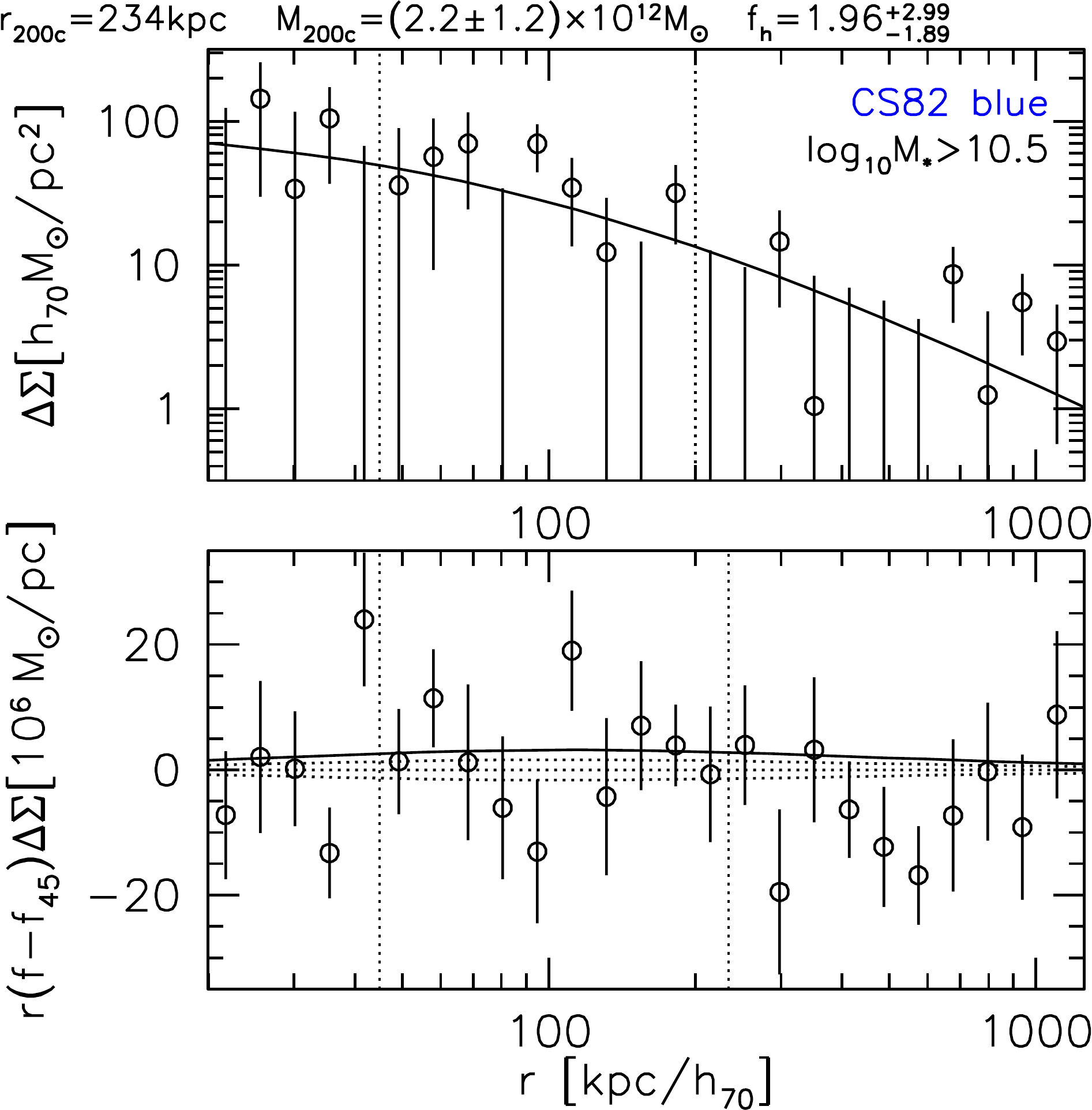}
  \includegraphics[width=7.4cm]{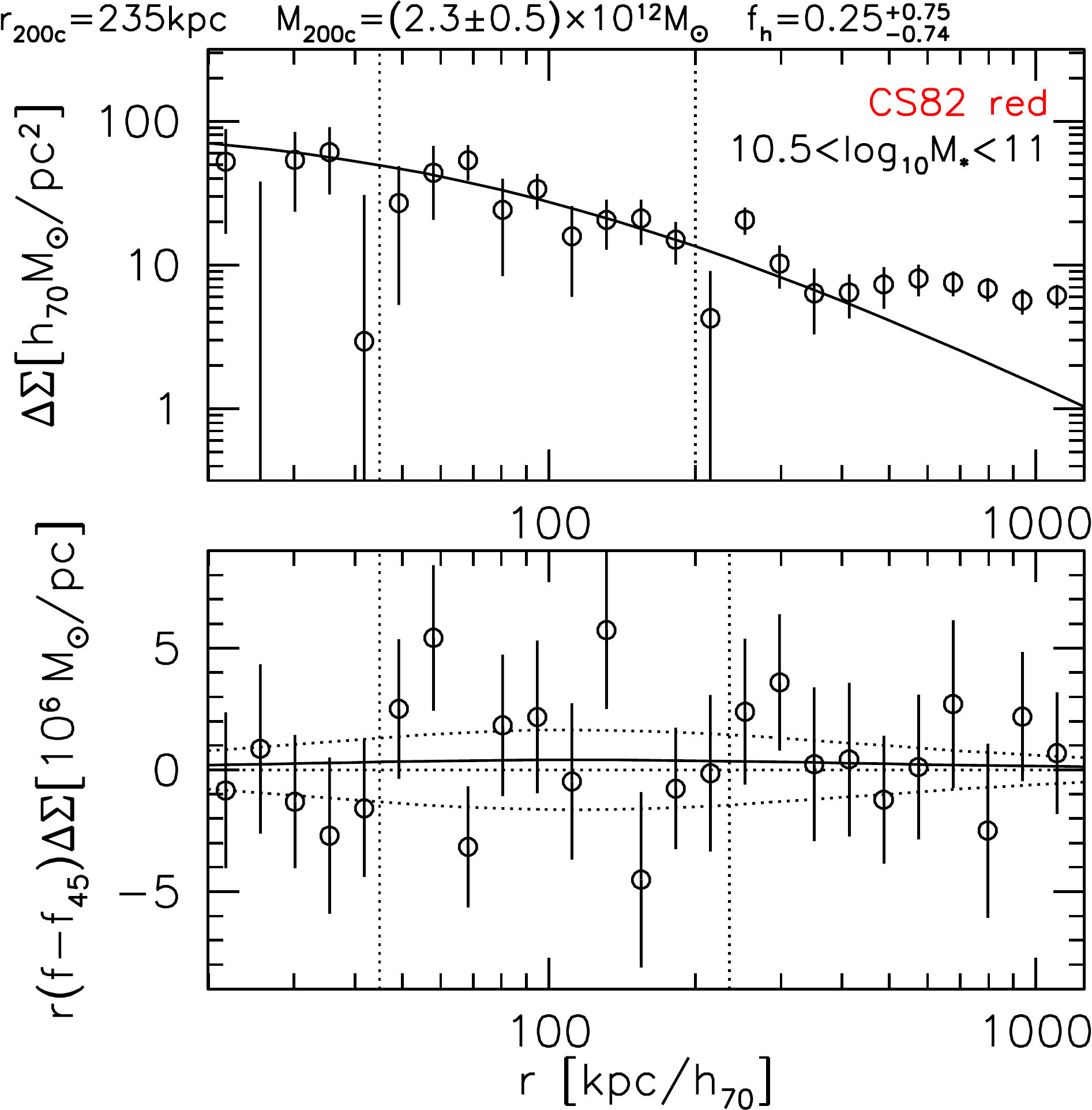}
  \includegraphics[width=7.4cm]{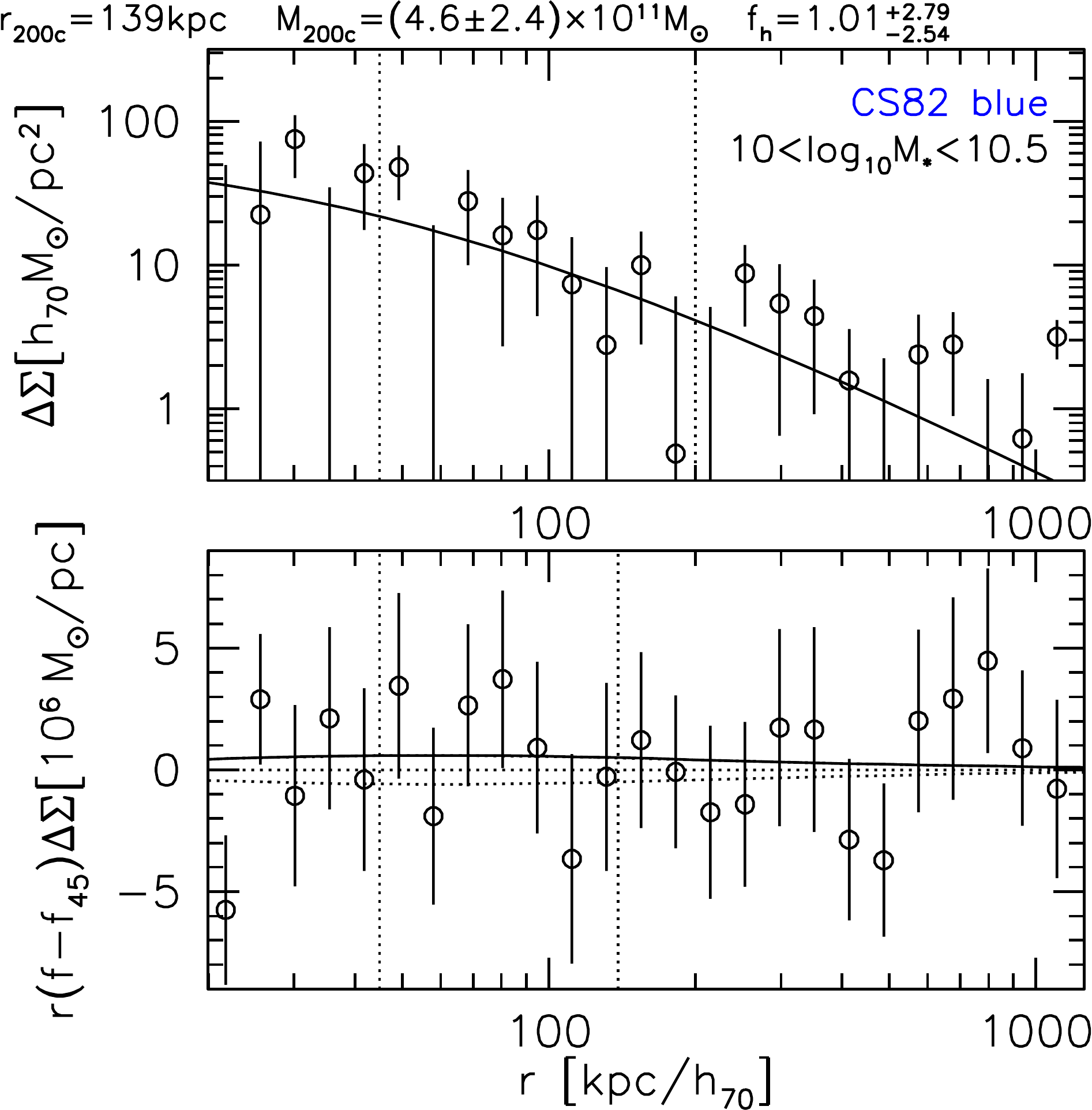}
  \includegraphics[width=7.4cm]{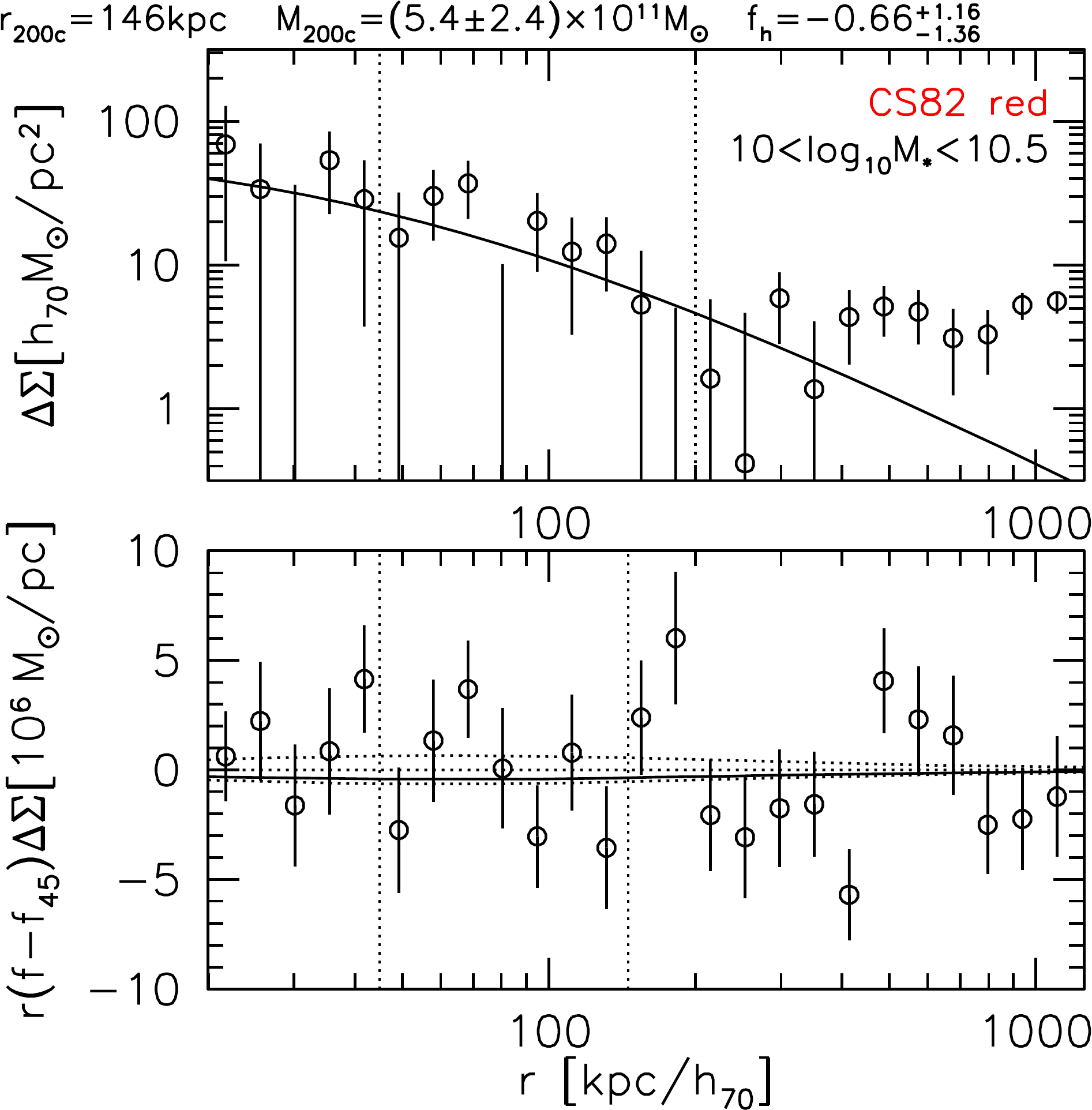}
  \includegraphics[width=7.4cm]{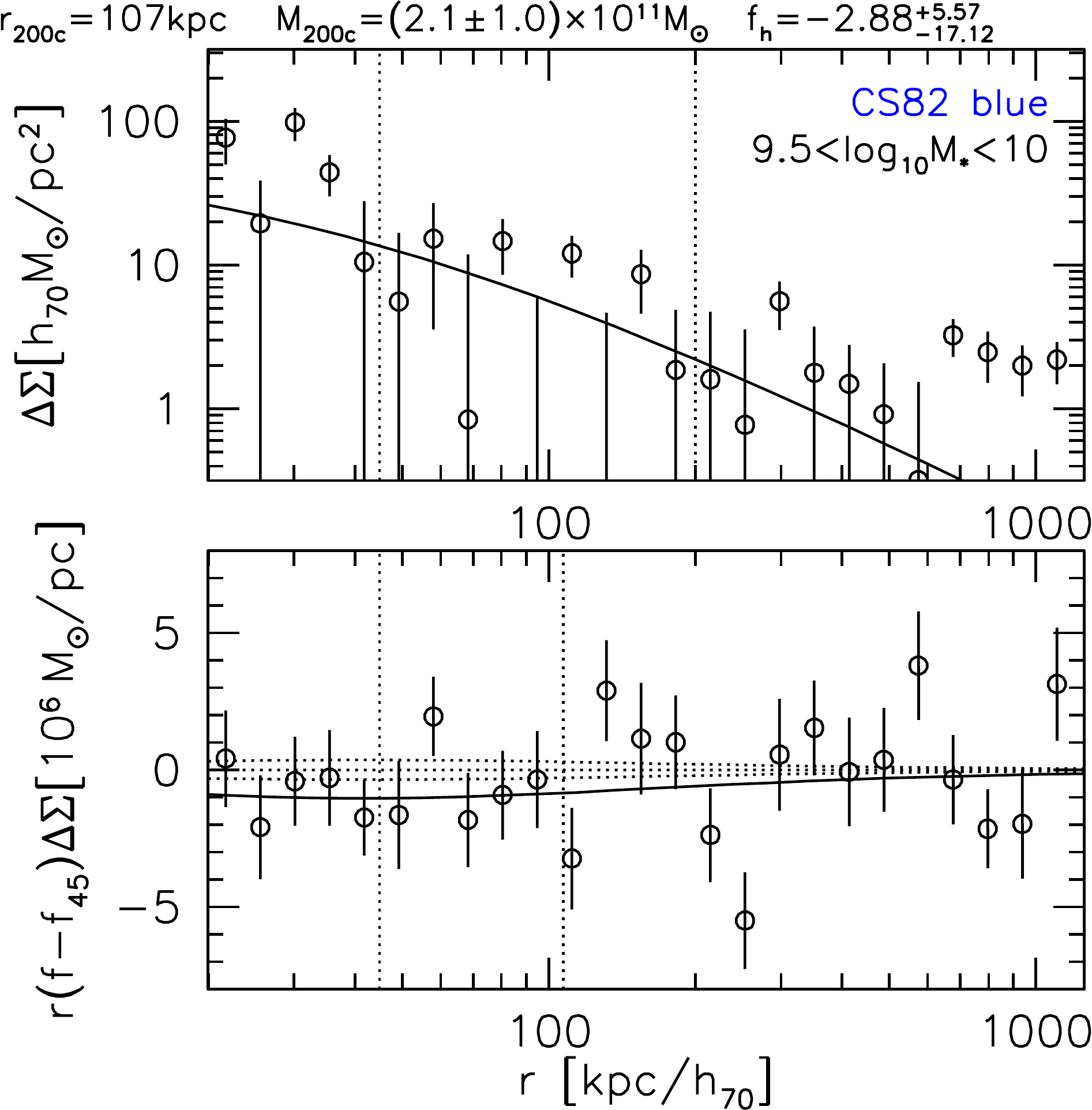}
  \caption{As Fig.\thinspace\ref{fi:shearfitsred_NGVS}, but computed from the CS82 data.
  }
   \label{fi:shearfitsred_CS82}
    \end{figure*}

\begin{figure*}
   \centering
  \includegraphics[width=7.4cm]{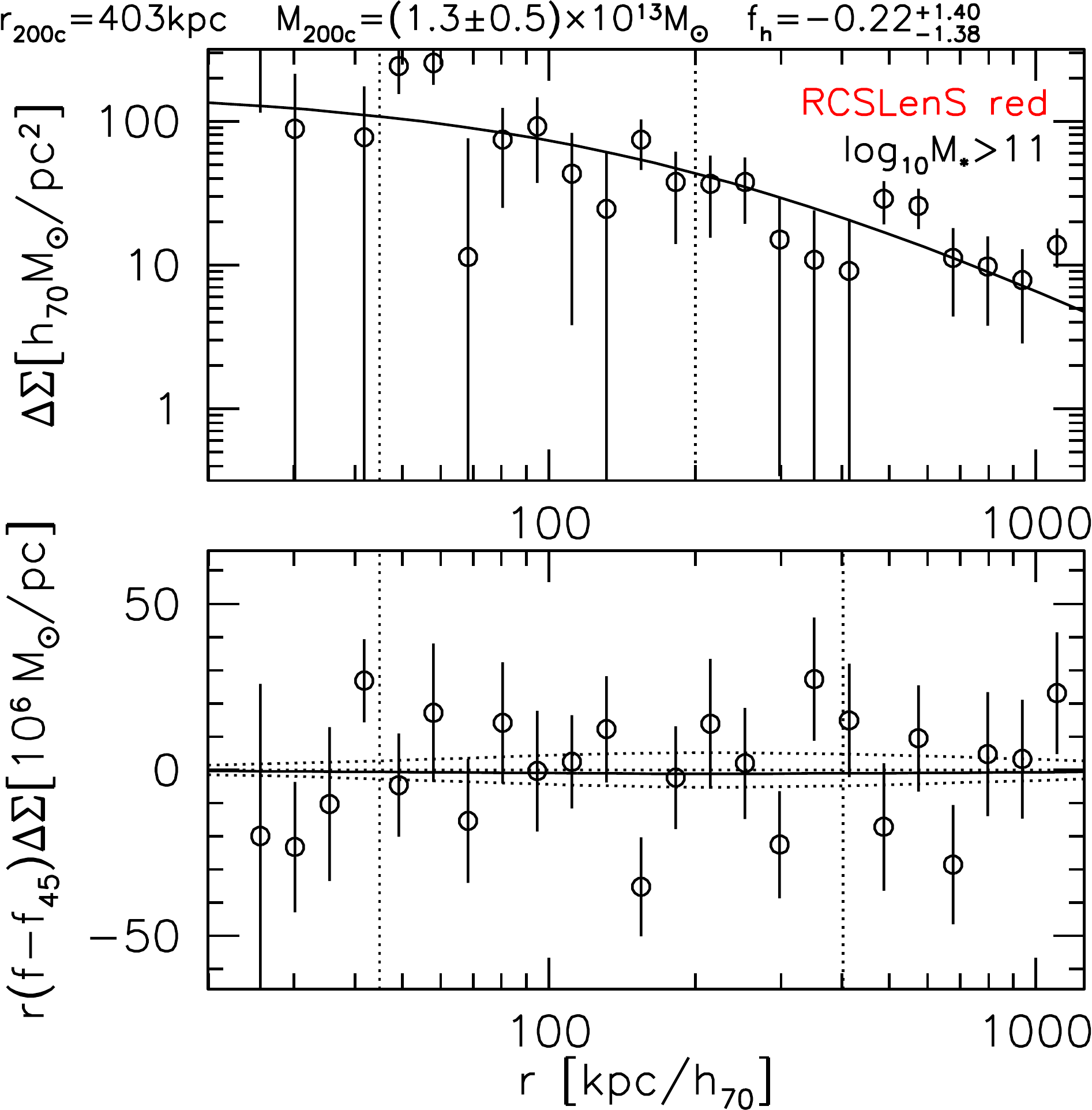}
  \includegraphics[width=7.4cm]{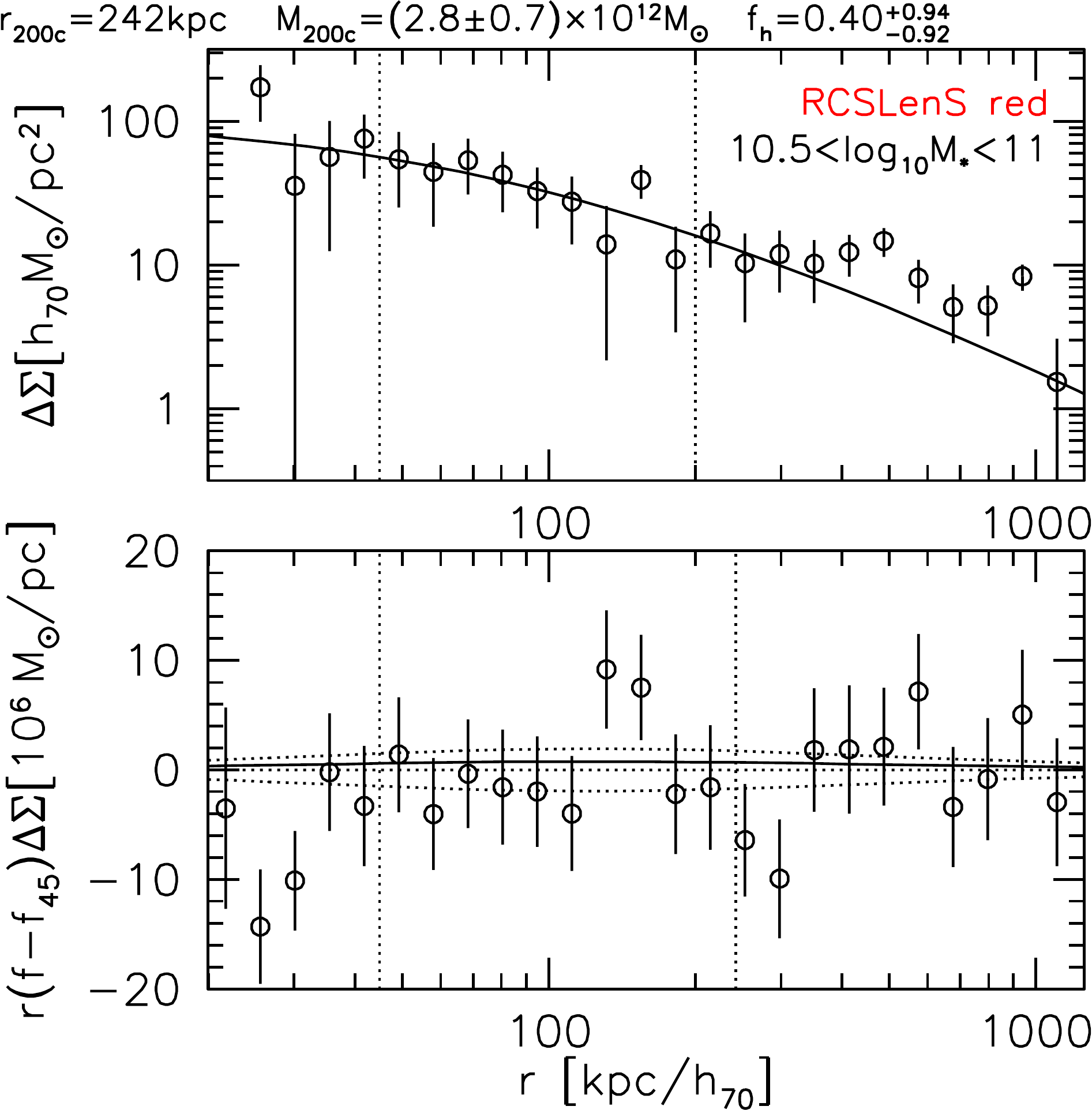}
  \includegraphics[width=7.4cm]{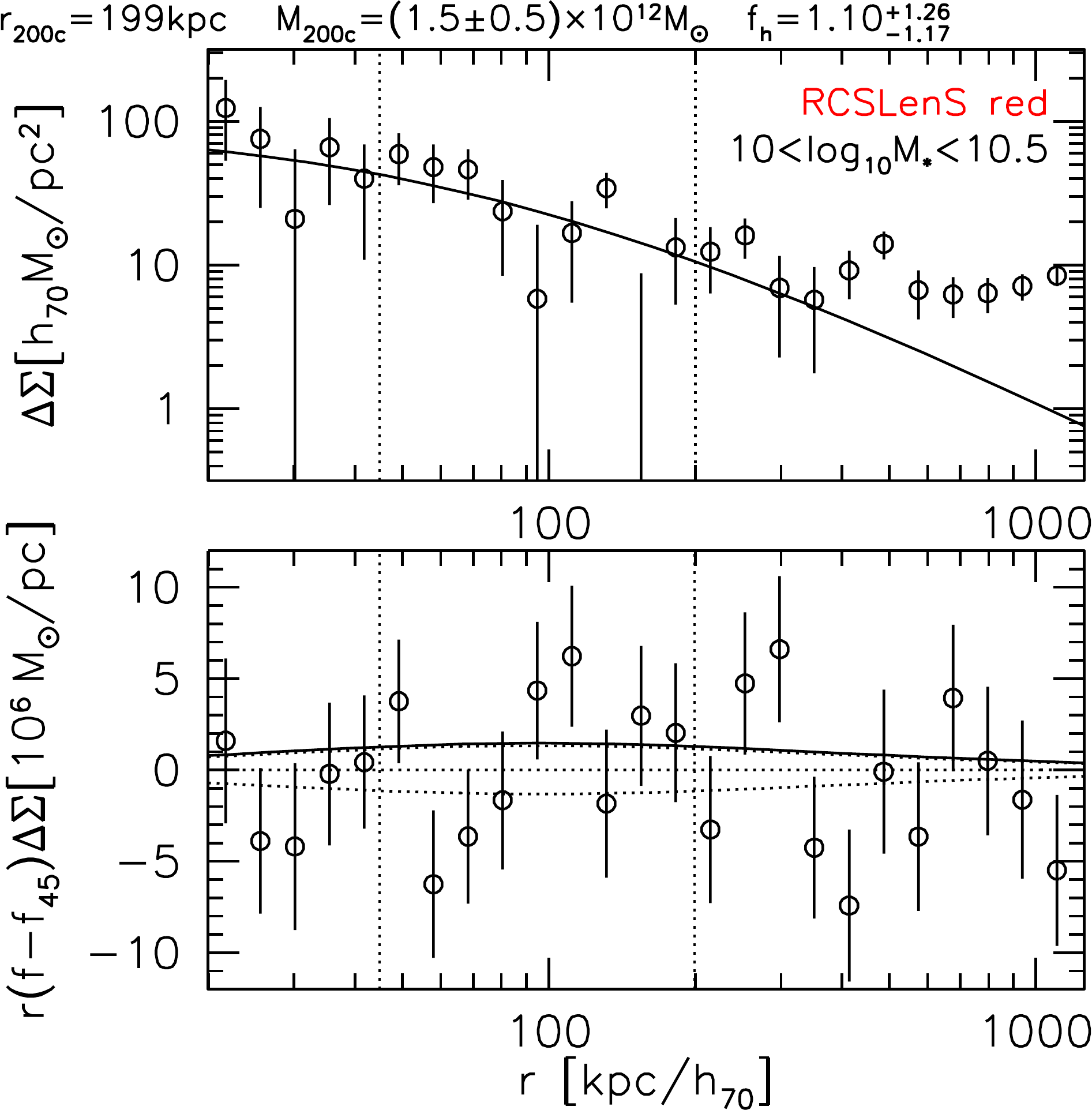}
  \caption{As Fig.\thinspace\ref{fi:shearfitsred_NGVS}, but computed from the RCSLenS data and showing only red lenses.
  }
   \label{fi:shearfitsred_RCS2}
    \end{figure*}

We show the  measured isotropic $\widehat{\Delta \Sigma}(r)$  and anisotropic  $\widehat{\left[f(r)-f_{45}(r)\right]\Delta
  \Sigma_\mathrm{iso}(r)}$ profiles for CS82, KV450, CFHTLenS, and RCSLenS in Figures \ref{fi:shearfitsred_KV450} to \ref{fi:shearfitsred_RCS2}.
The constraints for CFHTLenS differ slightly (well within the error-bars) from  the results reported by \citetalias{shh15}, which is caused by our inclusion of source galaxies with \mbox{$z_\mathrm{b}>1.3$}.

\end{document}